\documentclass[manuscript,authorversion]{acmart}
\usepackage{tabularray}
\pdfoutput=1
\AtBeginDocument{%
  }

\newcommand{\para}[1]{\vspace{3pt} 
\noindent {\em #1}}

\setcopyright{acmlicensed}
\copyrightyear{2018}
\acmDOI{XXXXXXX.XXXXXXX}

\usepackage{paralist}
\usepackage{soul}
\setcopyright{none}
\renewcommand\footnotetextcopyrightpermission[1]{}

\acmJournal{JACM}
\acmVolume{37}
\acmNumber{4}
\acmArticle{111}
\acmMonth{8}

\begin{document}
\title{{\em AI Automatons}: AI Systems Intended to Imitate Humans}

\author{Alexandra Olteanu}
\affiliation{%
 \institution{Microsoft Research}
 \country{Canada}}
\email{alexandra.olteanu@microsoft.com}
\authornote{Corresponding author.}

\author{Solon Barocas}
\affiliation{%
  \institution{Microsoft Research}
  \country{US}
}
\authornote{Key contributors in alphabetical order.}

\author{Su Lin Blodgett}
\affiliation{%
  \institution{Microsoft Research}
  \country{Canada}
}
\authornotemark[2]

\author{Lisa Egede}
\affiliation{
\institution{Carnegie Mellon University}
\country{US}}
\authornotemark[2]

\author{Alicia DeVrio}
\affiliation{
\institution{Carnegie Mellon University}
\country{US}}

\author{Myra Cheng} 
\affiliation{
\institution{Stanford University} 
\country{US}}

\renewcommand{\shortauthors}{Olteanu et al.}

\begin{abstract}
  There is a growing proliferation of AI systems designed to mimic people's behavior, work, abilities, likenesses, or humanness---systems we dub {\em AI automatons}. 
  Individuals, groups, or generic humans are being simulated to produce creative work in their styles, to respond to surveys in their places, to probe how they would use a new system before deployment, to provide users with assistance and companionship, and to anticipate their possible future behavior and interactions with others, just to name a few applications. 
  The research, design, deployment, and availability of such AI systems have, however, also prompted growing concerns about a wide range of possible legal, ethical, and other social impacts. 
  To both 1)~facilitate productive discussions about {\em whether}, {\em when}, and {\em how} to design and deploy such systems, and 2)~chart the current landscape of existing and prospective {\em AI automatons}, we need to tease apart determinant design axes and considerations that can aid our understanding of whether and how various design choices along these axes could {\em mitigate}---or instead {\em exacerbate}---potential adverse impacts that the development and use of {\em AI automatons} could give rise to.
  In this paper, through a synthesis of related literature and extensive examples of existing AI systems intended to mimic humans, we develop a conceptual framework to help foreground key axes of design variations and provide analytical scaffolding to foster greater recognition of the design choices available to developers, as well as the possible ethical implications these choices might have.
\end{abstract}

\maketitle

\section{Introduction}

There is a fast-growing number of examples where AI systems are developed or used to imitate humans. 
These include cases where the likenesses and voices of deceased or missing children were reenacted to help narrate their stories of abuse and violence~\cite{hutiri2024not}, such as that of a ``17-month-old [who] died in 2007 following months of physical abuse''~\cite{hassan2023ai}, or cases of AI characters meant to depict made-up intersectional identities, such as ```Liv' portraying a `proud Black queer momma of 2 \& truth-teller'\hphantom{}''~\cite{nbc2025Meta}. There is also ``[a]n array of popular apps [that] are offering AI companions [...] who are spinning up AI girlfriends, AI husbands, AI therapists---even AI parents''~\cite{washingtonpost2024}. Some developers seek to mimic specific individuals~\cite{mcilroy2022mimetic,lee2023speculating,leong2024dittos}, while others aim to imbue systems with more general human-like characteristics~\cite[e.g.,][]{kang2024nadine,strohmann2023toward}.\looseness=-1 

Indeed, growing beliefs that AI systems will be or are already human-like and able to replicate a wide range of human abilities or likenesses~\cite[e.g.,][]{fronsdal2024misr,cheng2024one,devrio2025taxonomy,salon2025AInurses,gratch2002creating,guardian2025AIempathy}---AI systems we dub {\em AI automatons}\footnote{An {\em AI automaton} is an AI system or model that ``is relatively self-operating'' or that is ``designed to follow automatically a predetermined sequence of operations or respond to encoded instructions''~\cite{automaton} in order to mimic humans or some of their characteristics.} to emphasize their {\em mechanical nature}---have both led to growing investment and interest in developing such systems~\cite[e.g.,][]{hofman2023steroids,washingtonpost2024,chandra2024lived,volante2016effects,sinatra2021social}, as well as growing concerns about these systems potentially replacing humans in various jobs, about the potential emotional toll from interacting with human-like systems, or about adverse impacts to humans in a wide range of other more or less anticipated ways~\cite[e.g.,][]{agnew2024illusion,byun2023dispensing,abercrombie2023mirages,kapania2024simulacrum,lee2023speculating,boine2023emotional,morris2024generative,chandra2024lived,ftc,hutiri2024not,washingtonpost2024,independent2025AIredline}.  
This trend is far from new, with the field of AI itself guided by the question of whether computational  systems are capable of thought or at least of faithfully imitating humans---also known as Turing's {\em imitation game}~\cite{gunderson1964imitation}---an aim which also has a long history in fields like gaming~\cite[e.g.,][]{maes1995artificial,laird2000creating,alexander2005gaming} and human-robot interaction~\cite[e.g.,][]{dautenhahn2007socially,fong2003survey}.\looseness=-1

What is perhaps {\em new} is the increasing feasibility and availability of AI systems that could be used to simulate specific individuals or highly realistic human-like entities that are able to engage in autonomous, open-ended interactions with others, 
coupled with a growing ubiquity of such systems across a wider and wider range of applications, 
both of which are propelled by expanding and increasingly pervasive claims about, and perceptions of, AI systems' capabilities. 
The growing possibility of developing highly accurate simulations of individuals, in particular, has led to recent examinations of their ethical and social implications~\cite[e.g.,][]{mcilroy2022mimetic,lee2023speculating,hutiri2024not}, and of the possibility of these simulations being used to replace humans in certain settings~\cite[e.g.,][]{agnew2024illusion,wang2024large,wu2023llms}. 
While critically important, these early efforts focus on either a specific class of AI automatons or a specific class of concerns, and do not examine how various design choices or aspects of these systems might amplify different types of concerns---for which we seek to provide an analytical foundation in this paper.\looseness=-1 

Our aims are two-fold: 
1) chart the current landscape of existing and prospective {\em AI automatons}, and 
2) provide analytical scaffolding for discussions about {\em whether}, {\em when}, and {\em how} to design and deploy such systems.
For this we developed a conceptual framework to help map key design considerations and distinctions when building and deploying various types of AI automatons---i.e., AI systems developed or believed to appear or act like humans---illustrating how these considerations can govern people's perceptions of and concerns about these systems~(\S\ref{framework-description}). 
In developing this framework, we focus on the {\em intended} goals for such AI automatons, rather than attempting to make claims or speculate about what these systems can or cannot do. In other words, we focus on what they are developed, deployed, or intended to be used for.
We do so to provide analytical structure to situate and understand various efforts to simulate or mimic humans, and help tease apart how differences in the design of these systems might either exacerbate or help mitigate growing concerns about the development, deployment, and use of AI automatons.


\section{Background \& Related Work}

AI systems are increasingly anthropomorphic~\cite{kuhne2023anthropomorphism,cheng2024one,devrio2025taxonomy}---i.e., described or perceived as human-like. Anthropomorphism can be by design, often by incorporating human-like features into systems (e.g., avatars with different skin colors or gendered hairstyles~\cite{harrington2023trust}, social robots with facial features~\cite{song2020trust} or producing human sounds and gestures~\cite{liu2023robots}).
Such design choices may be motivated by desires to increase users' engagement, comfort, familiarity, and trust~\cite{kang2024counseling,rossi2018socially,troshani2021we,renzullo2019anthropomorphized,wsj2025AIschoolcounselor}, improve user experiences~\cite{hernandez2023affective,wu2023deep}, or encourage consumer engagement~\cite{nguyen2023chatbots} and consumption~\cite{hamilton2021traveling,han2021impact,nicolescu2022human}, though anthropomorphization can also backfire~\cite{mende2019service}. 
AI systems may also be anthropomorphic even when not intentionally designed for. For example, language use, until relatively recently solely a human activity and made possible for AI systems by training on large quantities of human-produced language, can readily give rise to perceptions of human-likeness~\cite{devrio2025taxonomy}; \citet{devrio2025taxonomy} offer a taxonomy of linguistic expressions that can contribute to perceptions of human-likeness, such as expressions of identity, self-awareness, or emotion.\looseness=-1

\subsection{Existing \& Emerging Applications Simulating Humans}

\subsubsection{Simulating individuals.} 
Simulations have been developed to target a wide array of individuals, including models personalized to specific chess players~\cite{mcilroy2022learning} to predict their next moves, models simulating individual Supreme Court justices to predict future decisions~\cite{hamilton2023blind}, and models simulating users for sending emails, ``match[ing] the voice and tone in the emails you've already sent''
~\cite{SuperhumanAI}. 
Simulations may target people no longer alive, including loved ones as well as public figures~\cite{hutson2023life,morris2024generative}. Emerging applications also include those aimed at simulating many individuals at once, for example as participants for pilot studies~\cite{rothschild2024opportunities} and simulated polls~\cite{zhang2024electionsim}. 
\citet{park2024generative} simulate ``attitudes and behaviors of 1,052 real individuals.''
Other simulations may target fictional individuals---e.g., models role-playing as specific characters~\cite{wang-etal-2024-rolellm}---as well as entirely new characters, ``offering AI companions to millions of [...] users''~\cite{washingtonpost2024}.\looseness=-1

\subsubsection{Simulating groups.} Simulations may also target members of social groups, ranging from social or professional roles to demographic groups. For example,  \citet{qian-etal-2024-chatdev} develop agents in roles such as programmers and test engineers for software development, and \citet{sun2024lawluo} develop a legal consultation system with agents in roles as a receptionist, lawyer, secretary, and boss. \citet{argyle2023out} construct prompts with demographic information to investigate the degree to which a model reflects response distributions (for e.g., surveys) ``each aligned with a real human sub-population,'' while \citet{lee2024can} investigate whether LLMs conditioned on demographics can simulate responses to climate change surveys.
Other simulations of members of demographic groups include AI social media accounts, such as
``Grandpa Brian,'' a Meta account which ``described itself... as an African-American retired entrepreneur'' and whose bio is an ``entirely fictionalized biography based on a composite of real African American elders' lives''~\cite{meta2025}.\looseness=-1

\subsubsection{Simulating human phenomena \& interactions.} An increasing number of applications seek to simulate people in order to study human phenomena, including people's beliefs and attitudes~\cite{park2024generative}; social dynamics and interactions in simulated communities and social networks~\cite{gao2023s,ren2024emergence,tornberg2023simulating,yu2024researchtown,wang2024decoding}; and human decision-making across a range of settings, such as resource allocation~\cite{ji-etal-2024-srap} and government responses to public disaster~\cite{xiao2023simulating}. See \citet{mou2024individual} for a survey.\looseness=-1

\subsection{Growing Concerns}

As AI systems that are anthropomorphic or simulate humans have proliferated, work is also emerging interrogating and raising critical concerns about them. Scholars have long problematized ``relational artifacts [...] \emph{specifically designed to make people feel understood}, [artifacts that] are still without understanding''~\cite[][emphasis original]{turkle2007authenticity}, as they lack the human understanding required of the relationships that people try to create with them. 

Moreover, encouraging users to relate to AI systems as if they are human can also cause over-reliance on system outputs~\cite{abercrombie2023mirages}---leading to exaggerated perceptions of AI capabilities and performance as well as distorting moral judgments about responsibility \cite{placani2024anthropomorphism,shanahan2024talking}---and negatively impact critical thinking abilities~\cite{gerlich2025ai}. Such systems can also prevent ``users from assuming certain roles themselves, or [...] from questioning the need for certain roles in the first place''~\cite{maeda2024human}. 
In the long term, anthropomorphic systems may lead to long-term emotional dependence~\cite{gabriel2024ethics}, and their capacity to express feelings they cannot have may devalue expressions of genuine emotion and erode our collective ability to bond with each other~\cite{porra2020can}. Finally, anthropomorphic systems' perceived trustworthiness~\cite{araujo2018living}---potentially heightened by misuse of users' personal information~\cite{maeda2024human}---may enable increased user deception, manipulation, and exploitation~\cite{weidinger2022taxonomy}, and combined with their increased ubiquity risks gradual acclimatization that may facilitate surveillance capitalism and increase public acceptance of potentially unethical uses of AI, such as military use~\cite{renzullo2019anthropomorphized}.\looseness=-1

As they are designed to imitate humans, AI automatons are likely to be perceived as human-like and thus give rise to the same concerns as anthropomorphic systems more broadly. However, as systems explicitly designed to simulate people's behavior, work, abilities, or likenesses, they can give rise to additional concerns. Some of these concerns involve practical challenges; for example, \citet{agnew2024illusion} identify limitations of current systems aimed at replacing people, including their tendency to make mistakes and their propensity for reproducing dominant perspectives rather than those of the people they are intended to replace. \citet{wang2024large} argue that LLMs' tendencies to \emph{misportray} demographic groups (i.e., generate out-group rather than in-group members' perspectives) and \emph{flatten} groups (i.e., treat groups as monoliths, erasing heterogeneity and neglecting intersectional identities) make them unsuitable as replacements for human participants.\looseness=-1

Beyond these concerns---which could conceivably be overcome with modeling advancements---these and other works have identified a number of serious concerns fundamental to the act of simulating people. \citet{agnew2024illusion} and \citet{wang2024large} note that replacing human studies participants reproduces histories of minoritized groups' exclusion from decision-making and moves away from the meaningful sharing of power that is core to many visions of inclusion. \citet{wang2024large} point out that such simulations also risk essentializing identity by treating identities as ``rigid and innate.''
Addressing simulations of individuals in particular, \citet{mcilroy2022mimetic} propose a framework for characterizing normative concerns arising from \emph{mimetic models}---generative and interactive models simulating specific people. 
\citet{lee2023speculating} investigate users' and targets' perceptions of simulations (\emph{AI clones}), 
with participants raising concerns ranging from misrepresentation to replacement and exploitation. Finally, \citet{erscoi2023pygmalion} conceptualize \emph{Pygmalion displacement}---a ``pattern of sexist harm, whereby the humanization of AI and the dehumanization of women/the feminised go hand in hand.''\looseness=-1

\subsection{Concepts \& Working Terminology}

Drawing on~\citet{mcilroy2022mimetic} and others who have used similar terminologies to differentiate between different stakeholders~\cite{lee2023speculating,leong2024dittos,hutiri2024not}, our framework distinguishes between different types of stakeholders, including:\footnote{As opposed to \cite{mcilroy2022mimetic}, in our framework we do not distinguish between the builders of the systems and the operators of the systems, and by definition, spectators are different than operators and interactors. We also use a more expansive definition for the target, which does not need to be an individual.} 
\begin{inparaenum}
    \item {\em target} -- an individual, a group, a persona, or a generic human whose identity, characteristics, or behavior are being simulated; 
    \item {\em interactor} -- someone who is directly interacting with the simulation of the target;
    \item {\em spectator} -- someone who is observing the simulation of the target without directly interacting with it; 
    \item {\em operator} -- someone involved in the building, deployment, or operation of the system simulating the target.
\end{inparaenum}


\begin{table*}
    \scriptsize
    \def\arraystretch{0.4}
    \setlength{\tabcolsep}{0.23em}
    \centering
    \begin{tabular}{@{}p{1.55cm}p{8.4cm}p{5cm}@{}} 
        {\bf Dimension} & {\bf Description} & {\bf Examples} \\ \hline

        \multicolumn{3}{@{}l}{\bf Scope: what is the subject of the simulation?} \\ 
        Target & Who is being simulated? &  specific individuals~\cite{mcilroy2022mimetic}; identity groups~\cite{wang2024large,cheng2023compost}; subpopulation~\cite{argyle2023out}; human-like AI or chatbots~\cite{milani2023navigates,washingtonpost2024} \\ \cmidrule{3-3}
            & -- \underline{Fictitious}: when the target or their characteristics are fictional, or not true or real & counterfeit people~\cite{atlantic2024,favela2023ethics};  AI character profiles~\cite{404media2024}; AI generated users~\cite{financialtimes2024}; customized AI family~\cite{FamiliaAI} \\ \cmidrule{3-3}
            & -- \underline{Real}: when the target or their characteristics represent an actual, real person or group & scam victims~\cite{cbsnews2024-VoiceScam}; self~\cite{zheng2023self}; real individuals~\cite{park2024generative} \\
        \cmidrule{2-3}
        
        Characteristics & What about the target is being simulated? & \\ 
                        & -- \underline{Form}: the appearance  and style of the target & style transfer~\cite{cifka2020groove2groove}; human-like typos~\cite{newscientist2021}; childlike~\cite{hassan2023ai} \\ \cmidrule{3-3}
                        & -- \underline{Content}: what the target might say or do & video game navigation~\cite{milani2023navigates}; entire personality~\cite{techradar2025} \\ \cmidrule{3-3}
        
        Fidelity & How well or faithfully is the simulation intended to capture the target's characteristics? & algorithmic fidelity~\cite{argyle2023out}; precision or fidelity of a mimetic model~\cite{mcilroy2022mimetic}, AI clone~\cite{lee2023speculating}, or AI avatar~\cite{cao2023high} \\ \cmidrule{3-3}

        Specificity & To what degree the simulated characteristics are unique to the target and can identify the target? & accounting for unique, individualized needs, sensitivities, and characteristics~\cite{zheng2023self}\\ \cmidrule{3-3}

        Completeness & To what degree is the simulation intended to capture the target fully or in it's entirety? & only conversational mannerisms~\cite{abbasiantaeb2024let}; virtual~doppleganger~\cite{lucas2016effect}; digital~twins~\cite{techcrunch2025} \\ \cmidrule{3-3}

        Humanness & To what degree is the simulation intended to capture human-like characteristics? & communicate sounds like humans~\cite{mitnews2025}; humanizing AI~\cite{venturebeat2024-humanizing}; ability to convey human emotions~\cite{cnbc2024} \\ \cmidrule{3-3}

        Adaptability & To what degree is the simulation intended to evolve or adapt? & adaptability, versatility~\cite{Mitchell2025AI}; static clones~\cite{lee2023speculating} \\
        
        \specialrule{0.06em}{2pt}{-0.01em}
        \multicolumn{3}{@{}l}{\bf Intended Uses \& Goals: How is the simulation intended to be used?} \\ 
        To replace & Is the simulation intended to replace the target? For what purpose? & replace humans~\cite{clarke2024llm}; agents of replacement~\cite{gray2023psychology} \\  \cmidrule{3-3}
            & -- \underline{Relieve}: 
            relieve the target from drudgery or possible harm & automate drudgery~\cite{price2019artificial}; overtake meaningless jobs~\cite{nytimes2024meaningless} \\ \cmidrule{3-3}
            & -- \underline{Substitute}: 
            be a stand in or surrogate for the target when the target wants to delegate a task &  Dittos~\cite{leong2024dittos}; AI to talk to his wife for him~\cite{techcrunch2024} \\ \cmidrule{3-3}
            & -- \underline{Displace}: 
            take over the place, position, or role of the target to their detriment & replace human respondents in surveys~\cite{argyle2023out,agnew2024illusion}; AI employees~\cite{gizmodo2024};  professional healthcare job loss~\cite{zhan2024healthcare} \\ 
            \specialrule{0.06em}{2pt}{0.01em}

        To interact & Is the simulation intended to be interacted with? & interactive virtual humans~\cite{gratch2002creating} or deepfakes~\cite{horvitz2022horizon} \\ \cmidrule{3-3}
        
        \quad\quad -- {\em Modes} & How can the simulation be interacted with? & short vs. long-term~\cite{folstad2019different}; open-ended vs. structured~\cite{guo2024investigating} \\ \cmidrule{3-3}
        
        \quad\quad -- {\em Stakes} &  What the simulation is intended to do for the {\em interactor}? &  \\ 
            & \quad -- \underline{Enhance}: 
            enhance the interactor's current ability to carry out a task 
            & AI as steroids or sneakers~\cite{hofman2023steroids}; AI-assisted writing~\cite{biermann2022tool,hwang202480} \\ \cmidrule{3-3}
            
            & \quad -- \underline{Coach}: 
            help the interactor learn or improve skills & AI as coach~\cite{hofman2023steroids}; machine advisor~\cite{prahl2021out}; simulated practice partners~\cite{louie2024roleplay}; pedagogical agent~\cite{sinatra2021social} \\ \cmidrule{3-3}
            
            & \quad -- \underline{Serve}: 
            provide a service to the interactor & AI nurse/therapist~\cite{chavali2024ai,salon2025AInurses,kang2024counseling}; headshot generator~\cite{huffingtonpost2024}\looseness=-1 \\ \cmidrule{3-3}
            
            & \quad -- \underline{Accommodate}: 
            adapt to the interactor's needs or characteristics
            & ``adapt the agent’s demeanor''~\cite{leong2024dittos} \\ \cmidrule{3-3}
            
            & \quad -- \underline{Entertain}: entertain the interactor  & ``designed to make you laugh''~\cite{petapixel2024}
            \\ \cmidrule{3-3}
            
            & \quad -- \underline{Connect}: 
            provide social or emotional support to interactors & companionable agents~\cite{sidner2018creating}\\ \cmidrule{3-3}

            & \quad -- \underline{Collaborate}: act as a collaborator for the interactor &  AI and machine teammates~\cite{seeber2020collaborating,zhang2021ideal,zou2025rise}; co-pilot~\cite{sellen2024rise}\\ \cmidrule{3-3}
            
            & \quad -- \underline{Evaluate}: 
            evaluate or assess the interactor, without the goal of helping them improve  &  AI interviewer~\cite{forbes2024interviewer,theguardian2024Job} \\ 
            \cmidrule{3-3}
        
        \quad\quad -- {\em Affinity} &  
                    How similar to the {\em interactor} is the target intended to be? & affinity~\cite{cao2023high}; self-congruity~\cite{suh2011if}; self-dialogue~\cite{slater2019experimental}
                    \\ \cmidrule{1-3}

        To showcase & Is the simulation intended to be observed? & AI-made ads~\cite{nbcnews2024}; synthetic media~\cite{hassan2023ai} \\ \cmidrule{3-3}
        \quad\quad -- {\em Stakes} &  What the simulation is intended to do for the {\em spectator}? &  simulated voice narrating a story~\cite{goodereader2023}\\ \cmidrule{3-3}
        
        \quad\quad -- {\em Affinity} &  
                    How similar to the {\em spectator} is the target intended to be? &  observe self-clones~\cite{zheng2023self}
                    \\ \cmidrule{1-3}

        To study & Is the primary goal of the simulation to study a phenomenon or system performance? & \\
            & \quad -- Study human behavior & simulate social networks to study polarization~\cite{wang2024decoding}\\ \cmidrule{3-3}
            & \quad -- Study the simulations & ability to simulate research communities~\cite{yu2024researchtown}\\ \cmidrule{3-3}
            & \quad -- Study model/system performance & simulate users to test a product~\cite{ataei2024elicitron} \\ 

        \specialrule{0.06em}{2pt}{-0.01em}
        \multicolumn{3}{@{}l}{\bf Ownership \& Control: Who makes decisions about the simulations?} \\ 
        
        Who is in control? &  This can include the targets, the interactors, the spectators, the operators, or other 3rd parties & user-generated~\cite{washingtonpost2024}; developers~\cite{agnew2024illusion}; stakeholders~\cite{lee2023speculating} \\ \cmidrule{2-3}
        
        Type of control & What do they have control over? & \\ 
            & \quad -- Over who is being simulated & user-controlled representations of self~\cite{harrell2017reimagining} \\ \cmidrule{3-3}
            
            & \quad -- Over what about the target is being simulated & control over simulating identifying attributes~\cite{rothman2018right} 
             \\ \cmidrule{3-3}

            & \quad -- Over how the simulation is developed & chatbot trained on one's own diary entries~\cite{huang2022chatbot}\looseness=-1\\ \cmidrule{3-3}
            
            & \quad -- Over who can interact with the simulation & control over to whom their AI can reply to~\cite{petapixel2024}\\ \cmidrule{3-3}
            
            & \quad -- Over how the simulation can be used & control over downstream uses~\cite{widder2022limits} \\ \cmidrule{2-3}
            
        Degree of control & How much control do they have?  & \\ 
            & \quad -- \underline{No control}: no influence or control at any point in the development/deployment life-cycle & deepfakes of unsuspecting targets~\cite{cbsnews2024} \\ \cmidrule{3-3}
            & \quad -- \underline{Consulted}: some influence, but only at specific points in the development/deployment life-cycle & AI tutor personalized based on discrete feedback~\cite{park2024empowering}\\ \cmidrule{3-3}
            & \quad -- \underline{Included}: some influence at any point in the development/deployment life-cycle & messages sent on the target's behalf~\cite{liu2024leveraging}\\ \cmidrule{3-3}
            & \quad -- \underline{In control}: some control, but only at specific points in the development/deployment life-cycle & animation of historical family photos~\cite{DeepNostalgia}\\ \cmidrule{3-3}
            & \quad -- \underline{Ownership}: full control at any point in the development and deployment life-cycle & AI companion trained on developer's exes' profiles~\cite{katz2024she}\\ 

        \specialrule{0.06em}{2pt}{-0.01em}
        \multicolumn{3}{@{}l}{\bf Impacts: What are the impacts of simulating humans? } \\ 
        Stakeholders & Who is impacted?  & parents of the interactor~\cite{arstechnica2024} or target~\cite{hassan2023ai}; \\ \cmidrule{3-3}
        Adverse impacts & How are they impacted? & physical harm~\cite{arstechnica2024}; perceived substitution risk~\cite{zhan2024healthcare} \\ 
        \bottomrule
    \end{tabular}
\caption{Overview of our conceptual framework. 
The examples either highlight prior work noting the design consideration, or illustrate variation along that dimension.
}
\label{framework}
\end{table*}

\section{Conceptual Framework for AI Automatons}
\label{framework-description}

To tease apart design considerations that might both influence perceptions of and interactions with AI automatons, and be determining factors for the types of ethical concerns these systems may give rise to, we consider various aspects related to what a simulation is intended for, how these intended goals are or will be accomplished, and who is influencing these decisions. 
Our conceptual framework draws on design considerations initially identified through reviewing a purposive sample of related work studying specific types of AI automatons or their uses~\cite[e.g.,][]{mcilroy2022mimetic,lee2023speculating,agnew2024illusion,hofman2023steroids}, which we then iterated on and expanded through several discussion sessions with different subsets of our research team. 
During these sessions we examined further literature, as well as both existing and speculative applications and use scenarios to probe whether the framework captured known and possible design variations, and we resolved concerns and disagreements. 
Table~\ref{framework} overviews our framework, which is organized along four design axes: 1) what is simulated ({\em simulation scope},~\S\ref{scope}), 2) how the simulation is intended to be used ({\em simulation uses},~\S\ref{uses}), 3) who makes decisions about the simulations ({\em ownership and control},~\S\ref{control}), and 4) how the simulations might impact stakeholders ({\em adverse impacts},~\S\ref{impacts}).\looseness=-1

\subsection{Scope: What Is the Subject of the Simulation?}
\label{scope}

Aspects related to {\em who} is being simulated, {\em what} about them is being simulated, and with {\em how} much accuracy they are being simulated will govern people’s perceptions of and concerns about AI systems that simulate people's work, abilities, behavior, likenesses, or humanness~\cite[e.g.,][]{epley2007seeing,zhan2024healthcare,mcilroy2022mimetic,lee2023speculating}. 
For instance, what is being simulated and how the simulation is accomplished can influence perceptions of uncanniness and discomfort~\cite[e.g.][]{ventrurebeat2024,lee2023speculating,guingrich2023chatbots,forbes2024}, particularly 
when aspects that are unique to an individual are being simulated, 
when the simulation outputs appear eerily similar to what a human might do or look like, or 
when the simulation captures someone's characteristics with very high fidelity. 
Such properties of a simulation might also exacerbate other concerns such as those about privacy violations and lack of appropriate consent~\cite{kapania2024simulacrum,lee2023speculating,boine2023emotional}, or might threaten someone’s sense of identity and agency~\cite{lee2023speculating,friedman1992human,lee2024can,agnew2024illusion}.

\subsubsection{Target: Who is being simulated?} \enspace
Irrespective of the deployment context or possible use scenarios, the most foundational characteristics of an AI automaton---particularly as it pertains to determining whether the simulation should even exist or be developed, what the goals of the simulation should be, and how the simulation should be developed---are whether a person ({\em fictitious} or not) is simulated, and which of their characteristics or behaviors are simulated.
AI automatons can range from simulating specific, {\em real} individuals~\cite[e.g.,][]{vktr2025,lee2023speculating,mcilroy2022mimetic,kugler2024raising}, to cases where the simulations target real individuals but also aim to imbue the automaton with skills or characteristics the target might not have~\cite[e.g.,][]{lee2023speculating}, to simulating realistic, yet fictitious individuals~\cite[e.g.,][]{meta2025,financialtimes2024,nytimes2025,edwards2024reputation}, to targeting a group by simulating a target believed to be prototypical for the group~\cite[e.g.,][]{wang2021want,zhang2016data} or a population of targets that all belong to one or multiple groups~\cite[e.g.,][]{park2024generative}, or to aiming just to simulate something that might be or appear to be human in some way~\cite[e.g.,][]{muresan2019chats,mitnews2025,gratch2002creating}.\looseness=-1

\begin{figure*}
  \includegraphics[width=0.65\linewidth]{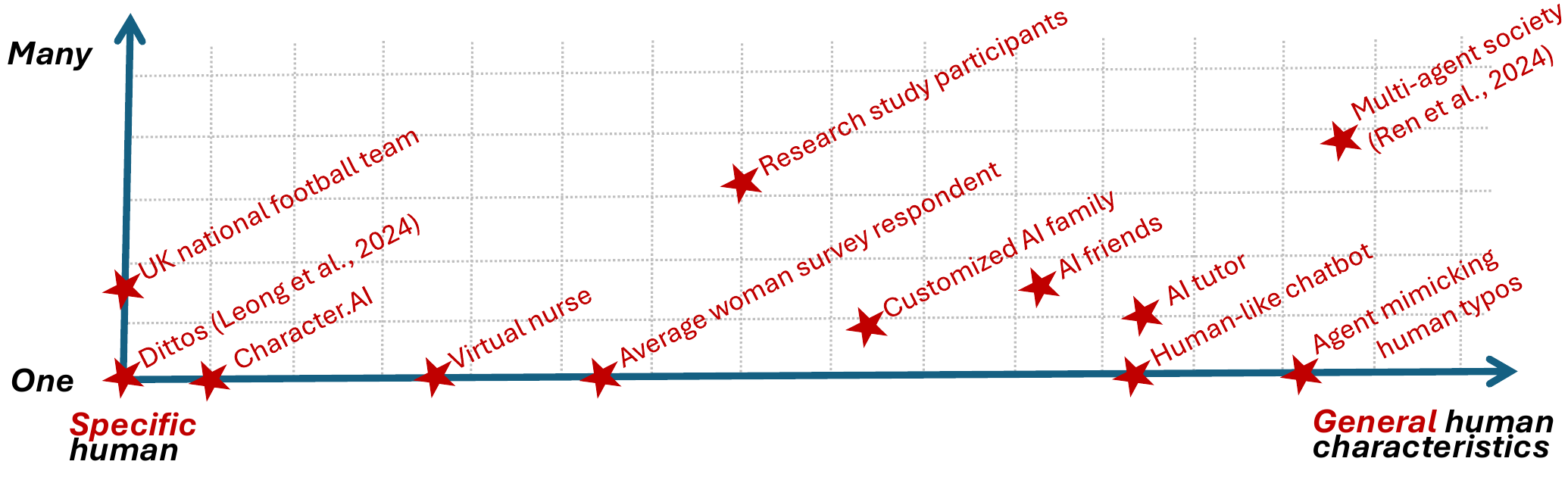}
  \vspace{-8pt}
  \caption{Simulations can range from targeting characteristics of specific individuals to more general group characteristics and even more general human characteristics (x-axis). They can also range from simulating single or multiple entities (y-axis).}
  \vspace{-8pt}
  \label{fig:target}
\end{figure*}

To help operationalize such distinctions in what the target of the simulation could be, at a high level we broadly differentiate between \begin{inparaenum}[i)] \item simulations where the intended target is a specific \underline{individual} (e.g., simulating someone's writing style) from \item simulations where the target is a \underline{group} or a persona representative of the group (e.g., simulating how the average woman would answer a survey question), or one that embodies characteristics specific to a group without necessarily being representative (e.g., simulating someone's voice for certain attributes e.g., ``a female voice with a North American accent'' and not for the voice actor's identity~\cite{hutiri2024not}), from \item simulations targeting more \underline{generic human} characteristics that are not seen as specific to any particular individual or group (e.g., AI avatars expressing human emotions~\cite{cnbc2024}).\end{inparaenum}\looseness=-1 

These distinctions also map to a design spectrum (illustrated in Figure~\ref{fig:target}) where the target can vary from specific individuals to general group characteristics to even more general human characteristics, and from representing a single entity to simulating multiple entities (e.g., a population of targets from the same group~\cite[e.g.,][]{lee2024can}). 
These distinctions are critical as concerns related to, for instance, deception, impersonation, and lack of consent~\cite[e.g.][]{futurism2024,cbsnews2024,Jones_2024,winkle2021assessing} are likely exacerbated when AI systems are designed to simulate the decisions and behaviors of specific, recognizable people---such as a specific AI researcher like Geoffrey Hinton or a specific singer-songwriter like Ed Sheeran---versus when they are designed to simulate a more generic AI researcher or singer-songwriter, versus when the aim is to embody generic, nonspecific human-like characteristics.\looseness=-1

Another critical aspect in specifying a simulation target is whether the target or some (or all) of their characteristics are \begin{inparaenum}[i)] \item \underline{fictitious} (e.g., chatbots simulating human or human-like characters from Game of Thrones or from popular games or anime~\cite{washingtonpost2024}) versus when they are \item \underline{real} (e.g., chatbots simulating real-life school shooters and their victims~\cite{futurism2024}, deceased loved ones~\cite{morris2024generative,vktr2025}), \end{inparaenum}  
since such differences in aims can also have drastically different ethical and social implications. Simulating fictitious characters that cannot reasonably be matched to a real person or group is less likely to raise concerns about identity fragmentation and objectification~\cite{lee2023speculating}, or about impersonation~\cite{newyorker2024}. 



\subsubsection{Characteristics: What about the target is being simulated?} \enspace
Furthermore, an AI automaton might be developed to only replicate or capture a subset of a target's characteristics, and might also aim to do so only for specific actions or tasks the target might undertake. 
An example could be an image generator developed to mimic someone's drawing style~\cite{wired2022}, but which does so only for specific elements in generated images but not for the rest of the image (e.g., for flowers, if present in an image), or 
a model developed to simulate only how someone would respond to a given set of survey questions.  
Various elements related to {\em which} of a target's characteristics are being simulated and {\em how} those characteristics are being simulated can, however, be more or less likely to lead to the output of a simulation being perceived as e.g., uncanny or unsettling~\cite{ventrurebeat2024,forbes2024} and govern people’s perceptions about and interactions with the overall system in different ways~\cite{alexander2005gaming}. 
Simulating someone's voice may heighten impersonation or deception risks~\cite{newyorker2024,ventrurebeat2024,winkle2021assessing,hutiri2024not} more than only simulating their written responses to a predefined set of questions would~\cite[e.g.,][]{argyle2023out}. 

Furthermore, simulating both what someone might say in response to a question and how they would say it (including their word choices, sentence structure, as well as their voice and accent) is more likely to lead to concerns about the system impersonating or even mocking the target~\cite[e.g.,][]{herbold2024large,giles1979accommodation,airaksinen2020mimetic}. 
Since simulating what a target does or might do (e.g., what they would say) as opposed to simulating how the target appears or does something (e.g., how they might say~it) may lead to different types of concerns, we distinguish between aiming to simulate characteristics of the target that are rather: \looseness=-1 
\begin{inparaenum}[i)] 
\item related to \underline{form}: when the simulation tries to mimic the likeness, appearance, or style of a target, versus
\item related to \underline{content}: when the simulation tries to mimic what the target might say or do.\looseness=-1 
\end{inparaenum} 


The choice of which characteristics to simulate also influences and is influenced by considerations related to how {\em coherent}, {\em realistic} or {\em naturalistic}, or {\em plausible} the simulation needs or is intended to be~\cite[e.g.,][]{lee2023speculating,mcilroy2022mimetic,achenbach2017fast,magnenat2005virtual,leong2024dittos,herbold2024large,dillion2023can}. 
For instance, the goal of simulating how a specific individual would respond to a question might require the answer be something the target might {\em plausibly} say, and that the way the response is formulated also {\em coherently} reflects their communication style~\cite{lee2023speculating,herbold2024large}.
Additionally, another important determinant factor constraining which of the target's characteristics are or could even be simulated by a given AI system is the {\em modality} of the system, or ``the domain [the system] operates in and the types of behaviors it is designed to reflect''~\cite{mcilroy2022mimetic}, which in turn can impact perceptions~\cite{laban2021perceptions}.
A system that outputs open-ended texts will be able to capture a different set of a target's characteristics than a system that outputs videos or one that only outputs answers to categorical survey questions. 

\subsubsection{Fidelity: How well is the simulation intended to capture the target characteristics?} \enspace
The fidelity or accuracy with which a system simulates a target's characteristics is believed to determine the value or usefulness of these systems~\cite[e.g.,][]{argyle2023out,mcilroy2022mimetic,volante2016effects,alexander2005gaming,jarrett2025language}---that is, the more faithfully such systems mimic a target, the better---and this, in turn, may require the simulation to be or appear {\em coherent}, {\em realistic}, and {\em plausible}.  
While low-fidelity simulations can give rise to concerns related to e.g., misrepresentation, deception, or reputational harms 
when the simulation drifts away from appropriately and accurately representing the target~\cite{mcilroy2022mimetic,edwards2024reputation,lee2023speculating}, other ethical or legal concerns are more likely to arise due to high-fidelity simulations~\cite[e.g.,][]{lee2023speculating,cao2023high}---particularly when the target has little control (\S\ref{control}) over whether and how their work, abilities, or likeness are being simulated~\cite[e.g.,][]{Jones_2024,lee2023speculating,jiang2023ai}.
For instance, high-fidelity simulations of an artist's style, work, or likeness are more likely to lead to copyright violations or infringement on someone's rights of publicity than low-fidelity ones (e.g., simulating a female-sounding voice versus Scarlett Johansson's voice~\cite{Jones_2024}).\looseness=-1 

\subsubsection{Specificity: To what degree are the simulated characteristics unique to the target and could identify the target?} \enspace
In addition to how faithfully a target or their characteristics are being simulated, how {\em unique} the simulated characteristics or abilities are to a target---e.g., in a way that uniquely represents them or that reproduces unique or rare abilities---is also critically important to consider, as the simulation of such characteristics raises questions about someone's ability to maintain their individuality~\cite{lee2023speculating}, their reputation (e.g., when a public figure's voice is simulated to spread defamatory content~\cite{hutiri2024not}), or their ability to capitalize on their own skills or talents (e.g., when an image generation tool reproduces an artist’s signature style~\cite{wired2022}). 
The mimicking of someone's unique characteristics can also exacerbate privacy or impersonation risks~\cite{wells2022s,beard2001clones,roberts2022you}, 
and both individuals and professional or cultural groups risk seeing their work devalued, losing part of their social capital or even their livelihood~\cite{mcilroy2022mimetic,hutiri2024not,wired2022}. 

\subsubsection{Completeness: To what degree is the simulation intended to fully capture the target?} \enspace 
How many of a target's characteristics are being simulated, or whether the target is intended to be {\em simulated in its entirety} is another design consideration that determines not only simulations' deployment settings, but also how the simulations are perceived and interacted with~\cite[e.g.,][]{lee2023speculating} and how {\em versatile} the resulting automaton is---e.g., how diverse the actions it can take are~\cite{Mitchell2025AI}. 
Simulations intended to be highly detailed and elaborate, that are meant to be complete and exhaustive, or that have high {\em generality}---i.e., there is a substantial ``breadth of scenarios and domains that a [system] can capture''~\cite{mcilroy2022mimetic}---will not only lead to a wider range of concerns, but also to heightened concerns, particularly when the reach of a system is more extensive. 
For instance, applications designed to ``visually resemble you, sound like you, and possess the knowledge you would want to carry into [a] meeting''~\cite{leong2024dittos}---requiring simulation of a wide range of characteristics---may give rise to different concerns than those simulating yes/no responses to survey questions~\cite{lee2024can}. Highly complex simulations of individuals are more likely to trigger concerns about their objectification, dehumanization, displacement, or loss of individuality~\cite{lee2023speculating}.\looseness=-1

\subsubsection{Adaptability: To what degree is the simulation intended to evolve or adapt?} \enspace 
While for some settings AI automatons may be intended to remain \underline{static} or reflect fixed snapshots of a given target (e.g., cloning one's younger self to talk to them~\cite{dazeddigital2023}), other settings may require automatons to \underline{evolve} 1)~based on interactions, feedback or new information (e.g., by learning from interactions~\cite{weber2022reflecting,namvarpour2024uncovering}), or 2)~according to the target's own evolving self (e.g., to maintain accurate representations of the target~\cite{lee2023speculating}). 
A static snapshot or one that evolves separately from the target, however, could for instance end up misrepresenting the target by presenting stale or inauthentic versions of the target~\cite{lee2023speculating,lu2022subverting}. 
Adaptation could also involve considerations about whether the automaton can adapt based on context (e.g., by changing how it presents the target depending on the role they need to engage in, such as a friend, mentor, or colleague~\cite{lee2023speculating}), or whether it is intended to mimic behavior or how it should present the target in {\em new} situations the target has not been in before~\cite{mcilroy2022mimetic}.\looseness=-1

\subsubsection{Humanness: To what degree is the simulation intended to capture human-like characteristics?} \enspace
The mimicry or the appearance of embodying human-like characteristics also has a important role in determining both how AI systems are perceived and interacted with, as well as critical ethical considerations that their deployment and use might give rise to~\cite[e.g.,][]{rapp2021human,gabriel2024ethics,turkle2007authenticity,de2016almost,kahn2007human,devrio2025taxonomy}. 
This is the case even when there is no identifiable person or group that a system is designed to mimic, or when the simulation is instead intended to capture only more general human-like attributes 
and behaviors~\cite{cheng2024one,cohn2024believing}, as imbuing non-human agents with human-like qualities---e.g., likenesses, intentions, motivations, and goals---may, for instance, end up objectifying and dehumanizing people~\cite{epley2007seeing,jack2013seeing,valenzuela2024artificial}, may lead to anthropomorphic deception where users might inadvertently believe they are talking to or interacting with a human rather than a machine~\cite{gros2022robots,winkle2021assessing,perry2023ai}, or may lead to people inappropriately developing material or emotional dependence on such agents~\cite{maeda2024human,laestadius2024too} or to a false sense of trust, safety and familiarity~\cite{mireshghallah2024trust,laestadius2024too}, all of which can have concerning social implications.\looseness=-1

\subsection{Intended Uses \& Goals: How Is the Simulation Intended to Be Used?} 
\label{uses}
The contexts AI automatons are developed for and deployed in, how they are being used, and which and how various stakeholders may benefit from interacting with these systems determine not only their usefulness and how people perceive and interact with them, but also what types of risks their use might bring about~\cite[e.g.,][]{hidalgo2021humans,hofman2023steroids,lee2023speculating}. 
To help foreground possible design decisions that influence and are being influenced by intended uses and goals, in our framework we consider four high-level design considerations related to \begin{inparaenum}[1)]
\item whether the simulation is intended to {\em replace} the target~(\S\ref{replace}), and 
\item whether the system is set up in a way that enables others to {\em interact} with the simulation~(\S\ref{interact}), 
\item {\em observe} the simulation~(\S\ref{observe}), or 
\item {\em study} the simulation~(\S\ref{study}). 
\end{inparaenum}
These high-level considerations are primarily meant to help make related design decisions more salient and are not mutually exclusive (e.g., a system could be designed both with the intention to replace the target as well as to allow the target to interact with their own simulation).

\subsubsection{To replace: Is the simulation intended to replace the target?} 
\label{replace}
Perhaps the most commonly mentioned concerns about AI automatons relate to how such systems could replace humans~\cite[e.g.,][]{hofman2023steroids,byun2023dispensing,chui2016machines,prahl2021out,brynjolfsson2023turing,lee2023speculating,wsj2024AI}, with some even declaring that ``the era of AI employees is here''---employees who ``won't complain about work-life balance''~\cite{gizmodo2024}. 
When and for what purposes the target of a simulation might be replaced, however, can color whether such replacement is seen as a {\em benefit} (e.g., when the simulation enables the target to delegate unwanted or harmful tasks or scale their work) or rather as a {\em concern} (e.g., when the simulation of a target's abilities is being used to do their paid job and replace them).
When one of the design goals is to replace the target, we distinguish between three different replacement goals:\looseness=-1
\begin{compactenum}[i)]
    \item to \underline{relieve}: when the goal is to relieve the target (or others) from drudgery, possible harm, or activities that would be unethical or unsafe for the target (or others) to carry out. This is typically done by the operator to mitigate harm or provide relief for the benefit of the target (e.g., using simulated study participants to protect human subjects from harm~\cite{agnew2024illusion,jo2024harmful}, or AI news anchors not to replace but to protect journalists from political retribution~\cite{cnn2024});\looseness=-1 
    
    \item to \underline{substitute}: when the simulation is intended to be used as a stand-in or surrogate for the target when the target is unavailable, the target wants to delegate their tasks, or when the activity is impractical or impossible for the target to do, typically initiated by the target and to the target's own benefit (e.g., responding to emails or messages on the target's behalf~\cite{techcrunch2024,liu2024leveraging,wired2024onlyfans} or as a stand-in when human expert annotators are scarce~\cite{jadhav2024limitations}). 
    This is typically initiated by the target or with their knowledge.
    For example, we consider the goal to be {\em substitution} when target delegates a data annotation task to an AI automaton instead of doing it themselves, but {\em relief} when such annotations are done without the target's involvement to protect them from possible harm from exposure to harmful content~\cite[e.g.,][]{kirk2022handling}, as is often the case for automated assessments of hateful content~\cite{jo2024harmful};\looseness=-1
    
    \item to \underline{displace}: when the simulation is intended to take over the place, position, or role traditionally occupied by the target to help an operator or interactor reduce costs, scale operations, increase speed, or enhance convenience, often to the detriment of the target (e.g., replacing human newscasters~\cite{tomsguide2023} or other human jobs~\cite{gizmodo2024} resulting in loss of opportunities or livelihood~\cite{wired2022,brynjolfsson2023turing}). 
    This is typically done by an operator or interactor, adversely impacting the target (or others) who may lack the means to mitigate the impacts (e.g., reduced wages, job loss, strained relationships).\looseness=-1

\end{compactenum}

\subsubsection{To interact: Is the simulation intended to be interacted with?}
\label{interact}
AI automatons are not only increasingly developed for interaction and engagement~\cite{mcilroy2022mimetic,pan2024multimodal,maeda2024human,stark2024animation,horvitz2022horizon,abramson2020imitating}, but they are also developed to support an increasingly varied set of modes of interaction~\cite{manzini2024code,pan2024multimodal,li2024situ,hutiri2024not}. These interactions have also led to diverse conceptualizations of AI automatons and the roles they play, from collaborators to companions to coaches to judges (to name a few), which in turn influence how and for what purposes interactive AI automatons are developed, deployed, and used.

\para{Interaction modes: How can the simulation be interacted with?} \enspace  
When and in what ways someone can interact with the simulation influences both the interaction dynamics as well as their perceptions of what is being simulated and the consequences of doing so~\cite{deshpande2024perceptions,jo2024understanding,leong2024dittos}. 
The interactional goals that operators aim for are guided by design considerations related to both the amount of freedom an interactor should have when engaging with a simulation (\S\ref{control}), as well as considerations about the types of actions the simulation is designed to carry out and for how long. 
The latter includes considerations about \begin{inparaenum}[1)]
\item whether the simulation supports \underline{open-ended} or instead only more \underline{constrained}, structured, or scripted interactions~\cite{guo2024investigating,mcilroy2022mimetic,leong2024dittos}; 
\item whether it allows only \underline{short-term} versus \underline{longer-term} interactions~\cite{folstad2019different,nytimes2025}; 
\item whether it can be used in \underline{new} situations rather than just reproducing past or \underline{known} behaviors or situations~\cite{mcilroy2022mimetic,leong2024dittos}; and 
\item whether it is intended to be \underline{generative} and produce new behaviors or rather be \underline{predictive or retrieval} in nature~\cite{mcilroy2022mimetic}. 
\end{inparaenum}
While facilitating open-ended, long-term, generative interactions are more likely to lead to self-disclosures of sensitive information and emotional dependence~\cite{jo2024understanding,nytimes2025,latimes2025AIdarkthoughts}, designing for more constrained, short-term interactions may also frustrate users for not recalling past interactions with them~\cite{de2024lessons}.\looseness=-1  

\para{Stakes: What is the simulation intended to do for the interactor?} 
When interaction is intended, a common leitmotif is wanting to support rather than replace humans~\cite[e.g.,][]{de2021ai,sellen2024rise,wilder2021learning}, with specific design choices motivated by varying aims for what the simulation is meant to do for an interactor~\cite[e.g.,][]{hofman2023steroids,maeda2024human,litt2016imagined,collins2024building}.
This is particularly well-illustrated by the distinction drawn by \citet{hofman2023steroids} between cases where AI systems are intended to help users attain certain goals by serving as \emph{steroids}---i.e., providing short-term performance boosts but risking deskilling in the longer term, \emph{sneakers}---i.e., by temporarily accelerating their abilities, and \emph{coaches}---i.e., by helping improve users' own capabilities rather than only helping them out in the moment. 
Such differences in how and what AI automatons are architected for can determine not only the impacts these systems might have on those interacting with them~\cite{hofman2023steroids}, but can also inform discussions about what trade-offs to strike between the value users are intended or believed to derive from these systems versus the adverse impacts these systems might have~\cite[e.g.,][]{latimes2025AIdarkthoughts,barnett2023ethical,cheng2024one}.

To foreground differences in what AI automatons are developed for, we distinguish between several interaction goals: 
\begin{inparaenum}[~i)] 
    \item to \underline{enhance}: %
    when aiming to improve or enhance the interactor's ability to complete a task or carry out an activity, without necessarily helping them also develop their skills (e.g., AI as steroids or as sneakers~\cite{hofman2023steroids}); \looseness=-1 
    \item to \underline{coach}: 
    when aiming to train or teach the interactor to help them learn or improve their abilities and skills (e.g., act as a chess coach~\cite{mcilroy2022mimetic}, practice with an automaton as an imagined audience~\cite{maeda2024human,litt2016imagined}); \looseness=-1 
    \item to \underline{serve}: 
    when aiming to provide specialized services to or for the interactor, typically services that someone else would perform for them (e.g., a virtual nurse providing medical services to patients~\cite{chavali2024ai,salon2025AInurses}); \looseness=-1  
    \item to \underline{connect}: 
    when aiming to provide social or emotional support to interactors (e.g., provide companionship or friendship~\cite{de2024ai,brandtzaeg2022my}); \looseness=-1  
    \item to \underline{entertain}:
    when aiming to entertain the interactor (e.g., gameplay~\cite{milani2023navigates}, AI characters as a ``new entertainment format''~\cite{financialtimes2024} or ``designed to make you laugh, generate memes''~\cite{petapixel2024}); 
    \item to \underline{accommodate}: 
    when aiming to adapt or customize an AI automaton's output or behavior to the target's characteristics or needs, typically to increase familiarity or comfort, to facilitate interactions, or to provide a personalized experience (e.g., ``Replika [...] learns [people's] texting styles to mimic them''~\cite{bloomberg2016}, ``adapt the agent’s demeanor''~\cite{leong2024dittos}, or customizing a generated voice depending on the interaction setting~\cite{byeon2022voice}); 
    \item to \underline{collaborate}: 
    when the automaton is intended to act or serve as a collaborator for the interactor (e.g., machine or AI teammates~\cite{zhang2021ideal,seeber2020collaborating} or AI systems as ``thought partners''~\cite{collins2024building}); and
    \item to \underline{evaluate}: 
    when aiming to assess the interactor, without necessarily being intended to help the interactor improve (e.g., a virtual interviewer developed to assess job applicants~\cite{forbes2024interviewer,theguardian2024Job}).\looseness=-1 
\end{inparaenum} 

\para{Affinity: How similar to the interactor is the simulation target intended or likely to be?}
AI automatons designed for interaction can also vary in how much the target is intended to adapt to or share some (or all) of the characteristics of those interacting with them. For instance, some simulations might be designed to take on the interactor's accent~\cite{huffingtonpost2024}, while in other settings the target {\em is} the interactor~\cite{dazeddigital2023,mcilroy2022mimetic}. 
However, whether and how much the target shares the characteristics of the interactor (or even those of the interactor's kith and kin)---either deliberately or accidentally (e.g., when an interactor happens to share some of the target's characteristics such as sharing the same profession or demographic attributes)---affects not only people's perceptions of these systems but also how they use and interact with them~\cite[e.g.,][]{nass2001does,hutiri2024not}. 
Similarity is often desirable as it can facilitate familiarity and likability~\cite{giles1980accommodation}, and people tend to prefer and respond more positively to systems and representations they perceive as reciprocating or sharing some of their characteristics~\cite{nass2000machines,nass2000does,maeda2024human,inkpen2011me,byeon2022voice} and even adjust their own behavior to virtual representations of themselves~\cite{yee2007proteus,szolin2023exploring}.
Nevertheless, while in certain scenarios using someone's accent or speaking style may help facilitate more high-quality interactions with a system, in other scenarios mimicking someone's mannerisms or accent risks being perceived as mocking or stereotyping them~\cite{giles1979accommodation,chaves2022chatbots}. Virtual representations that come across as too eerie, creepy, or self-like have also been found to trigger adverse reactions~\cite{shin2019uncanny}, and people may want to customize representations to distance themselves from them or blur certain characteristics (e.g., gender or age~\cite{byeon2022voice}), or to project an idealized version of themselves or others~\cite[e.g.,][]{dai2024beyond,zhang2021ideal}. \looseness=-1

\subsubsection{To showcase: Is the simulation intended to be observed by others?}
\label{observe} 
AI automatons can also be designed to provide {\em non-interactive} spectator experiences, such as watching or listening to AI-generated ads~\cite{nbcnews2024} or image, video, or audio deepfakes~\cite{mcilroy2022mimetic}. Even when there are no direct interactions with the simulation, however, concerns might still arise depending on what is being simulated, how it is simulated, and for what purposes.\looseness=-1

\para{Stakes: What is the simulation intended to do for the spectator?} \enspace
Existing non-interactive AI automatons have also been developed and deployed for a variety of goals and intended uses.
For instance, while in some non-interactive scenarios a target may be simulated in order to \underline{entertain} an audience of spectators (e.g., AI-generated music~\cite{fastcompany2024} or short films~\cite{theverge2024-shorts}, or a simulated voice narrating a story~\cite{hutiri2024not,goodereader2023}), in other scenarios they are meant to help \underline{train} (e.g., by providing a demonstration of how to perform a surgery \cite{small1999demonstration,vozenilek2004see}) or \underline{persuade} (e.g., generated ads for marketing campaigns \cite{whittaker2021rise,campbell2022preparing}) those listening or watching. 
Such differences can sharpen various risks in distinct ways: copying someone's voice would prompt concerns about impersonation, consent, or appropriate compensation in most scenarios~\cite[e.g.,][]{hutiri2024not,lee2023speculating}, yet these concerns might be especially heightened when this is done for malicious persuasion~\cite[e.g.][]{gijn2024}.\looseness=-1

\para{Affinity: How similar to the spectator is the simulation target intended or likely to be?}\enspace
Analogous to the question of similarity between an interactor and the target (\S\ref{interact}), the target can also vary in which and how many characteristics it shares with spectators, which can similarly color the spectators' perceptions of and concerns about AI automatons.  
For instance, people might respond differently to a deepfake of themselves compared to one of a public figure or of a different everyday person, or compared to a synthetic video of a fictitious individual or character~\cite[e.g.,][]{danry2022ai,brigham2024violation,dunn2020identity}.\looseness=-1  

\subsubsection{To study: Is the simulation developed to help study human or machine behavior or phenomena?}
\label{study}
Humans are also simulated for experimentation purposes to study theories about humans or the ability to simulate them. 
When the goal is to study either the targets or the automatons, we identify the following common goals: \looseness=-1
\begin{inparaenum}[i)] 
\item study \underline{human behavior}: when one or more targets and their possible interactions are simulated with the goal of understanding populations and their possible behaviors; understanding human beliefs, preferences, and values; or investigating any other human or social phenomena~\cite[e.g.,][]{hamalainen2023evaluating,hamilton2023blind,lee2024can,ji-etal-2024-srap,ren2024emergence,xiao2023simulating,gao2023s,tornberg2023simulating,wang2024decoding,yu2024researchtown};
\item study \underline{the simulation}: when one or more targets are simulated to test the ability to simulate these targets, or to understand the properties of the simulation (e.g., to probe how well language models simulating research participants align with human survey response distributions~\cite{argyle2023out} or reproduce well-known experiments~\cite{aher2023using}, or how well certain domain experts could be simulated~\cite{li2024agent}); 
\item study \underline{system or model performance}: when one or more targets are simulated to help anticipate how different stakeholders might interact with and use a system or model (e.g., simulating users interacting with a product to anticipate their needs and perspectives~\cite{ataei2024elicitron}).\looseness=-1 
\end{inparaenum} 

\subsection{Ownership \& Control: Who influences or makes decisions about the simulations?}
\label{control}

Critical considerations in the deployment of AI automatons are also related to {\em by whom}, {\em when}, and {\em how} decisions are made about what is being simulated. 
To help formalize how much {\em control} various stakeholders have over the scope and uses of the simulations, we adapt an existing conceptual framework for participation in AI~\cite{delgado2023participatory,suresh2024participation} as it helps tease apart some of these considerations, including
1) which stakeholders get to influence decisions, or {\em who is involved?}
2) what decisions do they get to influence, or {\em what is on the table?} and 
3) in what ways are they able to influence decisions, or {\em what form does participation take?} \enspace
The modes of participation~\citet{delgado2023participatory} derived from existing literature further echo the different degrees of control or decision-making power different stakeholders could be given over what AI automatons are intended to do and how they can be used. 
Drawing on this work, we consider the following dimensions along which both stakeholder control and related design considerations can vary:\looseness=-1

\subsubsection{Who is in control?} \enspace \label{who-controls}
Ethical stakes related to who and what about them is simulated, and how and by whom the simulation can be used, can feel different depending on {\em which stakeholders}---e.g., targets, interactors, spectators, operators, or others---get to participate in or influence these decisions. 
For instance, an {\em operator} deciding who the simulation {\em target} is without any input or consent from the {\em target} is more likely to raise concerns about issues with consent circumvention about how one's likeness or data might be used. Such concerns might be lessened when the {\em target} itself has full control over what about them is being simulated and when their simulations can be used. 
These distinctions are also important because the ability to influence or make decisions about the development, deployment, and use of AI automatons also determine {\em with whom responsibilities lie} if and when their development, deployment, or use lead to adverse impacts~\cite{hutiri2024not,lee2023speculating,leong2024dittos}, as well as the various stakeholders' abilities to mitigate such impacts~\cite[e.g.,][]{suresh2024participation,lee2023speculating,rothman2018right,widder2022limits,harrell2017reimagining}.\looseness=-1

\subsubsection{Type of control: What do they have control over?} \enspace
Different stakeholders might be able to influence or have control over different aspects of e.g., what is being simulated, how and what the simulation is being developed for, and even whether the simulation should be developed at all; such considerations about what stakeholders have control over---where the {\em locus of control} and responsibility lie---can help mitigate (or instead exacerbate) ethical concerns depending on how they limit or enable different stakeholders to influence how AI automatons are architected and used. 
For instance, if the {\em target} of a simulation only has control over what is being simulated about them, but not over all the ways in which the simulation of their likeness, work, or abilities is being used, the simulation target might still be worried about possible reputational or discrimination risks, and their ability to mitigate those risks. 
That is, a professional community or an actor might be comfortable with simulating their likeness to add or adjust a scene in a movie or documentary, but not with it being used in a video implying they are endorsing a political candidate or policy. People's preferences and concerns may depend on the context of use. 
Furthermore, because developing a simulation of a target is often data-intensive and requires a large corpus of digitized traces of a target's behavior and likeness, questions about consent and control over one's data have also become particularly acute~\cite{morreale2024unwitting,rapp2021human}.\looseness=-1 

To capture and operationalize these distinctions, we consider whether stakeholders have control over: 
\begin{inparaenum}[i)] 
\item \ul{whether a target is simulated}: when stakeholders can influence or make decisions about who the {\em target} of the simulation is and whether their simulation should be developed in the first place (e.g., users choose to build an AI version of themselves or a fictitious AI character~\cite{petapixel2024});
\item \ul{what about the target is being simulated}: when stakeholders can influence or make decisions about whether the {\em target} as a whole is being simulated or only certain specific characteristics of a target are being simulated (e.g., users retain control over what is being said on their behalf~\cite[e.g.,][]{robertson2021can});
\item \ul{how the simulation is being developed}: when stakeholders have influence or control over aspects related to how the simulation of the target is being accomplished (e.g., control over how the target's data can be used~\cite[e.g.,][]{hollanek2024griefbots}); 
\item \ul{who can interact with or observe the simulation}: when stakeholders can influence or make decisions about who might have access to the simulation in order to observe or interact with it (e.g., only the target can interact with their own simulation, or only adult users can interact with the simulation~\cite[e.g.,][]{hollanek2024griefbots}); and
\item \ul{how the simulation can be used}: when stakeholders can influence or make decisions about which deployment scenarios a simulation should be developed for or used in, how someone can interact with the simulation (if at all), or what tasks the simulation can perform (e.g., users can specify what ``topics to avoid''~\cite{petapixel2024}). 
This can also include considerations about whether stakeholders can refuse interactions with a simulation.\looseness=-1 
\end{inparaenum} 

\subsubsection{Degree of control: How much control do they have?} \enspace
Stakeholders' ability to influence the scope and use of simulations can also vary from no influence and control (e.g., fully autonomous human-like agents that act without input), to being able to provide only superficial feedback or input, all the way to having complete control---and thus being able to make decisions about all aspects related to who and what is being simulated, and when and how the simulations can be used. Stakeholders might also be able to influence or make decisions about the simulations only at certain points in the AI automaton's development and deployment life-cycle. 
We thus consider two main dimensions of variation: 
1) are stakeholders providing only feedback ({\em can influence}) or are they able to make decisions ({\em can control})? and 
2) when are stakeholders able to provide feedback or make decisions?
We operationalize these via five levels of control:\looseness=-1
\begin{compactenum}[i)]
\item has \underline{no control}: when stakeholders have no control or direct influence over the scope and use of the simulation---i.e., who and what about them is simulated, for what purpose are they simulated, how they or others can interact with the simulation, or how their or others' interactions with the simulation are used to adjust what is being simulated (e.g., deepfakes of unsuspecting targets~\cite{cbsnews2024}, 
or employees having no control over being replaced by AI automatons~\cite{gizmodo2024});\looseness=-1

\item was \underline{consulted}: when stakeholders have some influence over the scope and use of the simulations, typically by being able to express discrete preferences or provide input at specific points in the development and deployment life-cycle
(e.g., simulated chess coaches users can choose to use~\cite{mcilroy2022mimetic} but do not get to influence their design);\looseness=-1

\item was \underline{included}: when stakeholders can influence the scope and use of the simulations, typically through explicit feedback mechanisms implemented at most or all stages in the development and deployment life-cycle (e.g., indicate which type of messages the automaton can send on the target's behalf~\cite{liu2024leveraging});\looseness=-1

\item is \underline{in control}: when stakeholders are able to make some of the decisions about the scope and use of the simulations at specific points in the development and deployment life-cycle (e.g., a company allowing users to choose a sequence of gestures to animate historical family photos~\cite{DeepNostalgia} and specify who can interact with them~\cite[e.g.,][]{hollanek2024griefbots,meta2025});\looseness=-1

\item has \underline{ownership}: when stakeholders own the simulation and/or have full control over any decision related to any parts of the process used to create, deploy and use the simulation and at any points in the development and deployment life-cycle (e.g., companies offering users the ability to customize avatars for their own commercial purposes~\cite{AIStudios}, or maintain ``personal ownership and exclusive control over one’s digital image"~\cite{beard2001clones}).\looseness=-1
\end{compactenum}

\subsection{Impacts: What Are the Implications of Simulating Humans?}
\label{impacts}

Recently, Meta took down some of their AI accounts after they were deemed creepy, inaccurate, and disrespectful by users~\cite{meta2025}, while Replika restored some of theirs after users expressed anguish from being separated from their AI partners~\cite{de2024lessons,businessinsider2023} due to an update disallowing certain uses.
Indeed, concerns about how AI automatons might impact people and society govern both how people use and interact with AI automatons and the development of legal, ethical, and normative frameworks to guide and govern their use~\cite[e.g.,][]{hutiri2024not,mcilroy2022mimetic,cbsnews2024,zhang2024my,widder2022limits}, which in turn influence what is being built and deployed. Drawing on existing literature on harm anticipation and taxonomization~\cite{hutiri2024not,olteanu2020search,buccinca2023aha,boyarskaya2020overcoming}, we foreground two key areas of consideration that should govern design decisions: i) who might be affected by the development, deployment, and use of AI automatons ({\em impacted stakeholders}), and ii) in what ways the different stakeholders are being impacted ({\em possible adverse impacts}).\looseness=-1

\subsubsection{Impacted stakeholders: Who is being impacted?} 
As with any AI system, effectively reasoning about the implications of AI automatons requires careful consideration of all relevant stakeholders~\cite{buccinca2023aha}. 
The development, deployment, and use of AI automatons may impact not only {\em direct} stakeholders like those impacted due to interacting with or being the target of a simulation---e.g., family members who were deceived into believing their loved one was in a accident after interacting with a system imitating their loved one's voice~\cite{leaderpost2022}, or a target's identity being appropriated by third parties without consent~\cite{roberts2022you,hassan2023ai}---but also {\em indirect} stakeholders like individuals or communities associated with direct stakeholders even when they are not interactors (e.g., loved ones of a deceased target~\cite{morris2024generative}), or even society at large (e.g., due to erosion of public trust~\cite{hutiri2024not}). Furthermore, even when given two different interactors or operators with similar control over, for instance, an AI automaton designed to produce language in a certain minoritized variety, who the interactor or operator is might impact concerns differently---e.g., when the operator is a speaker of the minoritized variety it may constitute {\em reclamation} or just ordinary use, whereas when the operator is a corporation or a non-speaker it may be seen as linguistic {\em appropriation}. 
Thus, as with design considerations related to {\em who is in control} (\S\ref{who-controls}) of the development and deployment of AI automatons, differences in {\em which stakeholders}---e.g., targets, interactors, operators, or others---are involved and likely to be adversely impacted are influenced by {\em de facto} design decisions and {\em should} in turn influence those decisions.\looseness=-1   

\subsubsection{Adverse impacts: How are they being impacted?} 
The risks to different stakeholders are similarly influenced by and {\em should} in turn influence how AI automatons are being built and deployed~\cite[e.g.,][]{hutiri2024not,chandra2024lived,lee2023speculating,boine2023emotional}. 
For instance, having vulnerable individuals develop emotional attachment and trust towards an AI companion that results in them following harmful advice~\cite[][]{boine2023emotional,arstechnica2024,washingtonpost2024} {\em should} perhaps minimally lead to these systems being designed to provide appropriate disclosures and reminders of interacting with an AI system to users, among other guardrails~\cite{boine2023emotional}. 
Similarly, concerns about misrepresentation {\em should} result in allowing a target to control what their simulations say and do in autonomous interactions~\cite{lee2023speculating}. 
Adverse impacts are also determined by how and when those risks are likely to arise or by possible {\em pathways to harm}---i.e., ``causal chain[s] of events required for a harm to be realised''~\cite{connolly2022recommendations}. 
This includes considerations about how stakeholders get exposed to AI automatons (e.g., by being the target of, by interacting with, by operating, or by being denied access to an AI automaton~\cite[e.g.,][]{hutiri2024not}), which system behaviors are more likely to give rise to certain adverse impacts~\cite[e.g.,][]{devrio2025taxonomy,chandra2024lived,cheng2025dehumanizing}, as well as the role the simulation has in heightening the risk of these impacts (e.g., by being the ``perpetrator, instigator, facilitator, and enabler'' of harms~\cite{zhang2024my}).\looseness=-1


\section{Discussion and Concluding Remarks}

AI automatons are being developed and deployed in an ever-growing number of applications. We developed our framework with the goal of helping developers and other stakeholders recognize and analyze the design choices underpinning AI automatons. In so doing, we hope to support developers and other stakeholders in reflecting on the implications of those choices, including alternative choices possibly available to them. We believe this is a prerequisite for establishing a basis for discussions about what types of systems we should or should not build.\looseness=-1

\para{Being more explicit, reflexive, and intentional about design decisions.} 
As this paper demonstrates, there is a very wide range of design choices available to those seeking to develop AI automatons. Yet it is far from clear whether developers of these systems make such decisions by reflecting explicitly on the range of options available to them---and then intentionally adopting a set of choices that best serve their goals. It is even less clear the degree to which developers reflect on the ways that different design choices might affect the interests of those who serve as the target of the AI automaton, who interact with the AI automaton, and even those who do not get to be a target of or interact with the AI automaton. Our goal in developing this analytic framework was to foster greater recognition of the range of design choices available to developers such that they might make \emph{better} choices. We did not set out to provide developers with a {\em how-to} guide for navigating the ethical issues that might arise in developing and deploying AI automatons. Given the many dimensions of possible variation---and the additional complexity that arises from their interaction---it is unlikely that there are general principles that can guide decision making across all possible configurations. But mapping out the vast space of design choices reveals that there are many paths that developers can and should consider---and that no one path is preordained. Ideally, confronting the available alternatives can help developers recognize when a specific design choice is actually a choice, not a necessity, and help them make far more considered decisions.

\para{A foundation for more focused analyses of existing applications and for the design of empirical studies.} 
Our analytic framework can also serve as the foundation for more targeted analyses of specific types of already existing AI automatons, examining the degree to which seemingly similar applications actually vary along other dimensions---and whether this variation seems to affect our ethical intuitions about their relative desirability. It could also help to reveal when there are consistent patterns in the configurations of certain applications. Such findings might invite further study, looking into the possible reasons---e.g., technical, commercial, practical---why certain dimensions seem to vary consistently with each other. It can also highlight if developers have clustered in a particular part of the space of options, despite there being many other possibilities. Similarly, the dimensions of variation identified in our analytic framework can inform the design of empirical studies. Research subjects could be presented with different examples of AI automatons with carefully controlled variation along specific dimensions, with the goal of assessing how their reactions differ when manipulating discrete elements of the configuration. While in this work we do not provide a clear link for how variations along certain design axes have a determinative impact on normative concerns, such future empirical studies are crucial to both better understand people's reactions to the many possible ways of designing AI automatons and to provide evidence to better support researchers' (including our own) ethical intuitions about the kinds of configurations that people find more or less objectionable.\looseness=-1 

\para{Articulating design decisions in practice.} 
We argue that developers of AI automatons should clearly articulate their design decisions to help stakeholders better understand the concerns that can arise from those choices, particularly those choices that may not be evident based on limited use of a system. At the same time, this is complicated in at least two ways. First, AI and machine learning research communities' valorization of qualities such as generalizability~\cite{birhane2022values} means that systems are regularly accompanied by broad claims of their capabilities or other characteristics---e.g., ``general-purpose computational agents that replicate human behavior across domains''~\cite{park2024generative}---making it difficult for developers to precisely state, and other stakeholders to understand, who and which characteristics are simulated and for which purposes. Second, even if design decisions are well understood, it is often unclear what control current implementations of AI systems permit various stakeholders~\cite[e.g.,][]{young2024participation,lee2023speculating,geng2025control}. 
This is particularly salient when AI automatons are built on top of what are known as foundation models that are ``intended to be almost universally applicable''~\cite{suresh2024participation}, as this raises questions about how to prevent an AI automaton from having knowledge and capabilities that the target might not have. 
It remains an open question how illuminating design choices behind AI automatons can enable more effective intervention, governance, or resistance.\looseness=-1

\subsubsection*{Key References} 
We want to acknowledge a few references that have been particularly influential in shaping our thinking and informing our conceptual framework for AI automatons, which include:~\citet{mcilroy2022mimetic,agnew2024illusion,hofman2023steroids,delgado2023participatory,lee2023speculating,hutiri2024not}.

\section{Adverse Impacts \& Ethical Considerations} 

\noindent{\em Ethical considerations.} \enspace
In her commentary, \citet{suchman2023uncontroversial} argues that discussions of AI that hold it up as a self-evidently coherent, ``stable and agential'' entity elide important differences between various underlying techniques and between ``speculative [...] projects and technologies in widespread operation,'' uncritically reproducing beliefs in AI capabilities and making it difficult to carefully assess technologies and their impacts. 
While developing our framework, we thus {\em deliberately choose} not to focus on broad claims about AI automatons' capabilities or speculations about what they can or cannot do, which can reflect perceptions or illusions of intelligence, agency, vitality, or other human-like qualities~\cite{maeda2024human,stark2024animation,devrio2025taxonomy}; instead, we seek to address, as concretely and specifically as possible, the space of possible design goals and choices surrounding these AI automatons' development, deployment, and use. 
While we introduce AI automatons as a broad category of objects, through our framework we aim to make it clear that they are not a singular or stable object, but that it can in fact be configured in many different ways, with equally as many impacts. 

\para{Adverse impacts.} \enspace
This choice, however, may also risk two adverse impacts: 
first, focusing on possible design goals and choices---i.e., what a developer wants or intends to build---may suggest that some designs are possible or even desirable to implement in practice, even when they may not be. 
In other words, this focus may overlook or even obscure questions and assumptions about why to even consider certain design goals or uses, why these goals and uses are desirable or defensible, what problems AI automatons are intended to address, and why developers believe AI automatons are a solution to those problems instead of other alternative ways---including ways that possibly involve entirely non-tech ways---to address the same problems.  
Furthermore, foregrounding and speculating about a wider range of design choices might also risk drawing attention to possible system designs that might in fact heighten (rather than mitigate) existing concerns or even give rise to new ones---systems that perhaps should not be built. 
Second, not focusing on systems' actual capabilities and implementations---i.e., what AI automatons can do and how they do it---also limits our ability to speak to what capabilities and implementations are already present in practice, and what their attendant impacts might be.\looseness=-1

\section*{Acknowledgements}
We are grateful to Manohar Swaminathan and our colleagues in the MSR FATE group for thoughtful and insightful feedback and discussions. 

\bibliographystyle{ACM-Reference-Format}
\bibliography{references}


\begin{thebibliography}{254}


\ifx \showCODEN    \undefined \def \showCODEN     #1{\unskip}     \fi
\ifx \showDOI      \undefined \def \showDOI       #1{#1}\fi
\ifx \showISBNx    \undefined \def \showISBNx     #1{\unskip}     \fi
\ifx \showISBNxiii \undefined \def \showISBNxiii  #1{\unskip}     \fi
\ifx \showISSN     \undefined \def \showISSN      #1{\unskip}     \fi
\ifx \showLCCN     \undefined \def \showLCCN      #1{\unskip}     \fi
\ifx \shownote     \undefined \def \shownote      #1{#1}          \fi
\ifx \showarticletitle \undefined \def \showarticletitle #1{#1}   \fi
\ifx \showURL      \undefined \def \showURL       {\relax}        \fi
\providecommand\bibfield[2]{#2}
\providecommand\bibinfo[2]{#2}
\providecommand\natexlab[1]{#1}
\providecommand\showeprint[2][]{arXiv:#2}

\bibitem[Abbasiantaeb et~al\mbox{.}(2024)]%
        {abbasiantaeb2024let}
\bibfield{author}{\bibinfo{person}{Zahra Abbasiantaeb}, \bibinfo{person}{Yifei Yuan}, \bibinfo{person}{Evangelos Kanoulas}, {and} \bibinfo{person}{Mohammad Aliannejadi}.} \bibinfo{year}{2024}\natexlab{}.
\newblock \showarticletitle{Let the llms talk: Simulating human-to-human conversational qa via zero-shot llm-to-llm interactions}. In \bibinfo{booktitle}{\emph{Proceedings of the 17th ACM International Conference on Web Search and Data Mining}}. \bibinfo{pages}{8--17}.
\newblock


\bibitem[Abercrombie et~al\mbox{.}(2023)]%
        {abercrombie2023mirages}
\bibfield{author}{\bibinfo{person}{Gavin Abercrombie}, \bibinfo{person}{Amanda Cercas~Curry}, \bibinfo{person}{Tanvi Dinkar}, \bibinfo{person}{Verena Rieser}, {and} \bibinfo{person}{Zeerak Talat}.} \bibinfo{year}{2023}\natexlab{}.
\newblock \showarticletitle{Mirages. On Anthropomorphism in Dialogue Systems}. In \bibinfo{booktitle}{\emph{Proceedings of the 2023 Conference on Empirical Methods in Natural Language Processing}}, \bibfield{editor}{\bibinfo{person}{Houda Bouamor}, \bibinfo{person}{Juan Pino}, {and} \bibinfo{person}{Kalika Bali}} (Eds.). \bibinfo{publisher}{Association for Computational Linguistics}, \bibinfo{address}{Singapore}, \bibinfo{pages}{4776--4790}.
\newblock
\urldef\tempurl%
\url{https://doi.org/10.18653/v1/2023.emnlp-main.290}
\showDOI{\tempurl}


\bibitem[Abramson et~al\mbox{.}(2020)]%
        {abramson2020imitating}
\bibfield{author}{\bibinfo{person}{Josh Abramson}, \bibinfo{person}{Arun Ahuja}, \bibinfo{person}{Iain Barr}, \bibinfo{person}{Arthur Brussee}, \bibinfo{person}{Federico Carnevale}, \bibinfo{person}{Mary Cassin}, \bibinfo{person}{Rachita Chhaparia}, \bibinfo{person}{Stephen Clark}, \bibinfo{person}{Bogdan Damoc}, \bibinfo{person}{Andrew Dudzik}, {et~al\mbox{.}}} \bibinfo{year}{2020}\natexlab{}.
\newblock \showarticletitle{Imitating interactive intelligence}.
\newblock \bibinfo{journal}{\emph{arXiv preprint arXiv:2012.05672}} (\bibinfo{year}{2020}).
\newblock


\bibitem[Achenbach et~al\mbox{.}(2017)]%
        {achenbach2017fast}
\bibfield{author}{\bibinfo{person}{Jascha Achenbach}, \bibinfo{person}{Thomas Waltemate}, \bibinfo{person}{Marc~Erich Latoschik}, {and} \bibinfo{person}{Mario Botsch}.} \bibinfo{year}{2017}\natexlab{}.
\newblock \showarticletitle{Fast generation of realistic virtual humans}. In \bibinfo{booktitle}{\emph{Proceedings of the 23rd ACM symposium on virtual reality software and technology}}. \bibinfo{pages}{1--10}.
\newblock


\bibitem[Agnew et~al\mbox{.}(2024)]%
        {agnew2024illusion}
\bibfield{author}{\bibinfo{person}{William Agnew}, \bibinfo{person}{A~Stevie Bergman}, \bibinfo{person}{Jennifer Chien}, \bibinfo{person}{Mark D{\'\i}az}, \bibinfo{person}{Seliem El-Sayed}, \bibinfo{person}{Jaylen Pittman}, \bibinfo{person}{Shakir Mohamed}, {and} \bibinfo{person}{Kevin~R McKee}.} \bibinfo{year}{2024}\natexlab{}.
\newblock \showarticletitle{The illusion of artificial inclusion}. In \bibinfo{booktitle}{\emph{Proceedings of the CHI Conference on Human Factors in Computing Systems}}. \bibinfo{pages}{1--12}.
\newblock


\bibitem[Aher et~al\mbox{.}(2023)]%
        {aher2023using}
\bibfield{author}{\bibinfo{person}{Gati~V Aher}, \bibinfo{person}{Rosa~I Arriaga}, {and} \bibinfo{person}{Adam~Tauman Kalai}.} \bibinfo{year}{2023}\natexlab{}.
\newblock \showarticletitle{Using large language models to simulate multiple humans and replicate human subject studies}. In \bibinfo{booktitle}{\emph{International Conference on Machine Learning}}. PMLR, \bibinfo{pages}{337--371}.
\newblock


\bibitem[Airaksinen(2020)]%
        {airaksinen2020mimetic}
\bibfield{author}{\bibinfo{person}{Timo Airaksinen}.} \bibinfo{year}{2020}\natexlab{}.
\newblock \showarticletitle{Mimetic evil: A conceptual and ethical study}.
\newblock \bibinfo{journal}{\emph{Problemos}} \bibinfo{number}{98} (\bibinfo{year}{2020}), \bibinfo{pages}{58--70}.
\newblock


\bibitem[{Alex Shipps}(2025)]%
        {mitnews2025}
\bibfield{author}{\bibinfo{person}{{Alex Shipps}}.} \bibinfo{year}{2025}\natexlab{}.
\newblock \bibinfo{title}{{Teaching AI to communicate sounds like humans do}}.
\newblock \bibinfo{howpublished}{\url{https://news.mit.edu/2025/teaching-ai-communicate-sounds-humans-do-0109}}.
\newblock
\newblock
\shownote{[Online; last accessed January-2025]}.


\bibitem[Alexander et~al\mbox{.}(2005)]%
        {alexander2005gaming}
\bibfield{author}{\bibinfo{person}{Amy~L Alexander}, \bibinfo{person}{Tad Bruny{\'e}}, \bibinfo{person}{Jason Sidman}, \bibinfo{person}{Shawn~A Weil}, {et~al\mbox{.}}} \bibinfo{year}{2005}\natexlab{}.
\newblock \showarticletitle{From gaming to training: A review of studies on fidelity, immersion, presence, and buy-in and their effects on transfer in pc-based simulations and games}.
\newblock \bibinfo{journal}{\emph{DARWARS Training Impact Group}}  \bibinfo{volume}{5} (\bibinfo{year}{2005}), \bibinfo{pages}{1--14}.
\newblock


\bibitem[Araujo(2018)]%
        {araujo2018living}
\bibfield{author}{\bibinfo{person}{Theo Araujo}.} \bibinfo{year}{2018}\natexlab{}.
\newblock \showarticletitle{Living up to the chatbot hype: The influence of anthropomorphic design cues and communicative agency framing on conversational agent and company perceptions}.
\newblock \bibinfo{journal}{\emph{Computers in human behavior}}  \bibinfo{volume}{85} (\bibinfo{year}{2018}), \bibinfo{pages}{183--189}.
\newblock


\bibitem[Argyle et~al\mbox{.}(2023)]%
        {argyle2023out}
\bibfield{author}{\bibinfo{person}{Lisa~P Argyle}, \bibinfo{person}{Ethan~C Busby}, \bibinfo{person}{Nancy Fulda}, \bibinfo{person}{Joshua~R Gubler}, \bibinfo{person}{Christopher Rytting}, {and} \bibinfo{person}{David Wingate}.} \bibinfo{year}{2023}\natexlab{}.
\newblock \showarticletitle{Out of one, many: Using language models to simulate human samples}.
\newblock \bibinfo{journal}{\emph{Political Analysis}} \bibinfo{volume}{31}, \bibinfo{number}{3} (\bibinfo{year}{2023}), \bibinfo{pages}{337--351}.
\newblock


\bibitem[{Ashley Belanger}(2024)]%
        {arstechnica2024}
\bibfield{author}{\bibinfo{person}{{Ashley Belanger}}.} \bibinfo{year}{2024}\natexlab{}.
\newblock \bibinfo{title}{{Chatbots urged teen to self-harm, suggested murdering parents, lawsuit says}}.
\newblock \bibinfo{howpublished}{\url{https://arstechnica.com/tech-policy/2024/12/chatbots-urged-teen-to-self-harm-suggested-murdering-parents-lawsuit-says/}}.
\newblock
\newblock
\shownote{[Online; last accessed January-2025]}.


\bibitem[Ataei et~al\mbox{.}(2025)]%
        {ataei2024elicitron}
\bibfield{author}{\bibinfo{person}{Mohammadmehdi Ataei}, \bibinfo{person}{Hyunmin Cheong}, \bibinfo{person}{Daniele Grandi}, \bibinfo{person}{Ye Wang}, \bibinfo{person}{Nigel Morris}, {and} \bibinfo{person}{Alexander Tessier}.} \bibinfo{year}{2025}\natexlab{}.
\newblock \showarticletitle{Elicitron: A Large Language Model Agent-Based Simulation Framework for Design Requirements Elicitation}.
\newblock \bibinfo{journal}{\emph{Journal of Computing and Information Science in Engineering}} \bibinfo{volume}{25}, \bibinfo{number}{2} (\bibinfo{year}{2025}).
\newblock


\bibitem[Axelrod(2024)]%
        {cbsnews2024}
\bibfield{author}{\bibinfo{person}{Jim Axelrod}.} \bibinfo{year}{2024}\natexlab{}.
\newblock \bibinfo{title}{{Teen victim of AI-generated "deepfake pornography" urges Congress to pass "Take It Down Act"}}.
\newblock \bibinfo{howpublished}{\url{https://www.cbsnews.com/news/deepfake-pornography-victim-congress/}}.
\newblock
\newblock
\shownote{[Online; last accessed January-2025]}.


\bibitem[Bandara(2024)]%
        {petapixel2024}
\bibfield{author}{\bibinfo{person}{Pesala Bandara}.} \bibinfo{year}{2024}\natexlab{}.
\newblock \bibinfo{title}{{Instagram Now Lets You Create an AI Version of Yourself}}.
\newblock \bibinfo{howpublished}{\url{https://petapixel.com/2024/07/31/instagram-now-lets-you-create-an-ai-version-of-yourself/}}.
\newblock
\newblock
\shownote{[Online; last accessed January-2025]}.


\bibitem[Barnett(2023)]%
        {barnett2023ethical}
\bibfield{author}{\bibinfo{person}{Julia Barnett}.} \bibinfo{year}{2023}\natexlab{}.
\newblock \showarticletitle{The ethical implications of generative audio models: A systematic literature review}. In \bibinfo{booktitle}{\emph{Proceedings of the 2023 AAAI/ACM Conference on AI, Ethics, and Society}}. \bibinfo{pages}{146--161}.
\newblock


\bibitem[Beard(2001)]%
        {beard2001clones}
\bibfield{author}{\bibinfo{person}{Joseph~J Beard}.} \bibinfo{year}{2001}\natexlab{}.
\newblock \showarticletitle{Clones, bones and twilight zones: protecting the digital persona of the quick, the dead and the imaginary}.
\newblock \bibinfo{journal}{\emph{J. Copyright Soc'y USA}}  \bibinfo{volume}{49} (\bibinfo{year}{2001}), \bibinfo{pages}{441}.
\newblock


\bibitem[{Bernard Marr}(2024)]%
        {forbes2024}
\bibfield{author}{\bibinfo{person}{{Bernard Marr}}.} \bibinfo{year}{2024}\natexlab{}.
\newblock \bibinfo{title}{{The Uncanny Valley: Advancements And Anxieties Of AI That Mimics Life}}.
\newblock \bibinfo{howpublished}{\url{https://www.forbes.com/sites/bernardmarr/2024/02/07/the-uncanny-valley-advancements-and-anxieties-of-ai-that-mimics-life/}}.
\newblock
\newblock
\shownote{[Online; last accessed January-2025]}.


\bibitem[Bethea(2024a)]%
        {atlantic2024}
\bibfield{author}{\bibinfo{person}{Charles Bethea}.} \bibinfo{year}{2024}\natexlab{a}.
\newblock \bibinfo{title}{{The Problem With Counterfeit People}}.
\newblock \bibinfo{howpublished}{\url{https://www.theatlantic.com/technology/archive/2023/05/problem-counterfeit-people/674075/}}.
\newblock
\newblock
\shownote{[Online; last accessed January-2025]}.


\bibitem[Bethea(2024b)]%
        {newyorker2024}
\bibfield{author}{\bibinfo{person}{Charles Bethea}.} \bibinfo{year}{2024}\natexlab{b}.
\newblock \bibinfo{title}{{The Terrifying A.I. Scam That Uses Your Loved One’s Voice}}.
\newblock \bibinfo{howpublished}{\url{https://www.newyorker.com/science/annals-of-artificial-intelligence/the-terrifying-ai-scam-that-uses-your-loved-ones-voice}}.
\newblock
\newblock
\shownote{[Online; last accessed January-2025]}.


\bibitem[Biermann et~al\mbox{.}(2022)]%
        {biermann2022tool}
\bibfield{author}{\bibinfo{person}{Oloff~C Biermann}, \bibinfo{person}{Ning~F Ma}, {and} \bibinfo{person}{Dongwook Yoon}.} \bibinfo{year}{2022}\natexlab{}.
\newblock \showarticletitle{From tool to companion: Storywriters want AI writers to respect their personal values and writing strategies}. In \bibinfo{booktitle}{\emph{Proceedings of the 2022 ACM Designing Interactive Systems Conference}}. \bibinfo{pages}{1209--1227}.
\newblock


\bibitem[Binns and Edwards(2025)]%
        {edwards2024reputation}
\bibfield{author}{\bibinfo{person}{Reuben Binns} {and} \bibinfo{person}{Lilian Edwards}.} \bibinfo{year}{2025}\natexlab{}.
\newblock \showarticletitle{Reputation Management in the ChatGPT Era}.
\newblock \bibinfo{journal}{\emph{Forthcoming Oxford Handbook on the Foundations and Regulation of Generative AI}} (\bibinfo{year}{2025}).
\newblock
\newblock
\shownote{Available at SSRN: \url{https://ssrn.com/abstract=5026615} or \url{http://dx.doi.org/10.2139/ssrn.5026615}}.


\bibitem[Birhane et~al\mbox{.}(2022)]%
        {birhane2022values}
\bibfield{author}{\bibinfo{person}{Abeba Birhane}, \bibinfo{person}{Pratyusha Kalluri}, \bibinfo{person}{Dallas Card}, \bibinfo{person}{William Agnew}, \bibinfo{person}{Ravit Dotan}, {and} \bibinfo{person}{Michelle Bao}.} \bibinfo{year}{2022}\natexlab{}.
\newblock \showarticletitle{The values encoded in machine learning research}. In \bibinfo{booktitle}{\emph{Proceedings of the 2022 ACM Conference on Fairness, Accountability, and Transparency}}. \bibinfo{pages}{173--184}.
\newblock


\bibitem[Boine(2023)]%
        {boine2023emotional}
\bibfield{author}{\bibinfo{person}{Claire Boine}.} \bibinfo{year}{2023}\natexlab{}.
\newblock \showarticletitle{Emotional {Attachment} to {AI} {Companions} and {European} {Law}}.
\newblock \bibinfo{journal}{\emph{MIT Case Studies in Social and Ethical Responsibilities of Computing}} \bibinfo{number}{Winter 2023} (\bibinfo{date}{feb 27} \bibinfo{year}{2023}).
\newblock
\newblock
\shownote{https://mit-serc.pubpub.org/pub/ai-companions-eu-law}.


\bibitem[Boyarskaya et~al\mbox{.}(2020)]%
        {boyarskaya2020overcoming}
\bibfield{author}{\bibinfo{person}{Margarita Boyarskaya}, \bibinfo{person}{Alexandra Olteanu}, {and} \bibinfo{person}{Kate Crawford}.} \bibinfo{year}{2020}\natexlab{}.
\newblock \showarticletitle{Overcoming Failures of Imagination in AI Infused System Development and Deployment}. In \bibinfo{booktitle}{\emph{In the Navigating the Broader Impacts of AI Research Workshop at NeurIPS 2020}}.
\newblock
\urldef\tempurl%
\url{https://www.microsoft.com/en-us/research/publication/overcoming-failures-of-imagination-in-ai-infused-system-development-and-deployment/}
\showURL{%
\tempurl}


\bibitem[Brandtzaeg et~al\mbox{.}(2022)]%
        {brandtzaeg2022my}
\bibfield{author}{\bibinfo{person}{Petter~Bae Brandtzaeg}, \bibinfo{person}{Marita Skjuve}, {and} \bibinfo{person}{Asbj{\o}rn F{\o}lstad}.} \bibinfo{year}{2022}\natexlab{}.
\newblock \showarticletitle{My AI friend: How users of a social chatbot understand their human--AI friendship}.
\newblock \bibinfo{journal}{\emph{Human Communication Research}} \bibinfo{volume}{48}, \bibinfo{number}{3} (\bibinfo{year}{2022}), \bibinfo{pages}{404--429}.
\newblock


\bibitem[Brigham et~al\mbox{.}(2024)]%
        {brigham2024violation}
\bibfield{author}{\bibinfo{person}{Natalie~Grace Brigham}, \bibinfo{person}{Miranda Wei}, \bibinfo{person}{Tadayoshi Kohno}, {and} \bibinfo{person}{Elissa~M. Redmiles}.} \bibinfo{year}{2024}\natexlab{}.
\newblock \showarticletitle{"Violation of my body": perceptions of AI-generated non-consensual (intimate) imagery}. In \bibinfo{booktitle}{\emph{Proceedings of the Twentieth USENIX Conference on Usable Privacy and Security}} (Philadelphia, PA, USA) \emph{(\bibinfo{series}{SOUPS '24})}. \bibinfo{publisher}{USENIX Association}, \bibinfo{address}{USA}, Article \bibinfo{articleno}{20}, \bibinfo{numpages}{20}~pages.
\newblock
\showISBNx{978-1-939133-42-7}


\bibitem[Browne(2024)]%
        {cnbc2024}
\bibfield{author}{\bibinfo{person}{Ryan Browne}.} \bibinfo{year}{2024}\natexlab{}.
\newblock \bibinfo{title}{{Nvidia-backed startup Synthesia unveils AI avatars that can convey human emotions}}.
\newblock \bibinfo{howpublished}{\url{https://www.cnbc.com/2024/04/25/nvidia-backed-synthesia-unveils-ai-avatars-that-can-be-generated-from-text.html}}.
\newblock
\newblock
\shownote{[Online; last accessed January-2025]}.


\bibitem[{Bruna Horvath}(2024)]%
        {nbcnews2024}
\bibfield{author}{\bibinfo{person}{{Bruna Horvath}}.} \bibinfo{year}{2024}\natexlab{}.
\newblock \bibinfo{title}{{Coca-Cola causes controversy with AI-made ad}}.
\newblock \bibinfo{howpublished}{\url{https://www.nbcnews.com/tech/innovation/coca-cola-causes-controversy-ai-made-ad-rcna180665}}.
\newblock
\newblock
\shownote{[Online; last accessed January-2025]}.


\bibitem[Brynjolfsson(2023)]%
        {brynjolfsson2023turing}
\bibfield{author}{\bibinfo{person}{Erik Brynjolfsson}.} \bibinfo{year}{2023}\natexlab{}.
\newblock \showarticletitle{The turing trap: The promise \& peril of human-like artificial intelligence}.
\newblock In \bibinfo{booktitle}{\emph{Augmented education in the global age}}. \bibinfo{publisher}{Routledge}, \bibinfo{pages}{103--116}.
\newblock


\bibitem[Bu{\c{c}}inca et~al\mbox{.}(2023)]%
        {buccinca2023aha}
\bibfield{author}{\bibinfo{person}{Zana Bu{\c{c}}inca}, \bibinfo{person}{Chau~Minh Pham}, \bibinfo{person}{Maurice Jakesch}, \bibinfo{person}{Marco~Tulio Ribeiro}, \bibinfo{person}{Alexandra Olteanu}, {and} \bibinfo{person}{Saleema Amershi}.} \bibinfo{year}{2023}\natexlab{}.
\newblock \showarticletitle{Aha!: Facilitating ai impact assessment by generating examples of harms}.
\newblock \bibinfo{journal}{\emph{arXiv preprint arXiv:2306.03280}} (\bibinfo{year}{2023}).
\newblock


\bibitem[Byeon et~al\mbox{.}(2022)]%
        {byeon2022voice}
\bibfield{author}{\bibinfo{person}{Hyeon~Jeong Byeon}, \bibinfo{person}{Chaerin Lee}, \bibinfo{person}{Jeemin Lee}, {and} \bibinfo{person}{Uran Oh}.} \bibinfo{year}{2022}\natexlab{}.
\newblock \showarticletitle{“A voice that suits the situation”: Understanding the needs and challenges for supporting end-user voice customization}. In \bibinfo{booktitle}{\emph{Proceedings of the 2022 CHI Conference on Human Factors in Computing Systems}}. \bibinfo{pages}{1--10}.
\newblock


\bibitem[Byun et~al\mbox{.}(2023)]%
        {byun2023dispensing}
\bibfield{author}{\bibinfo{person}{Courtni Byun}, \bibinfo{person}{Piper Vasicek}, {and} \bibinfo{person}{Kevin Seppi}.} \bibinfo{year}{2023}\natexlab{}.
\newblock \showarticletitle{Dispensing with humans in human-computer interaction research}. In \bibinfo{booktitle}{\emph{Extended Abstracts of the 2023 CHI Conference on Human Factors in Computing Systems}}. \bibinfo{pages}{1--26}.
\newblock


\bibitem[Campbell et~al\mbox{.}(2022)]%
        {campbell2022preparing}
\bibfield{author}{\bibinfo{person}{Colin Campbell}, \bibinfo{person}{Kirk Plangger}, \bibinfo{person}{Sean Sands}, {and} \bibinfo{person}{Jan Kietzmann}.} \bibinfo{year}{2022}\natexlab{}.
\newblock \showarticletitle{Preparing for an era of deepfakes and AI-generated ads: A framework for understanding responses to manipulated advertising}.
\newblock \bibinfo{journal}{\emph{Journal of Advertising}} \bibinfo{volume}{51}, \bibinfo{number}{1} (\bibinfo{year}{2022}), \bibinfo{pages}{22--38}.
\newblock


\bibitem[Cao et~al\mbox{.}(2023)]%
        {cao2023high}
\bibfield{author}{\bibinfo{person}{Qiongdan Cao}, \bibinfo{person}{Hui Yu}, \bibinfo{person}{Paul Charisse}, \bibinfo{person}{Si Qiao}, {and} \bibinfo{person}{Brett Stevens}.} \bibinfo{year}{2023}\natexlab{}.
\newblock \showarticletitle{Is high-fidelity important for human-like virtual avatars in human computer interactions?}
\newblock \bibinfo{journal}{\emph{International Journal of Network Dynamics and Intelligence}} (\bibinfo{year}{2023}), \bibinfo{pages}{15--23}.
\newblock


\bibitem[Center(2024)]%
        {ftc}
\bibfield{author}{\bibinfo{person}{FTC Consumer~Response Center}.} \bibinfo{year}{2024}\natexlab{}.
\newblock \bibinfo{title}{{FTC Proposes New Protections to Combat AI Impersonation of Individuals}}.
\newblock \bibinfo{howpublished}{\url{https://www.ftc.gov/news-events/news/press-releases/2024/02/ftc-proposes-new-protections-combat-ai-impersonation-individuals}}.
\newblock
\newblock
\shownote{[Online; accessed 14-August-2024]}.


\bibitem[Chandra et~al\mbox{.}(2024)]%
        {chandra2024lived}
\bibfield{author}{\bibinfo{person}{Mohit Chandra}, \bibinfo{person}{Suchismita Naik}, \bibinfo{person}{Denae Ford}, \bibinfo{person}{Ebele Okoli}, \bibinfo{person}{Munmun De~Choudhury}, \bibinfo{person}{Mahsa Ershadi}, \bibinfo{person}{Gonzalo Ramos}, \bibinfo{person}{Javier Hernandez}, \bibinfo{person}{Ananya Bhattacharjee}, \bibinfo{person}{Shahed Warreth}, {et~al\mbox{.}}} \bibinfo{year}{2024}\natexlab{}.
\newblock \showarticletitle{From Lived Experience to Insight: Unpacking the Psychological Risks of Using AI Conversational Agents}.
\newblock \bibinfo{journal}{\emph{arXiv preprint arXiv:2412.07951}} (\bibinfo{year}{2024}).
\newblock


\bibitem[Chavali et~al\mbox{.}(2024)]%
        {chavali2024ai}
\bibfield{author}{\bibinfo{person}{Durga Chavali}, \bibinfo{person}{Vinod~Kumar Dhiman}, {and} \bibinfo{person}{Siri~Chandana Katari}.} \bibinfo{year}{2024}\natexlab{}.
\newblock \showarticletitle{AI-Powered Virtual Health Assistants: Transforming Patient Engagement Through Virtual Nursing}.
\newblock \bibinfo{journal}{\emph{Int. J. of Pharm. Sci}}  \bibinfo{volume}{2} (\bibinfo{year}{2024}), \bibinfo{pages}{613--624}.
\newblock


\bibitem[Chaves et~al\mbox{.}(2022)]%
        {chaves2022chatbots}
\bibfield{author}{\bibinfo{person}{Ana~Paula Chaves}, \bibinfo{person}{Jesse Egbert}, \bibinfo{person}{Toby Hocking}, \bibinfo{person}{Eck Doerry}, {and} \bibinfo{person}{Marco~Aurelio Gerosa}.} \bibinfo{year}{2022}\natexlab{}.
\newblock \showarticletitle{Chatbots language design: The influence of language variation on user experience with tourist assistant chatbots}.
\newblock \bibinfo{journal}{\emph{ACM Transactions on Computer-Human Interaction}} \bibinfo{volume}{29}, \bibinfo{number}{2} (\bibinfo{year}{2022}), \bibinfo{pages}{1--38}.
\newblock


\bibitem[Cheng et~al\mbox{.}(2025)]%
        {cheng2025dehumanizing}
\bibfield{author}{\bibinfo{person}{Myra Cheng}, \bibinfo{person}{Su~Lin Blodgett}, \bibinfo{person}{Alicia DeVrio}, \bibinfo{person}{Lisa Egede}, {and} \bibinfo{person}{Alexandra Olteanu}.} \bibinfo{year}{2025}\natexlab{}.
\newblock \showarticletitle{Dehumanizing Machines: Mitigating Anthropomorphic Behaviors in Text Generation Systems}.
\newblock \bibinfo{journal}{\emph{arXiv preprint arXiv:2502.14019}} (\bibinfo{year}{2025}).
\newblock


\bibitem[Cheng et~al\mbox{.}(2024)]%
        {cheng2024one}
\bibfield{author}{\bibinfo{person}{Myra Cheng}, \bibinfo{person}{Alicia DeVrio}, \bibinfo{person}{Lisa Egede}, \bibinfo{person}{Su~Lin Blodgett}, {and} \bibinfo{person}{Alexandra Olteanu}.} \bibinfo{year}{2024}\natexlab{}.
\newblock \showarticletitle{“I Am the One and Only, Your Cyber BFF”: Understanding the Impact of GenAI Requires Understanding the Impact of Anthropomorphic AI}.
\newblock \bibinfo{journal}{\emph{arXiv preprint arXiv:2410.08526}} (\bibinfo{year}{2024}).
\newblock


\bibitem[Cheng et~al\mbox{.}(2023)]%
        {cheng2023compost}
\bibfield{author}{\bibinfo{person}{Myra Cheng}, \bibinfo{person}{Tiziano Piccardi}, {and} \bibinfo{person}{Diyi Yang}.} \bibinfo{year}{2023}\natexlab{}.
\newblock \showarticletitle{CoMPosT: Characterizing and Evaluating Caricature in LLM Simulations}. In \bibinfo{booktitle}{\emph{Proceedings of the 2023 Conference on Empirical Methods in Natural Language Processing}}. \bibinfo{pages}{10853--10875}.
\newblock


\bibitem[{Chris Stokel-Walker}(2021)]%
        {newscientist2021}
\bibfield{author}{\bibinfo{person}{{Chris Stokel-Walker}}.} \bibinfo{year}{2021}\natexlab{}.
\newblock \bibinfo{title}{{AI that mimics human typos on a smartphone could improve keyboards}}.
\newblock \bibinfo{howpublished}{\url{https://www.newscientist.com/article/2277214-ai-that-mimics-human-typos-on-a-smartphone-could-improve-keyboards/}}.
\newblock
\newblock
\shownote{[Online; last accessed January-2025]}.


\bibitem[{Chris Stokel-Walker}(2024)]%
        {fastcompany2024}
\bibfield{author}{\bibinfo{person}{{Chris Stokel-Walker}}.} \bibinfo{year}{2024}\natexlab{}.
\newblock \bibinfo{title}{{Spotify is full of AI music, and some say it’s ruining the platform}}.
\newblock \bibinfo{howpublished}{\url{https://www.fastcompany.com/91170296/spotify-ai-music}}.
\newblock
\newblock
\shownote{[Online; last accessed January-2025]}.


\bibitem[Chui et~al\mbox{.}(2016)]%
        {chui2016machines}
\bibfield{author}{\bibinfo{person}{Michael Chui}, \bibinfo{person}{James Manyika}, {and} \bibinfo{person}{Mehdi Miremadi}.} \bibinfo{year}{2016}\natexlab{}.
\newblock \showarticletitle{Where machines could replace humans-and where they can't (yet)}.
\newblock \bibinfo{journal}{\emph{The McKinsey Quarterly}} (\bibinfo{year}{2016}), \bibinfo{pages}{1--12}.
\newblock


\bibitem[C{\'\i}fka et~al\mbox{.}(2020)]%
        {cifka2020groove2groove}
\bibfield{author}{\bibinfo{person}{Ond{\v{r}}ej C{\'\i}fka}, \bibinfo{person}{Umut {\c{S}}im{\c{s}}ekli}, {and} \bibinfo{person}{Ga{\"e}l Richard}.} \bibinfo{year}{2020}\natexlab{}.
\newblock \showarticletitle{Groove2groove: One-shot music style transfer with supervision from synthetic data}.
\newblock \bibinfo{journal}{\emph{IEEE/ACM Transactions on Audio, Speech, and Language Processing}}  \bibinfo{volume}{28} (\bibinfo{year}{2020}), \bibinfo{pages}{2638--2650}.
\newblock


\bibitem[Clarke and Dietz(2024)]%
        {clarke2024llm}
\bibfield{author}{\bibinfo{person}{Charles~LA Clarke} {and} \bibinfo{person}{Laura Dietz}.} \bibinfo{year}{2024}\natexlab{}.
\newblock \showarticletitle{LLM-based relevance assessment still can't replace human relevance assessment}.
\newblock \bibinfo{journal}{\emph{arXiv preprint arXiv:2412.17156}} (\bibinfo{year}{2024}).
\newblock


\bibitem[Cohn et~al\mbox{.}(2024)]%
        {cohn2024believing}
\bibfield{author}{\bibinfo{person}{Michelle Cohn}, \bibinfo{person}{Mahima Pushkarna}, \bibinfo{person}{Gbolahan~O Olanubi}, \bibinfo{person}{Joseph~M Moran}, \bibinfo{person}{Daniel Padgett}, \bibinfo{person}{Zion Mengesha}, {and} \bibinfo{person}{Courtney Heldreth}.} \bibinfo{year}{2024}\natexlab{}.
\newblock \showarticletitle{Believing Anthropomorphism: Examining the Role of Anthropomorphic Cues on Trust in Large Language Models}. In \bibinfo{booktitle}{\emph{Extended Abstracts of the CHI Conference on Human Factors in Computing Systems}}. \bibinfo{pages}{1--15}.
\newblock


\bibitem[Collins et~al\mbox{.}(2024)]%
        {collins2024building}
\bibfield{author}{\bibinfo{person}{Katherine~M Collins}, \bibinfo{person}{Ilia Sucholutsky}, \bibinfo{person}{Umang Bhatt}, \bibinfo{person}{Kartik Chandra}, \bibinfo{person}{Lionel Wong}, \bibinfo{person}{Mina Lee}, \bibinfo{person}{Cedegao~E Zhang}, \bibinfo{person}{Tan Zhi-Xuan}, \bibinfo{person}{Mark Ho}, \bibinfo{person}{Vikash Mansinghka}, {et~al\mbox{.}}} \bibinfo{year}{2024}\natexlab{}.
\newblock \showarticletitle{Building machines that learn and think with people}.
\newblock \bibinfo{journal}{\emph{Nature human behaviour}} \bibinfo{volume}{8}, \bibinfo{number}{10} (\bibinfo{year}{2024}), \bibinfo{pages}{1851--1863}.
\newblock


\bibitem[Connolly et~al\mbox{.}(2022)]%
        {connolly2022recommendations}
\bibfield{author}{\bibinfo{person}{John~B Connolly}, \bibinfo{person}{John~D Mumford}, \bibinfo{person}{Debora~CM Glandorf}, \bibinfo{person}{Sarah Hartley}, \bibinfo{person}{Owen~T Lewis}, \bibinfo{person}{Sam~Weiss Evans}, \bibinfo{person}{Geoff Turner}, \bibinfo{person}{Camilla Beech}, \bibinfo{person}{Naima Sykes}, \bibinfo{person}{Mamadou~B Coulibaly}, {et~al\mbox{.}}} \bibinfo{year}{2022}\natexlab{}.
\newblock \showarticletitle{Recommendations for environmental risk assessment of gene drive applications for malaria vector control}.
\newblock \bibinfo{journal}{\emph{Malaria journal}} \bibinfo{volume}{21}, \bibinfo{number}{1} (\bibinfo{year}{2022}), \bibinfo{pages}{152}.
\newblock


\bibitem[Criddle" and Murphy"(2024)]%
        {financialtimes2024}
\bibfield{author}{\bibinfo{person}{"Cristina Criddle"} {and} \bibinfo{person}{"Hannah Murphy"}.} \bibinfo{year}{2024}\natexlab{}.
\newblock \bibinfo{title}{{Meta envisages social media filled with AI-generated users}}.
\newblock \bibinfo{howpublished}{\url{https://www.ft.com/content/91183cbb-50f9-464a-9d2e-96063825bfcf}}.
\newblock
\newblock
\shownote{[Online; last accessed January-2025]}.


\bibitem[Crockett(2025)]%
        {guardian2025AIempathy}
\bibfield{author}{\bibinfo{person}{MJ Crockett}.} \bibinfo{year}{2025}\natexlab{}.
\newblock \bibinfo{title}{{AI is ‘beating’ humans at empathy and creativity. But these games are rigged}}.
\newblock \bibinfo{howpublished}{\url{https://www.theguardian.com/commentisfree/2025/feb/28/ai-empathy-humans}}.
\newblock
\newblock
\shownote{[Online; last accessed February-2025]}.


\bibitem[Cuthbertson(2025)]%
        {independent2025AIredline}
\bibfield{author}{\bibinfo{person}{Anthony Cuthbertson}.} \bibinfo{year}{2025}\natexlab{}.
\newblock \bibinfo{title}{{AI crosses `red line' after learning to replicate itself}}.
\newblock \bibinfo{howpublished}{\url{https://www.independent.co.uk/tech/ai-red-line-b2687013.html}}.
\newblock
\newblock
\shownote{[Online; last accessed February-2025]}.


\bibitem[Dai(2024)]%
        {dai2024beyond}
\bibfield{author}{\bibinfo{person}{Jessica Dai}.} \bibinfo{year}{2024}\natexlab{}.
\newblock \showarticletitle{Beyond Personhood: Agency, Accountability, and the Limits of Anthropomorphic Ethical Analysis}.
\newblock \bibinfo{journal}{\emph{arXiv preprint arXiv:2404.13861}} (\bibinfo{year}{2024}).
\newblock


\bibitem[Danry et~al\mbox{.}(2022)]%
        {danry2022ai}
\bibfield{author}{\bibinfo{person}{Valdemar Danry}, \bibinfo{person}{Joanne Leong}, \bibinfo{person}{Pat Pataranutaporn}, \bibinfo{person}{Pulkit Tandon}, \bibinfo{person}{Yimeng Liu}, \bibinfo{person}{Roy Shilkrot}, \bibinfo{person}{Parinya Punpongsanon}, \bibinfo{person}{Tsachy Weissman}, \bibinfo{person}{Pattie Maes}, {and} \bibinfo{person}{Misha Sra}.} \bibinfo{year}{2022}\natexlab{}.
\newblock \showarticletitle{AI-generated characters: putting deepfakes to good use}. In \bibinfo{booktitle}{\emph{CHI Conference on Human Factors in Computing Systems Extended Abstracts}}. \bibinfo{pages}{1--5}.
\newblock


\bibitem[Dautenhahn(2007)]%
        {dautenhahn2007socially}
\bibfield{author}{\bibinfo{person}{Kerstin Dautenhahn}.} \bibinfo{year}{2007}\natexlab{}.
\newblock \showarticletitle{Socially intelligent robots: dimensions of human--robot interaction}.
\newblock \bibinfo{journal}{\emph{Philosophical transactions of the royal society B: Biological sciences}} \bibinfo{volume}{362}, \bibinfo{number}{1480} (\bibinfo{year}{2007}), \bibinfo{pages}{679--704}.
\newblock


\bibitem[De~Cremer and Kasparov(2021)]%
        {de2021ai}
\bibfield{author}{\bibinfo{person}{David De~Cremer} {and} \bibinfo{person}{Garry Kasparov}.} \bibinfo{year}{2021}\natexlab{}.
\newblock \showarticletitle{AI should augment human intelligence, not replace it}.
\newblock \bibinfo{journal}{\emph{Harvard Business Review}} \bibinfo{volume}{18}, \bibinfo{number}{1} (\bibinfo{year}{2021}).
\newblock


\bibitem[De~Freitas et~al\mbox{.}(2024a)]%
        {de2024lessons}
\bibfield{author}{\bibinfo{person}{Julian De~Freitas}, \bibinfo{person}{Noah Castelo}, \bibinfo{person}{Ahmet~Kaan Uğuralp}, {and} \bibinfo{person}{Zeliha Uğuralp}.} \bibinfo{year}{2024}\natexlab{a}.
\newblock \showarticletitle{Lessons From an App Update at Replika AI: Identity Discontinuity in Human-AI Relationships}.
\newblock  (\bibinfo{date}{December 4} \bibinfo{year}{2024}).
\newblock
\urldef\tempurl%
\url{https://doi.org/10.2139/ssrn.4976449}
\showDOI{\tempurl}
\newblock
\shownote{Harvard Business Working Paper No. 25-018}.


\bibitem[De~Freitas et~al\mbox{.}(2024b)]%
        {de2024ai}
\bibfield{author}{\bibinfo{person}{Julian De~Freitas}, \bibinfo{person}{Ahmet~Kaan U{\u{g}}uralp}, \bibinfo{person}{Zeliha U{\u{g}}uralp}, {and} \bibinfo{person}{Stefano Puntoni}.} \bibinfo{year}{2024}\natexlab{b}.
\newblock \showarticletitle{Ai companions reduce loneliness}.
\newblock  (\bibinfo{year}{2024}).
\newblock


\bibitem[De~Visser et~al\mbox{.}(2016)]%
        {de2016almost}
\bibfield{author}{\bibinfo{person}{Ewart~J De~Visser}, \bibinfo{person}{Samuel~S Monfort}, \bibinfo{person}{Ryan McKendrick}, \bibinfo{person}{Melissa~AB Smith}, \bibinfo{person}{Patrick~E McKnight}, \bibinfo{person}{Frank Krueger}, {and} \bibinfo{person}{Raja Parasuraman}.} \bibinfo{year}{2016}\natexlab{}.
\newblock \showarticletitle{Almost human: Anthropomorphism increases trust resilience in cognitive agents.}
\newblock \bibinfo{journal}{\emph{Journal of Experimental Psychology: Applied}} \bibinfo{volume}{22}, \bibinfo{number}{3} (\bibinfo{year}{2016}), \bibinfo{pages}{331}.
\newblock


\bibitem[Delgado et~al\mbox{.}(2023)]%
        {delgado2023participatory}
\bibfield{author}{\bibinfo{person}{Fernando Delgado}, \bibinfo{person}{Stephen Yang}, \bibinfo{person}{Michael Madaio}, {and} \bibinfo{person}{Qian Yang}.} \bibinfo{year}{2023}\natexlab{}.
\newblock \showarticletitle{The participatory turn in ai design: Theoretical foundations and the current state of practice}. In \bibinfo{booktitle}{\emph{Proceedings of the 3rd ACM Conference on Equity and Access in Algorithms, Mechanisms, and Optimization}}. \bibinfo{pages}{1--23}.
\newblock


\bibitem[Deshpande et~al\mbox{.}(2024)]%
        {deshpande2024perceptions}
\bibfield{author}{\bibinfo{person}{Manoj Deshpande}, \bibinfo{person}{Jisu Park}, \bibinfo{person}{Supratim Pait}, {and} \bibinfo{person}{Brian Magerko}.} \bibinfo{year}{2024}\natexlab{}.
\newblock \showarticletitle{Perceptions of Interaction Dynamics in Co-Creative AI: A Comparative Study of Interaction Modalities in Drawcto}. In \bibinfo{booktitle}{\emph{Proceedings of the 16th Conference on Creativity \& Cognition}}. \bibinfo{pages}{102--116}.
\newblock


\bibitem[DeVrio et~al\mbox{.}(2025)]%
        {devrio2025taxonomy}
\bibfield{author}{\bibinfo{person}{Alicia DeVrio}, \bibinfo{person}{Myra Cheng}, \bibinfo{person}{Lisa Egede}, \bibinfo{person}{Alexandra Olteanu}, {and} \bibinfo{person}{Su~Lin Blodgett}.} \bibinfo{year}{2025}\natexlab{}.
\newblock \showarticletitle{A Taxonomy of Linguistic Expressions That Contribute To Anthropomorphism of Language Technologies}. In \bibinfo{booktitle}{\emph{Proceedings of the 2025 CHI Conference on Human Factors in Computing Systems}} (Yokohama, Japan) \emph{(\bibinfo{series}{CHI '25})}. \bibinfo{publisher}{Association for Computing Machinery}.
\newblock
\urldef\tempurl%
\url{https://doi.org/10.1145/3706598.3714038}
\showDOI{\tempurl}


\bibitem[Dillion et~al\mbox{.}(2023)]%
        {dillion2023can}
\bibfield{author}{\bibinfo{person}{Danica Dillion}, \bibinfo{person}{Niket Tandon}, \bibinfo{person}{Yuling Gu}, {and} \bibinfo{person}{Kurt Gray}.} \bibinfo{year}{2023}\natexlab{}.
\newblock \showarticletitle{Can AI language models replace human participants?}
\newblock \bibinfo{journal}{\emph{Trends in Cognitive Sciences}} \bibinfo{volume}{27}, \bibinfo{number}{7} (\bibinfo{year}{2023}), \bibinfo{pages}{597--600}.
\newblock


\bibitem[Dunn(2020)]%
        {dunn2020identity}
\bibfield{author}{\bibinfo{person}{Suzie Dunn}.} \bibinfo{year}{2020}\natexlab{}.
\newblock \showarticletitle{Identity manipulation: Responding to advances in artificial intelligence and robotics}. In \bibinfo{booktitle}{\emph{Suzie Dunn,“Identity Manipulation: Responding to Advances in Artificial Intelligence and Robotics”(2020) WeRobot, 2020, Conference Paper}}.
\newblock


\bibitem[{Emily Barnes}(2025)]%
        {vktr2025}
\bibfield{author}{\bibinfo{person}{{Emily Barnes}}.} \bibinfo{year}{2025}\natexlab{}.
\newblock \bibinfo{title}{{When AI Brings Back the Dead: Balancing Comfort and Consequences}}.
\newblock \bibinfo{howpublished}{\url{https://www.vktr.com/ai-ethics-law-risk/when-ai-brings-back-the-dead-balancing-comfort-and-consequences/}}.
\newblock
\newblock
\shownote{[Online; last accessed January-2025]}.


\bibitem[{Emma Goldberg}(2024)]%
        {nytimes2024meaningless}
\bibfield{author}{\bibinfo{person}{{Emma Goldberg}}.} \bibinfo{year}{2024}\natexlab{}.
\newblock \bibinfo{title}{{Will A.I. Kill Meaningless Jobs?}}
\newblock \bibinfo{howpublished}{\url{https://www.nytimes.com/2024/08/03/business/ai-replacing-jobs.html}}.
\newblock
\newblock
\shownote{[Online; last accessed January-2025]}.


\bibitem[{Emma Roth}(2024)]%
        {theverge2024-shorts}
\bibfield{author}{\bibinfo{person}{{Emma Roth}}.} \bibinfo{year}{2024}\natexlab{}.
\newblock \bibinfo{title}{{TCL’s new AI short films range from bad comedy to existential horror}}.
\newblock \bibinfo{howpublished}{\url{https://www.theverge.com/2024/12/21/24319502/tcl-new-ai-films-bad-comedy-existential-horror-ranked}}.
\newblock
\newblock
\shownote{[Online; last accessed January-2025]}.


\bibitem[Epley et~al\mbox{.}(2007)]%
        {epley2007seeing}
\bibfield{author}{\bibinfo{person}{Nicholas Epley}, \bibinfo{person}{Adam Waytz}, {and} \bibinfo{person}{John~T Cacioppo}.} \bibinfo{year}{2007}\natexlab{}.
\newblock \showarticletitle{On seeing human: a three-factor theory of anthropomorphism.}
\newblock \bibinfo{journal}{\emph{Psychological review}} \bibinfo{volume}{114}, \bibinfo{number}{4} (\bibinfo{year}{2007}), \bibinfo{pages}{864}.
\newblock


\bibitem[{Eric Hal Schwartz}(2025)]%
        {techradar2025}
\bibfield{author}{\bibinfo{person}{{Eric Hal Schwartz}}.} \bibinfo{year}{2025}\natexlab{}.
\newblock \bibinfo{title}{{One conversation is all it takes for this AI to deepfake your entire personality}}.
\newblock \bibinfo{howpublished}{\url{https://www.techradar.com/computing/artificial-intelligence/one-conversation-is-all-it-takes-for-this-ai-to-deepfake-your-entire-personality}}.
\newblock
\newblock
\shownote{[Online; last accessed January-2025]}.


\bibitem[Erscoi et~al\mbox{.}(2023)]%
        {erscoi2023pygmalion}
\bibfield{author}{\bibinfo{person}{Lelia Erscoi}, \bibinfo{person}{Annelies Kleinherenbrink}, {and} \bibinfo{person}{Olivia Guest}.} \bibinfo{year}{2023}\natexlab{}.
\newblock \showarticletitle{{Pygmalion displacement: when humanising AI dehumanises women}}.
\newblock  (\bibinfo{year}{2023}).
\newblock


\bibitem[{Familia AI}(2024)]%
        {FamiliaAI}
\bibfield{author}{\bibinfo{person}{{Familia AI}}.} \bibinfo{year}{2024}\natexlab{}.
\newblock \bibinfo{title}{{Your AI Family}}.
\newblock \bibinfo{howpublished}{\url{https://familia.ai/}}.
\newblock
\newblock
\shownote{[Online; last accessed January-2025]}.


\bibitem[Favela and Amon(2023)]%
        {favela2023ethics}
\bibfield{author}{\bibinfo{person}{Luis~H Favela} {and} \bibinfo{person}{Mary~Jean Amon}.} \bibinfo{year}{2023}\natexlab{}.
\newblock \showarticletitle{The ethics of human digital twins: Counterfeit people, personhood, and the right to privacy}. In \bibinfo{booktitle}{\emph{2023 IEEE 3rd International Conference on Digital Twins and Parallel Intelligence (DTPI)}}. IEEE, \bibinfo{pages}{1--6}.
\newblock


\bibitem[F{\o}lstad et~al\mbox{.}(2019)]%
        {folstad2019different}
\bibfield{author}{\bibinfo{person}{Asbj{\o}rn F{\o}lstad}, \bibinfo{person}{Marita Skjuve}, {and} \bibinfo{person}{Petter~Bae Brandtzaeg}.} \bibinfo{year}{2019}\natexlab{}.
\newblock \showarticletitle{Different chatbots for different purposes: towards a typology of chatbots to understand interaction design}. In \bibinfo{booktitle}{\emph{Internet Science: INSCI 2018 International Workshops, St. Petersburg, Russia, October 24--26, 2018, Revised Selected Papers 5}}. Springer, \bibinfo{pages}{145--156}.
\newblock


\bibitem[Fong et~al\mbox{.}(2003)]%
        {fong2003survey}
\bibfield{author}{\bibinfo{person}{Terrence Fong}, \bibinfo{person}{Illah Nourbakhsh}, {and} \bibinfo{person}{Kerstin Dautenhahn}.} \bibinfo{year}{2003}\natexlab{}.
\newblock \showarticletitle{A survey of socially interactive robots}.
\newblock \bibinfo{journal}{\emph{Robotics and autonomous systems}} \bibinfo{volume}{42}, \bibinfo{number}{3-4} (\bibinfo{year}{2003}), \bibinfo{pages}{143--166}.
\newblock


\bibitem[Franzen(2024)]%
        {ventrurebeat2024}
\bibfield{author}{\bibinfo{person}{Carl Franzen}.} \bibinfo{year}{2024}\natexlab{}.
\newblock \bibinfo{title}{{'Uncanny': ChatGPT's Advanced Voice Mode is blowing minds}}.
\newblock \bibinfo{howpublished}{\url{https://venturebeat.com/ai/uncanny-chatgpts-advanced-voice-mode-is-blowing-minds/}}.
\newblock
\newblock
\shownote{[Online; last accessed January-2025]}.


\bibitem[Friedman and Kahn~Jr(1992)]%
        {friedman1992human}
\bibfield{author}{\bibinfo{person}{Batya Friedman} {and} \bibinfo{person}{Peter~H Kahn~Jr}.} \bibinfo{year}{1992}\natexlab{}.
\newblock \showarticletitle{Human agency and responsible computing: Implications for computer system design}.
\newblock \bibinfo{journal}{\emph{Journal of Systems and Software}} \bibinfo{volume}{17}, \bibinfo{number}{1} (\bibinfo{year}{1992}), \bibinfo{pages}{7--14}.
\newblock


\bibitem[Fronsdal and Lindner(2024)]%
        {fronsdal2024misr}
\bibfield{author}{\bibinfo{person}{Kai Fronsdal} {and} \bibinfo{person}{David Lindner}.} \bibinfo{year}{2024}\natexlab{}.
\newblock \showarticletitle{MISR: Measuring Instrumental Self-Reasoning in Frontier Models}.
\newblock \bibinfo{journal}{\emph{arXiv preprint arXiv:2412.03904}} (\bibinfo{year}{2024}).
\newblock


\bibitem[Gabriel et~al\mbox{.}(2024)]%
        {gabriel2024ethics}
\bibfield{author}{\bibinfo{person}{Iason Gabriel}, \bibinfo{person}{Arianna Manzini}, \bibinfo{person}{Geoff Keeling}, \bibinfo{person}{Lisa~Anne Hendricks}, \bibinfo{person}{Verena Rieser}, \bibinfo{person}{Hasan Iqbal}, \bibinfo{person}{Nenad Tomašev}, \bibinfo{person}{Ira Ktena}, \bibinfo{person}{Zachary Kenton}, \bibinfo{person}{Mikel Rodriguez}, \bibinfo{person}{Seliem El-Sayed}, \bibinfo{person}{Sasha Brown}, \bibinfo{person}{Canfer Akbulut}, \bibinfo{person}{Andrew Trask}, \bibinfo{person}{Edward Hughes}, \bibinfo{person}{A.~Stevie Bergman}, \bibinfo{person}{Renee Shelby}, \bibinfo{person}{Nahema Marchal}, \bibinfo{person}{Conor Griffin}, \bibinfo{person}{Juan Mateos-Garcia}, \bibinfo{person}{Laura Weidinger}, \bibinfo{person}{Winnie Street}, \bibinfo{person}{Benjamin Lange}, \bibinfo{person}{Alex Ingerman}, \bibinfo{person}{Alison Lentz}, \bibinfo{person}{Reed Enger}, \bibinfo{person}{Andrew Barakat}, \bibinfo{person}{Victoria Krakovna}, \bibinfo{person}{John~Oliver Siy}, \bibinfo{person}{Zeb
  Kurth-Nelson}, \bibinfo{person}{Amanda McCroskery}, \bibinfo{person}{Vijay Bolina}, \bibinfo{person}{Harry Law}, \bibinfo{person}{Murray Shanahan}, \bibinfo{person}{Lize Alberts}, \bibinfo{person}{Borja Balle}, \bibinfo{person}{Sarah de Haas}, \bibinfo{person}{Yetunde Ibitoye}, \bibinfo{person}{Allan Dafoe}, \bibinfo{person}{Beth Goldberg}, \bibinfo{person}{Sébastien Krier}, \bibinfo{person}{Alexander Reese}, \bibinfo{person}{Sims Witherspoon}, \bibinfo{person}{Will Hawkins}, \bibinfo{person}{Maribeth Rauh}, \bibinfo{person}{Don Wallace}, \bibinfo{person}{Matija Franklin}, \bibinfo{person}{Josh~A. Goldstein}, \bibinfo{person}{Joel Lehman}, \bibinfo{person}{Michael Klenk}, \bibinfo{person}{Shannon Vallor}, \bibinfo{person}{Courtney Biles}, \bibinfo{person}{Meredith~Ringel Morris}, \bibinfo{person}{Helen King}, \bibinfo{person}{Blaise~Agüera y Arcas}, \bibinfo{person}{William Isaac}, {and} \bibinfo{person}{James Manyika}.} \bibinfo{year}{2024}\natexlab{}.
\newblock \bibinfo{title}{The Ethics of Advanced AI Assistants}.
\newblock
\newblock
\showeprint[arxiv]{2404.16244}~[cs.CY]
\urldef\tempurl%
\url{https://arxiv.org/abs/2404.16244}
\showURL{%
\tempurl}


\bibitem[Gao et~al\mbox{.}(2023)]%
        {gao2023s}
\bibfield{author}{\bibinfo{person}{Chen Gao}, \bibinfo{person}{Xiaochong Lan}, \bibinfo{person}{Zhi jie Lu}, \bibinfo{person}{Jinzhu Mao}, \bibinfo{person}{Jing Piao}, \bibinfo{person}{Huandong Wang}, \bibinfo{person}{Depeng Jin}, {and} \bibinfo{person}{Yong Li}.} \bibinfo{year}{2023}\natexlab{}.
\newblock \showarticletitle{S3: Social-network Simulation System with Large Language Model-Empowered Agents}.
\newblock \bibinfo{journal}{\emph{ArXiv}}  \bibinfo{volume}{abs/2307.14984} (\bibinfo{year}{2023}).
\newblock
\urldef\tempurl%
\url{https://api.semanticscholar.org/CorpusID:260202947}
\showURL{%
\tempurl}


\bibitem[Geng et~al\mbox{.}(2025)]%
        {geng2025control}
\bibfield{author}{\bibinfo{person}{Yilin Geng}, \bibinfo{person}{Haonan Li}, \bibinfo{person}{Honglin Mu}, \bibinfo{person}{Xudong Han}, \bibinfo{person}{Timothy Baldwin}, \bibinfo{person}{Omri Abend}, \bibinfo{person}{Eduard Hovy}, {and} \bibinfo{person}{Lea Frermann}.} \bibinfo{year}{2025}\natexlab{}.
\newblock \showarticletitle{Control Illusion: The Failure of Instruction Hierarchies in Large Language Models}.
\newblock \bibinfo{journal}{\emph{arXiv preprint arXiv:2502.15851}} (\bibinfo{year}{2025}).
\newblock


\bibitem[Gerlich(2025)]%
        {gerlich2025ai}
\bibfield{author}{\bibinfo{person}{Michael Gerlich}.} \bibinfo{year}{2025}\natexlab{}.
\newblock \showarticletitle{AI Tools in Society: Impacts on Cognitive Offloading and the Future of Critical Thinking}.
\newblock \bibinfo{journal}{\emph{Societies}} \bibinfo{volume}{15}, \bibinfo{number}{1} (\bibinfo{year}{2025}), \bibinfo{pages}{6}.
\newblock


\bibitem[Giles(1979)]%
        {giles1979accommodation}
\bibfield{author}{\bibinfo{person}{H Giles}.} \bibinfo{year}{1979}\natexlab{}.
\newblock \showarticletitle{Accommodation Theory: Optimal Levels of Convergence}.
\newblock \bibinfo{journal}{\emph{Language and social psychology/University Park}} (\bibinfo{year}{1979}).
\newblock


\bibitem[Giles et~al\mbox{.}(1980)]%
        {giles1980accommodation}
\bibfield{author}{\bibinfo{person}{Howard Giles} {et~al\mbox{.}}} \bibinfo{year}{1980}\natexlab{}.
\newblock \showarticletitle{Accommodation theory: Some new directions}.
\newblock \bibinfo{journal}{\emph{York papers in Linguistics}} \bibinfo{volume}{9}, \bibinfo{number}{450} (\bibinfo{year}{1980}), \bibinfo{pages}{105--136}.
\newblock


\bibitem[Glover(2024)]%
        {huffingtonpost2024}
\bibfield{author}{\bibinfo{person}{Amy Glover}.} \bibinfo{year}{2024}\natexlab{}.
\newblock \bibinfo{title}{{I Used An AI Headshot Generator, And Here Are My Honest Thoughts}}.
\newblock \bibinfo{howpublished}{\url{https://www.huffingtonpost.co.uk/entry/ai-headshot-generator-portraitpal-review_uk_66e1ab2be4b0e13c292dcede}}.
\newblock
\newblock
\shownote{[Online; last accessed January-2025]}.


\bibitem[{Godofredo A. Vásquez}(2025)]%
        {nbc2025Meta}
\bibfield{author}{\bibinfo{person}{{Godofredo A. Vásquez}}.} \bibinfo{year}{2025}\natexlab{}.
\newblock \bibinfo{title}{{Meta removes AI character accounts after users criticize them as ‘creepy and unnecessary’}}.
\newblock \bibinfo{howpublished}{\url{https://www.nbcnews.com/tech/social-media/meta-ai-insta-shuts-character-instagram-fb-accounts-user-outcry-rcna186177}}.
\newblock
\newblock
\shownote{[Online; last accessed January-2025]}.


\bibitem[Gratch et~al\mbox{.}(2002)]%
        {gratch2002creating}
\bibfield{author}{\bibinfo{person}{Jonathan Gratch}, \bibinfo{person}{Jeff Rickel}, \bibinfo{person}{Elisabeth Andr{\'e}}, \bibinfo{person}{Justine Cassell}, \bibinfo{person}{Eric Petajan}, {and} \bibinfo{person}{Norman Badler}.} \bibinfo{year}{2002}\natexlab{}.
\newblock \showarticletitle{Creating interactive virtual humans: Some assembly required}.
\newblock \bibinfo{journal}{\emph{IEEE Intelligent systems}} \bibinfo{volume}{17}, \bibinfo{number}{4} (\bibinfo{year}{2002}), \bibinfo{pages}{54--63}.
\newblock


\bibitem[Gray et~al\mbox{.}(2023)]%
        {gray2023psychology}
\bibfield{author}{\bibinfo{person}{Kurt Gray}, \bibinfo{person}{Kai~Chi Yam}, \bibinfo{person}{Alexander~Eng Zhen’An}, \bibinfo{person}{Danica Wilbanks}, {and} \bibinfo{person}{Adam Waytz}.} \bibinfo{year}{2023}\natexlab{}.
\newblock \showarticletitle{The psychology of robots and artificial intelligence}.
\newblock \bibinfo{journal}{\emph{The handbook of social psychology}} (\bibinfo{year}{2023}).
\newblock


\bibitem[Gros et~al\mbox{.}(2022)]%
        {gros2022robots}
\bibfield{author}{\bibinfo{person}{David Gros}, \bibinfo{person}{Yu Li}, {and} \bibinfo{person}{Zhou Yu}.} \bibinfo{year}{2022}\natexlab{}.
\newblock \showarticletitle{Robots-Dont-Cry: Understanding Falsely Anthropomorphic Utterances in Dialog Systems}. In \bibinfo{booktitle}{\emph{Proceedings of the 2022 Conference on Empirical Methods in Natural Language Processing}}, \bibfield{editor}{\bibinfo{person}{Yoav Goldberg}, \bibinfo{person}{Zornitsa Kozareva}, {and} \bibinfo{person}{Yue Zhang}} (Eds.). \bibinfo{publisher}{Association for Computational Linguistics}, \bibinfo{address}{Abu Dhabi, United Arab Emirates}, \bibinfo{pages}{3266--3284}.
\newblock
\urldef\tempurl%
\url{https://doi.org/10.18653/v1/2022.emnlp-main.215}
\showDOI{\tempurl}


\bibitem[Guingrich and Graziano(2023)]%
        {guingrich2023chatbots}
\bibfield{author}{\bibinfo{person}{Rose~E Guingrich} {and} \bibinfo{person}{Michael~SA Graziano}.} \bibinfo{year}{2023}\natexlab{}.
\newblock \showarticletitle{Chatbots as social companions: How people perceive consciousness, human likeness, and social health benefits in machines}.
\newblock \bibinfo{journal}{\emph{arXiv preprint arXiv:2311.10599}} (\bibinfo{year}{2023}).
\newblock


\bibitem[Gunderson(1964)]%
        {gunderson1964imitation}
\bibfield{author}{\bibinfo{person}{Keith Gunderson}.} \bibinfo{year}{1964}\natexlab{}.
\newblock \showarticletitle{The imitation game}.
\newblock \bibinfo{journal}{\emph{Mind}} \bibinfo{volume}{73}, \bibinfo{number}{290} (\bibinfo{year}{1964}), \bibinfo{pages}{234--245}.
\newblock


\bibitem[Guo et~al\mbox{.}(2024)]%
        {guo2024investigating}
\bibfield{author}{\bibinfo{person}{Jiajing Guo}, \bibinfo{person}{Vikram Mohanty}, \bibinfo{person}{Jorge~H Piazentin~Ono}, \bibinfo{person}{Hongtao Hao}, \bibinfo{person}{Liang Gou}, {and} \bibinfo{person}{Liu Ren}.} \bibinfo{year}{2024}\natexlab{}.
\newblock \showarticletitle{Investigating Interaction Modes and User Agency in Human-LLM Collaboration for Domain-Specific Data Analysis}. In \bibinfo{booktitle}{\emph{Extended Abstracts of the CHI Conference on Human Factors in Computing Systems}}. \bibinfo{pages}{1--9}.
\newblock


\bibitem[H\"{a}m\"{a}l\"{a}inen et~al\mbox{.}(2023)]%
        {hamalainen2023evaluating}
\bibfield{author}{\bibinfo{person}{Perttu H\"{a}m\"{a}l\"{a}inen}, \bibinfo{person}{Mikke Tavast}, {and} \bibinfo{person}{Anton Kunnari}.} \bibinfo{year}{2023}\natexlab{}.
\newblock \showarticletitle{{Evaluating Large Language Models in Generating Synthetic HCI Research Data: a Case Study}}. In \bibinfo{booktitle}{\emph{Proceedings of the 2023 CHI Conference on Human Factors in Computing Systems}} (Hamburg, Germany) \emph{(\bibinfo{series}{CHI '23})}. \bibinfo{publisher}{Association for Computing Machinery}, \bibinfo{address}{New York, NY, USA}, Article \bibinfo{articleno}{433}, \bibinfo{numpages}{19}~pages.
\newblock
\showISBNx{9781450394215}
\urldef\tempurl%
\url{https://doi.org/10.1145/3544548.3580688}
\showDOI{\tempurl}


\bibitem[Hamilton et~al\mbox{.}(2021)]%
        {hamilton2021traveling}
\bibfield{author}{\bibinfo{person}{Ryan Hamilton}, \bibinfo{person}{Rosellina Ferraro}, \bibinfo{person}{Kelly~L Haws}, {and} \bibinfo{person}{Anirban Mukhopadhyay}.} \bibinfo{year}{2021}\natexlab{}.
\newblock \showarticletitle{Traveling with companions: The social customer journey}.
\newblock \bibinfo{journal}{\emph{Journal of Marketing}} \bibinfo{volume}{85}, \bibinfo{number}{1} (\bibinfo{year}{2021}), \bibinfo{pages}{68--92}.
\newblock


\bibitem[Hamilton(2023)]%
        {hamilton2023blind}
\bibfield{author}{\bibinfo{person}{Sil Hamilton}.} \bibinfo{year}{2023}\natexlab{}.
\newblock \showarticletitle{Blind judgement: Agent-based supreme court modelling with gpt}.
\newblock \bibinfo{journal}{\emph{arXiv preprint arXiv:2301.05327}} (\bibinfo{year}{2023}).
\newblock


\bibitem[Han(2021)]%
        {han2021impact}
\bibfield{author}{\bibinfo{person}{Min~Chung Han}.} \bibinfo{year}{2021}\natexlab{}.
\newblock \showarticletitle{The impact of anthropomorphism on consumers’ purchase decision in chatbot commerce}.
\newblock \bibinfo{journal}{\emph{Journal of Internet Commerce}} \bibinfo{volume}{20}, \bibinfo{number}{1} (\bibinfo{year}{2021}), \bibinfo{pages}{46--65}.
\newblock


\bibitem[Harrell and Lim(2017)]%
        {harrell2017reimagining}
\bibfield{author}{\bibinfo{person}{D~Fox Harrell} {and} \bibinfo{person}{Chong-U Lim}.} \bibinfo{year}{2017}\natexlab{}.
\newblock \showarticletitle{Reimagining the avatar dream: Modeling social identity in digital media}.
\newblock \bibinfo{journal}{\emph{Commun. ACM}} \bibinfo{volume}{60}, \bibinfo{number}{7} (\bibinfo{year}{2017}), \bibinfo{pages}{50--61}.
\newblock


\bibitem[Harrington and Egede(2023)]%
        {harrington2023trust}
\bibfield{author}{\bibinfo{person}{Christina~N. Harrington} {and} \bibinfo{person}{Lisa Egede}.} \bibinfo{year}{2023}\natexlab{}.
\newblock \showarticletitle{Trust, Comfort and Relatability: Understanding Black Older Adults’ Perceptions of Chatbot Design for Health Information Seeking}. In \bibinfo{booktitle}{\emph{Proceedings of the 2023 CHI Conference on Human Factors in Computing Systems}} (Hamburg, Germany) \emph{(\bibinfo{series}{CHI '23})}. \bibinfo{publisher}{Association for Computing Machinery}, \bibinfo{address}{New York, NY, USA}, Article \bibinfo{articleno}{120}, \bibinfo{numpages}{18}~pages.
\newblock
\showISBNx{9781450394215}
\urldef\tempurl%
\url{https://doi.org/10.1145/3544548.3580719}
\showDOI{\tempurl}


\bibitem[Hassan(2023)]%
        {hassan2023ai}
\bibfield{author}{\bibinfo{person}{Jennifer Hassan}.} \bibinfo{year}{2023}\natexlab{}.
\newblock \showarticletitle{AI is being used to give dead, missing kids a voice they didn’t ask for}.
\newblock \bibinfo{journal}{\emph{The Washington Post}} (\bibinfo{year}{2023}).
\newblock


\bibitem[Herbold et~al\mbox{.}(2024)]%
        {herbold2024large}
\bibfield{author}{\bibinfo{person}{Steffen Herbold}, \bibinfo{person}{Alexander Trautsch}, \bibinfo{person}{Zlata Kikteva}, {and} \bibinfo{person}{Annette Hautli-Janisz}.} \bibinfo{year}{2024}\natexlab{}.
\newblock \showarticletitle{Large Language Models can impersonate politicians and other public figures}.
\newblock \bibinfo{journal}{\emph{arXiv preprint arXiv:2407.12855}} (\bibinfo{year}{2024}).
\newblock


\bibitem[Hernandez et~al\mbox{.}(2023)]%
        {hernandez2023affective}
\bibfield{author}{\bibinfo{person}{Javier Hernandez}, \bibinfo{person}{Jina Suh}, \bibinfo{person}{Judith Amores}, \bibinfo{person}{Kael Rowan}, \bibinfo{person}{Gonzalo Ramos}, {and} \bibinfo{person}{Mary Czerwinski}.} \bibinfo{year}{2023}\natexlab{}.
\newblock \bibinfo{title}{Affective Conversational Agents: Understanding Expectations and Personal Influences}.  (\bibinfo{date}{October} \bibinfo{year}{2023}).
\newblock
\urldef\tempurl%
\url{https://www.microsoft.com/en-us/research/publication/affective-conversational-agents-understanding-expectations-and-personal-influences/}
\showURL{%
\tempurl}
\newblock
\shownote{ArXiv}.


\bibitem[Hidalgo et~al\mbox{.}(2021)]%
        {hidalgo2021humans}
\bibfield{author}{\bibinfo{person}{C{\'e}sar~A Hidalgo}, \bibinfo{person}{Diana Orghian}, \bibinfo{person}{Jordi~Albo Canals}, \bibinfo{person}{Filipa De~Almeida}, {and} \bibinfo{person}{Natalia Martin}.} \bibinfo{year}{2021}\natexlab{}.
\newblock \bibinfo{booktitle}{\emph{How humans judge machines}}.
\newblock \bibinfo{publisher}{MIT Press}.
\newblock


\bibitem[Hofman et~al\mbox{.}(2023)]%
        {hofman2023steroids}
\bibfield{author}{\bibinfo{person}{Jake~M Hofman}, \bibinfo{person}{Daniel~G Goldstein}, {and} \bibinfo{person}{David~M Rothschild}.} \bibinfo{year}{2023}\natexlab{}.
\newblock \showarticletitle{Steroids, Sneakers, Coach: The Spectrum of Human-AI Relationships}.
\newblock \bibinfo{journal}{\emph{Available at SSRN 4578180}} (\bibinfo{year}{2023}).
\newblock


\bibitem[Hollanek and Nowaczyk-Basi{\'n}ska(2024)]%
        {hollanek2024griefbots}
\bibfield{author}{\bibinfo{person}{Tomasz Hollanek} {and} \bibinfo{person}{Katarzyna Nowaczyk-Basi{\'n}ska}.} \bibinfo{year}{2024}\natexlab{}.
\newblock \showarticletitle{Griefbots, deadbots, postmortem avatars: On responsible applications of generative AI in the digital afterlife industry}.
\newblock \bibinfo{journal}{\emph{Philosophy \& Technology}} \bibinfo{volume}{37}, \bibinfo{number}{2} (\bibinfo{year}{2024}), \bibinfo{pages}{63}.
\newblock


\bibitem[Horvitz(2022)]%
        {horvitz2022horizon}
\bibfield{author}{\bibinfo{person}{Eric Horvitz}.} \bibinfo{year}{2022}\natexlab{}.
\newblock \showarticletitle{On the horizon: Interactive and compositional deepfakes}. In \bibinfo{booktitle}{\emph{Proceedings of the 2022 International Conference on Multimodal Interaction}}. \bibinfo{pages}{653--661}.
\newblock


\bibitem[Huang(2022)]%
        {huang2022chatbot}
\bibfield{author}{\bibinfo{person}{Michelle Huang}.} \bibinfo{year}{2022}\natexlab{}.
\newblock \bibinfo{title}{{i trained an ai chatbot on my childhood journal entries - so that i could engage in real-time dialogue with my ``inner child''}}.
\newblock \bibinfo{howpublished}{\url{https://x.com/michellehuang42/status/1597005489413713921}}.
\newblock
\newblock
\shownote{[Online; last accessed January-2025]}.


\bibitem[Huet(2016)]%
        {bloomberg2016}
\bibfield{author}{\bibinfo{person}{Ellen Huet}.} \bibinfo{year}{2016}\natexlab{}.
\newblock \bibinfo{title}{{Pushing the Boundaries of AI to Talk to the Dead}}.
\newblock \bibinfo{howpublished}{\url{https://www.bloomberg.com/news/articles/2016-10-20/pushing-the-boundaries-of-ai-to-talk-to-the-dead}}.
\newblock
\newblock
\shownote{[Online; accessed December-2024]}.


\bibitem[Hutiri et~al\mbox{.}(2024)]%
        {hutiri2024not}
\bibfield{author}{\bibinfo{person}{Wiebke Hutiri}, \bibinfo{person}{Orestis Papakyriakopoulos}, {and} \bibinfo{person}{Alice Xiang}.} \bibinfo{year}{2024}\natexlab{}.
\newblock \showarticletitle{Not My Voice! A Taxonomy of Ethical and Safety Harms of Speech Generators}. In \bibinfo{booktitle}{\emph{The 2024 ACM Conference on Fairness, Accountability, and Transparency}}. \bibinfo{pages}{359--376}.
\newblock


\bibitem[Hutson and Ratican(2023)]%
        {hutson2023life}
\bibfield{author}{\bibinfo{person}{James Hutson} {and} \bibinfo{person}{Jay Ratican}.} \bibinfo{year}{2023}\natexlab{}.
\newblock \showarticletitle{Life, death, and AI: Exploring digital necromancy in popular culture—Ethical considerations, technological limitations, and the pet cemetery conundrum}.
\newblock \bibinfo{journal}{\emph{Metaverse}} \bibinfo{volume}{4}, \bibinfo{number}{1} (\bibinfo{year}{2023}).
\newblock


\bibitem[Hwang et~al\mbox{.}(2024)]%
        {hwang202480}
\bibfield{author}{\bibinfo{person}{Angel Hsing-Chi Hwang}, \bibinfo{person}{Q~Vera Liao}, \bibinfo{person}{Su~Lin Blodgett}, \bibinfo{person}{Alexandra Olteanu}, {and} \bibinfo{person}{Adam Trischler}.} \bibinfo{year}{2024}\natexlab{}.
\newblock \showarticletitle{" It was 80\% me, 20\% AI": Seeking Authenticity in Co-Writing with Large Language Models}.
\newblock \bibinfo{journal}{\emph{arXiv preprint arXiv:2411.13032}} (\bibinfo{year}{2024}).
\newblock


\bibitem[Inkpen and Sedlins(2011)]%
        {inkpen2011me}
\bibfield{author}{\bibinfo{person}{Kori~M Inkpen} {and} \bibinfo{person}{Mara Sedlins}.} \bibinfo{year}{2011}\natexlab{}.
\newblock \showarticletitle{Me and my avatar: exploring users' comfort with avatars for workplace communication}. In \bibinfo{booktitle}{\emph{Proceedings of the ACM 2011 conference on Computer supported cooperative work}}. \bibinfo{pages}{383--386}.
\newblock


\bibitem[Jack et~al\mbox{.}(2013)]%
        {jack2013seeing}
\bibfield{author}{\bibinfo{person}{Anthony~I Jack}, \bibinfo{person}{Abigail~J Dawson}, {and} \bibinfo{person}{Megan~E Norr}.} \bibinfo{year}{2013}\natexlab{}.
\newblock \showarticletitle{Seeing human: Distinct and overlapping neural signatures associated with two forms of dehumanization}.
\newblock \bibinfo{journal}{\emph{NeuroImage}}  \bibinfo{volume}{79} (\bibinfo{year}{2013}), \bibinfo{pages}{313--328}.
\newblock


\bibitem[{Jack Kelly}(2024)]%
        {forbes2024interviewer}
\bibfield{author}{\bibinfo{person}{{Jack Kelly}}.} \bibinfo{year}{2024}\natexlab{}.
\newblock \bibinfo{title}{{Your Next Job Interview May Be With ‘Alex,’ The AI Interviewer}}.
\newblock \bibinfo{howpublished}{\url{https://www.forbes.com/sites/jackkelly/2024/05/10/your-next-job-interview-may-be-with-alex-the-ai-interviewer/}}.
\newblock
\newblock
\shownote{[Online; last accessed January-2025]}.


\bibitem[Jadhav et~al\mbox{.}(2024)]%
        {jadhav2024limitations}
\bibfield{author}{\bibinfo{person}{Suramya Jadhav}, \bibinfo{person}{Abhay Shanbhag}, \bibinfo{person}{Amogh Thakurdesai}, \bibinfo{person}{Ridhima Sinare}, {and} \bibinfo{person}{Raviraj Joshi}.} \bibinfo{year}{2024}\natexlab{}.
\newblock \showarticletitle{On Limitations of LLM as Annotator for Low Resource Languages}.
\newblock \bibinfo{journal}{\emph{arXiv preprint arXiv:2411.17637}} (\bibinfo{year}{2024}).
\newblock


\bibitem[{James Thomason}(2024)]%
        {venturebeat2024-humanizing}
\bibfield{author}{\bibinfo{person}{{James Thomason}}.} \bibinfo{year}{2024}\natexlab{}.
\newblock \bibinfo{title}{{Confronting the ethical issues of human-like AI}}.
\newblock \bibinfo{howpublished}{\url{https://venturebeat.com/ai/confronting-the-ethical-issues-of-human-like-ai/}}.
\newblock
\newblock
\shownote{[Online; last accessed January-2025]}.


\bibitem[Jargon(2025)]%
        {wsj2025AIschoolcounselor}
\bibfield{author}{\bibinfo{person}{Julie Jargon}.} \bibinfo{year}{2025}\natexlab{}.
\newblock \bibinfo{title}{{When There’s No School Counselor, There’s a Bot}}.
\newblock \bibinfo{howpublished}{\url{https://www.wsj.com/tech/ai/student-mental-health-ai-chat-bots-school-4eb1ba55}}.
\newblock
\newblock
\shownote{[Online; last accessed February-2025]}.


\bibitem[Jarrett et~al\mbox{.}(2025)]%
        {jarrett2025language}
\bibfield{author}{\bibinfo{person}{Daniel Jarrett}, \bibinfo{person}{Miruna Pislar}, \bibinfo{person}{Michiel~A Bakker}, \bibinfo{person}{Michael~Henry Tessler}, \bibinfo{person}{Raphael K{\"o}ster}, \bibinfo{person}{Jan Balaguer}, \bibinfo{person}{Romuald Elie}, \bibinfo{person}{Christopher Summerfield}, {and} \bibinfo{person}{Andrea Tacchetti}.} \bibinfo{year}{2025}\natexlab{}.
\newblock \showarticletitle{Language agents as digital representatives in collective decision-making}.
\newblock \bibinfo{journal}{\emph{arXiv preprint arXiv:2502.09369}} (\bibinfo{year}{2025}).
\newblock


\bibitem[{Jennifer Ackerman}(2022)]%
        {leaderpost2022}
\bibfield{author}{\bibinfo{person}{{Jennifer Ackerman}}.} \bibinfo{year}{2022}\natexlab{}.
\newblock \bibinfo{title}{{Regina couple says possible AI voice scam nearly cost them \$9,400}}.
\newblock \bibinfo{howpublished}{\url{https://leaderpost.com/news/local-news/regina-couple-says-possible-ai-voice-scam-nearly-cost-them-9400}}.
\newblock
\newblock
\shownote{[Online; last accessed January-2025]}.


\bibitem[Ji et~al\mbox{.}(2024)]%
        {ji-etal-2024-srap}
\bibfield{author}{\bibinfo{person}{Jiarui Ji}, \bibinfo{person}{Yang Li}, \bibinfo{person}{Hongtao Liu}, \bibinfo{person}{Zhicheng Du}, \bibinfo{person}{Zhewei Wei}, \bibinfo{person}{Qi Qi}, \bibinfo{person}{Weiran Shen}, {and} \bibinfo{person}{Yankai Lin}.} \bibinfo{year}{2024}\natexlab{}.
\newblock \showarticletitle{{SRAP}-Agent: Simulating and Optimizing Scarce Resource Allocation Policy with {LLM}-based Agent}. In \bibinfo{booktitle}{\emph{Findings of the Association for Computational Linguistics: EMNLP 2024}}, \bibfield{editor}{\bibinfo{person}{Yaser Al-Onaizan}, \bibinfo{person}{Mohit Bansal}, {and} \bibinfo{person}{Yun-Nung Chen}} (Eds.). \bibinfo{publisher}{Association for Computational Linguistics}, \bibinfo{address}{Miami, Florida, USA}, \bibinfo{pages}{267--293}.
\newblock
\urldef\tempurl%
\url{https://doi.org/10.18653/v1/2024.findings-emnlp.15}
\showDOI{\tempurl}


\bibitem[Jiang et~al\mbox{.}(2023)]%
        {jiang2023ai}
\bibfield{author}{\bibinfo{person}{Harry~H Jiang}, \bibinfo{person}{Lauren Brown}, \bibinfo{person}{Jessica Cheng}, \bibinfo{person}{Mehtab Khan}, \bibinfo{person}{Abhishek Gupta}, \bibinfo{person}{Deja Workman}, \bibinfo{person}{Alex Hanna}, \bibinfo{person}{Johnathan Flowers}, {and} \bibinfo{person}{Timnit Gebru}.} \bibinfo{year}{2023}\natexlab{}.
\newblock \showarticletitle{AI Art and its Impact on Artists}. In \bibinfo{booktitle}{\emph{Proceedings of the 2023 AAAI/ACM Conference on AI, Ethics, and Society}}. \bibinfo{pages}{363--374}.
\newblock


\bibitem[Jo et~al\mbox{.}(2024b)]%
        {jo2024harmful}
\bibfield{author}{\bibinfo{person}{Claire~Wonjeong Jo}, \bibinfo{person}{Miki Weso{\l}owska}, {and} \bibinfo{person}{Magdalena Wojcieszak}.} \bibinfo{year}{2024}\natexlab{b}.
\newblock \showarticletitle{Harmful YouTube Video Detection: A Taxonomy of Online Harm and MLLMs as Alternative Annotators}.
\newblock \bibinfo{journal}{\emph{arXiv preprint arXiv:2411.05854}} (\bibinfo{year}{2024}).
\newblock


\bibitem[Jo et~al\mbox{.}(2024a)]%
        {jo2024understanding}
\bibfield{author}{\bibinfo{person}{Eunkyung Jo}, \bibinfo{person}{Yuin Jeong}, \bibinfo{person}{SoHyun Park}, \bibinfo{person}{Daniel~A Epstein}, {and} \bibinfo{person}{Young-Ho Kim}.} \bibinfo{year}{2024}\natexlab{a}.
\newblock \showarticletitle{Understanding the Impact of Long-Term Memory on Self-Disclosure with Large Language Model-Driven Chatbots for Public Health Intervention}. In \bibinfo{booktitle}{\emph{Proceedings of the CHI Conference on Human Factors in Computing Systems}}. \bibinfo{pages}{1--21}.
\newblock


\bibitem[Jones(2024)]%
        {Jones_2024}
\bibfield{author}{\bibinfo{person}{Nicola Jones}.} \bibinfo{year}{2024}\natexlab{}.
\newblock \bibinfo{title}{Who owns your voice? Scarlett Johansson openai complaint raises questions}.
\newblock
\newblock
\urldef\tempurl%
\url{https://www.nature.com/articles/d41586-024-01578-4}
\showURL{%
\tempurl}


\bibitem[Kahn~Jr et~al\mbox{.}(2007)]%
        {kahn2007human}
\bibfield{author}{\bibinfo{person}{Peter~H Kahn~Jr}, \bibinfo{person}{Hiroshi Ishiguro}, \bibinfo{person}{Batya Friedman}, \bibinfo{person}{Takayuki Kanda}, \bibinfo{person}{Nathan~G Freier}, \bibinfo{person}{Rachel~L Severson}, {and} \bibinfo{person}{Jessica Miller}.} \bibinfo{year}{2007}\natexlab{}.
\newblock \showarticletitle{What is a human?: Toward psychological benchmarks in the field of human--robot interaction}.
\newblock \bibinfo{journal}{\emph{Interaction Studies}} \bibinfo{volume}{8}, \bibinfo{number}{3} (\bibinfo{year}{2007}), \bibinfo{pages}{363--390}.
\newblock


\bibitem[Kang and Kang(2024)]%
        {kang2024counseling}
\bibfield{author}{\bibinfo{person}{Eunbin Kang} {and} \bibinfo{person}{Youn~Ah Kang}.} \bibinfo{year}{2024}\natexlab{}.
\newblock \showarticletitle{Counseling chatbot design: The effect of anthropomorphic chatbot characteristics on user self-disclosure and companionship}.
\newblock \bibinfo{journal}{\emph{International Journal of Human--Computer Interaction}} \bibinfo{volume}{40}, \bibinfo{number}{11} (\bibinfo{year}{2024}), \bibinfo{pages}{2781--2795}.
\newblock


\bibitem[Kang et~al\mbox{.}(2024)]%
        {kang2024nadine}
\bibfield{author}{\bibinfo{person}{Hangyeol Kang}, \bibinfo{person}{Maher~Ben Moussa}, {and} \bibinfo{person}{Nadia Magnenat-Thalmann}.} \bibinfo{year}{2024}\natexlab{}.
\newblock \showarticletitle{Nadine: An LLM-driven Intelligent Social Robot with Affective Capabilities and Human-like Memory}.
\newblock \bibinfo{journal}{\emph{arXiv preprint arXiv:2405.20189}} (\bibinfo{year}{2024}).
\newblock


\bibitem[Kapania et~al\mbox{.}(2024)]%
        {kapania2024simulacrum}
\bibfield{author}{\bibinfo{person}{Shivani Kapania}, \bibinfo{person}{William Agnew}, \bibinfo{person}{Motahhare Eslami}, \bibinfo{person}{Hoda Heidari}, {and} \bibinfo{person}{Sarah Fox}.} \bibinfo{year}{2024}\natexlab{}.
\newblock \showarticletitle{'Simulacrum of Stories': Examining Large Language Models as Qualitative Research Participants}.
\newblock \bibinfo{journal}{\emph{arXiv preprint arXiv:2409.19430}} (\bibinfo{year}{2024}).
\newblock


\bibitem[{Kashmir Hill}(2025)]%
        {nytimes2025}
\bibfield{author}{\bibinfo{person}{{Kashmir Hill}}.} \bibinfo{year}{2025}\natexlab{}.
\newblock \bibinfo{title}{{She Is in Love With ChatGPT}}.
\newblock \bibinfo{howpublished}{\url{https://www.nytimes.com/2025/01/15/technology/ai-chatgpt-boyfriend-companion.html}}.
\newblock
\newblock
\shownote{[Online; last accessed January-2025]}.


\bibitem[Katz(2024)]%
        {katz2024she}
\bibfield{author}{\bibinfo{person}{Leslie Katz}.} \bibinfo{year}{2024}\natexlab{}.
\newblock \bibinfo{title}{{She Just Married Her Perfect Man: A `Calm, Caring' AI Hologram}}.
\newblock \bibinfo{howpublished}{\url{https://www.forbes.com/sites/lesliekatz/2024/11/07/shell-say-i-do---to-an-ai-hologram/}}.
\newblock
\newblock
\shownote{[Online; last accessed January 2025]}.


\bibitem[Kirk et~al\mbox{.}(2022)]%
        {kirk2022handling}
\bibfield{author}{\bibinfo{person}{Hannah Kirk}, \bibinfo{person}{Abeba Birhane}, \bibinfo{person}{Bertie Vidgen}, {and} \bibinfo{person}{Leon Derczynski}.} \bibinfo{year}{2022}\natexlab{}.
\newblock \showarticletitle{Handling and Presenting Harmful Text in NLP Research}. In \bibinfo{booktitle}{\emph{Findings of the Association for Computational Linguistics: EMNLP 2022}}. \bibinfo{pages}{497--510}.
\newblock


\bibitem[Knibbs(2024)]%
        {wired2024onlyfans}
\bibfield{author}{\bibinfo{person}{Kate Knibbs}.} \bibinfo{year}{2024}\natexlab{}.
\newblock \bibinfo{title}{{Onlyfans Models Are Using AI Impersonators to Keep Up with Their DMs}}.
\newblock \bibinfo{howpublished}{\url{https://www.wired.com/story/onlyfans-models-are-using-ai-impersonators-to-keep-up-with-their-dms/}}.
\newblock
\newblock
\shownote{[Online; last accessed January-2025]}.


\bibitem[Koebler(2024)]%
        {404media2024}
\bibfield{author}{\bibinfo{person}{Jason Koebler}.} \bibinfo{year}{2024}\natexlab{}.
\newblock \bibinfo{title}{{Meta's AI Profiles Are Indistinguishable From Terrible Spam That Took Over Facebook}}.
\newblock \bibinfo{howpublished}{\url{https://www.404media.co/metas-ai-profiles-are-indistinguishable-from-terrible-spam-that-took-over-facebook/}}.
\newblock
\newblock
\shownote{[Online; last accessed January-2025]}.


\bibitem[Kugler(2024)]%
        {kugler2024raising}
\bibfield{author}{\bibinfo{person}{Logan Kugler}.} \bibinfo{year}{2024}\natexlab{}.
\newblock \bibinfo{title}{Raising the Dead with AI}.
\newblock
\newblock


\bibitem[K{\"u}hne and Peter(2023)]%
        {kuhne2023anthropomorphism}
\bibfield{author}{\bibinfo{person}{Rinaldo K{\"u}hne} {and} \bibinfo{person}{Jochen Peter}.} \bibinfo{year}{2023}\natexlab{}.
\newblock \showarticletitle{Anthropomorphism in human--robot interactions: a multidimensional conceptualization}.
\newblock \bibinfo{journal}{\emph{Communication Theory}} \bibinfo{volume}{33}, \bibinfo{number}{1} (\bibinfo{year}{2023}), \bibinfo{pages}{42--52}.
\newblock


\bibitem[Laban(2021)]%
        {laban2021perceptions}
\bibfield{author}{\bibinfo{person}{Guy Laban}.} \bibinfo{year}{2021}\natexlab{}.
\newblock \showarticletitle{Perceptions of anthropomorphism in a chatbot dialogue: the role of animacy and intelligence}. In \bibinfo{booktitle}{\emph{Proceedings of the 9th international conference on human-agent interaction}}. \bibinfo{pages}{305--310}.
\newblock


\bibitem[Laestadius et~al\mbox{.}(2024)]%
        {laestadius2024too}
\bibfield{author}{\bibinfo{person}{Linnea Laestadius}, \bibinfo{person}{Andrea Bishop}, \bibinfo{person}{Michael Gonzalez}, \bibinfo{person}{Diana Illen{\v{c}}{\'\i}k}, {and} \bibinfo{person}{Celeste Campos-Castillo}.} \bibinfo{year}{2024}\natexlab{}.
\newblock \showarticletitle{Too human and not human enough: A grounded theory analysis of mental health harms from emotional dependence on the social chatbot Replika}.
\newblock \bibinfo{journal}{\emph{New Media \& Society}} \bibinfo{volume}{26}, \bibinfo{number}{10} (\bibinfo{year}{2024}), \bibinfo{pages}{5923--5941}.
\newblock


\bibitem[Laird and Duchi(2000)]%
        {laird2000creating}
\bibfield{author}{\bibinfo{person}{John~E Laird} {and} \bibinfo{person}{John~C Duchi}.} \bibinfo{year}{2000}\natexlab{}.
\newblock \showarticletitle{Creating human-like synthetic characters with multiple skill levels: A case study using the soar quakebot}. In \bibinfo{booktitle}{\emph{AAAI 2000 Fall Symposium Series: Simulating Human Agents}}, Vol.~\bibinfo{volume}{1001}. AAAI Press Palo Alto, CA, \bibinfo{pages}{48109--2110}.
\newblock


\bibitem[Lee et~al\mbox{.}(2023)]%
        {lee2023speculating}
\bibfield{author}{\bibinfo{person}{Patrick Yung~Kang Lee}, \bibinfo{person}{Ning~F Ma}, \bibinfo{person}{Ig-Jae Kim}, {and} \bibinfo{person}{Dongwook Yoon}.} \bibinfo{year}{2023}\natexlab{}.
\newblock \showarticletitle{Speculating on risks of AI clones to selfhood and relationships: Doppelganger-phobia, identity fragmentation, and living memories}.
\newblock \bibinfo{journal}{\emph{Proceedings of the ACM on Human-Computer Interaction}} \bibinfo{volume}{7}, \bibinfo{number}{CSCW1} (\bibinfo{year}{2023}), \bibinfo{pages}{1--28}.
\newblock


\bibitem[Lee et~al\mbox{.}(2024)]%
        {lee2024can}
\bibfield{author}{\bibinfo{person}{Sanguk Lee}, \bibinfo{person}{Tai-Quan Peng}, \bibinfo{person}{Matthew~H Goldberg}, \bibinfo{person}{Seth~A Rosenthal}, \bibinfo{person}{John~E Kotcher}, \bibinfo{person}{Edward~W Maibach}, {and} \bibinfo{person}{Anthony Leiserowitz}.} \bibinfo{year}{2024}\natexlab{}.
\newblock \showarticletitle{Can large language models estimate public opinion about global warming? An empirical assessment of algorithmic fidelity and bias}.
\newblock \bibinfo{journal}{\emph{PLOS Climate}} \bibinfo{volume}{3}, \bibinfo{number}{8} (\bibinfo{year}{2024}), \bibinfo{pages}{e0000429}.
\newblock


\bibitem[Leong et~al\mbox{.}(2024)]%
        {leong2024dittos}
\bibfield{author}{\bibinfo{person}{Joanne Leong}, \bibinfo{person}{John Tang}, \bibinfo{person}{Edward Cutrell}, \bibinfo{person}{Sasa Junuzovic}, \bibinfo{person}{Gregory~Paul Baribault}, {and} \bibinfo{person}{Kori Inkpen}.} \bibinfo{year}{2024}\natexlab{}.
\newblock \showarticletitle{Dittos: Personalized, Embodied Agents That Participate in Meetings When You Are Unavailable}.
\newblock \bibinfo{journal}{\emph{Proceedings of the ACM on Human-Computer Interaction}} \bibinfo{volume}{8}, \bibinfo{number}{CSCW2} (\bibinfo{year}{2024}), \bibinfo{pages}{1--28}.
\newblock


\bibitem[Li et~al\mbox{.}(2024a)]%
        {li2024agent}
\bibfield{author}{\bibinfo{person}{Junkai Li}, \bibinfo{person}{Siyu Wang}, \bibinfo{person}{Meng Zhang}, \bibinfo{person}{Weitao Li}, \bibinfo{person}{Yunghwei Lai}, \bibinfo{person}{Xinhui Kang}, \bibinfo{person}{Weizhi Ma}, {and} \bibinfo{person}{Yang Liu}.} \bibinfo{year}{2024}\natexlab{a}.
\newblock \showarticletitle{Agent hospital: A simulacrum of hospital with evolvable medical agents}.
\newblock \bibinfo{journal}{\emph{arXiv preprint arXiv:2405.02957}} (\bibinfo{year}{2024}).
\newblock


\bibitem[Li et~al\mbox{.}(2024b)]%
        {li2024situ}
\bibfield{author}{\bibinfo{person}{Yongming Li}, \bibinfo{person}{Hangyue Zhang}, \bibinfo{person}{Andrea~Yaoyun Cui}, \bibinfo{person}{Zisong Ma}, \bibinfo{person}{Yunpeng Song}, \bibinfo{person}{Zhongmin Cai}, {and} \bibinfo{person}{Yun Huang}.} \bibinfo{year}{2024}\natexlab{b}.
\newblock \showarticletitle{In-Situ Mode: Generative AI-Driven Characters Transforming Art Engagement Through Anthropomorphic Narratives}.
\newblock \bibinfo{journal}{\emph{arXiv preprint arXiv:2409.15769}} (\bibinfo{year}{2024}).
\newblock


\bibitem[Litt and Hargittai(2016)]%
        {litt2016imagined}
\bibfield{author}{\bibinfo{person}{Eden Litt} {and} \bibinfo{person}{Eszter Hargittai}.} \bibinfo{year}{2016}\natexlab{}.
\newblock \showarticletitle{The imagined audience on social network sites}.
\newblock \bibinfo{journal}{\emph{Social Media+ Society}} \bibinfo{volume}{2}, \bibinfo{number}{1} (\bibinfo{year}{2016}), \bibinfo{pages}{2056305116633482}.
\newblock


\bibitem[Liu et~al\mbox{.}(2024)]%
        {liu2024leveraging}
\bibfield{author}{\bibinfo{person}{Siru Liu}, \bibinfo{person}{Allison~B McCoy}, \bibinfo{person}{Aileen~P Wright}, \bibinfo{person}{Babatunde Carew}, \bibinfo{person}{Julian~Z Genkins}, \bibinfo{person}{Sean~S Huang}, \bibinfo{person}{Josh~F Peterson}, \bibinfo{person}{Bryan Steitz}, {and} \bibinfo{person}{Adam Wright}.} \bibinfo{year}{2024}\natexlab{}.
\newblock \showarticletitle{Leveraging large language models for generating responses to patient messages—a subjective analysis}.
\newblock \bibinfo{journal}{\emph{Journal of the American Medical Informatics Association}} \bibinfo{volume}{31}, \bibinfo{number}{6} (\bibinfo{year}{2024}), \bibinfo{pages}{1367--1379}.
\newblock


\bibitem[Liu et~al\mbox{.}(2023)]%
        {liu2023robots}
\bibfield{author}{\bibinfo{person}{Xiaozhen Liu}, \bibinfo{person}{Jiayuan Dong}, {and} \bibinfo{person}{Myounghoon Jeon}.} \bibinfo{year}{2023}\natexlab{}.
\newblock \showarticletitle{{Robots’“Woohoo” and “Argh” Can Enhance Users’ Emotional and Social Perceptions: An Exploratory Study on Non-lexical Vocalizations and Non-linguistic Sounds}}.
\newblock \bibinfo{journal}{\emph{ACM Transactions on Human-Robot Interaction}} \bibinfo{volume}{12}, \bibinfo{number}{4} (\bibinfo{year}{2023}), \bibinfo{pages}{1--20}.
\newblock


\bibitem[Louie et~al\mbox{.}(2024)]%
        {louie2024roleplay}
\bibfield{author}{\bibinfo{person}{Ryan Louie}, \bibinfo{person}{Ananjan Nandi}, \bibinfo{person}{William Fang}, \bibinfo{person}{Cheng Chang}, \bibinfo{person}{Emma Brunskill}, {and} \bibinfo{person}{Diyi Yang}.} \bibinfo{year}{2024}\natexlab{}.
\newblock \showarticletitle{Roleplay-doh: Enabling Domain-Experts to Create LLM-simulated Patients via Eliciting and Adhering to Principles}. In \bibinfo{booktitle}{\emph{Proceedings of the 2024 Conference on Empirical Methods in Natural Language Processing}}. \bibinfo{pages}{10570--10603}.
\newblock


\bibitem[Lu et~al\mbox{.}(2022)]%
        {lu2022subverting}
\bibfield{author}{\bibinfo{person}{Christina Lu}, \bibinfo{person}{Jackie Kay}, {and} \bibinfo{person}{Kevin McKee}.} \bibinfo{year}{2022}\natexlab{}.
\newblock \showarticletitle{Subverting machines, fluctuating identities: Re-learning human categorization}. In \bibinfo{booktitle}{\emph{Proceedings of the 2022 ACM Conference on Fairness, Accountability, and Transparency}}. \bibinfo{pages}{1005--1015}.
\newblock


\bibitem[Lucas et~al\mbox{.}(2016)]%
        {lucas2016effect}
\bibfield{author}{\bibinfo{person}{Gale Lucas}, \bibinfo{person}{Evan Szablowski}, \bibinfo{person}{Jonathan Gratch}, \bibinfo{person}{Andrew Feng}, \bibinfo{person}{Tiffany Huang}, \bibinfo{person}{Jill Boberg}, {and} \bibinfo{person}{Ari Shapiro}.} \bibinfo{year}{2016}\natexlab{}.
\newblock \showarticletitle{The effect of operating a virtual doppleganger in a 3D simulation}. In \bibinfo{booktitle}{\emph{Proceedings of the 9th International Conference on Motion in Games}}. \bibinfo{pages}{167--174}.
\newblock


\bibitem[Maeda and Quan-Haase(2024)]%
        {maeda2024human}
\bibfield{author}{\bibinfo{person}{Takuya Maeda} {and} \bibinfo{person}{Anabel Quan-Haase}.} \bibinfo{year}{2024}\natexlab{}.
\newblock \showarticletitle{When Human-AI Interactions Become Parasocial: Agency and Anthropomorphism in Affective Design}. In \bibinfo{booktitle}{\emph{The 2024 ACM Conference on Fairness, Accountability, and Transparency}}. \bibinfo{pages}{1068--1077}.
\newblock


\bibitem[Maes(1995)]%
        {maes1995artificial}
\bibfield{author}{\bibinfo{person}{Pattie Maes}.} \bibinfo{year}{1995}\natexlab{}.
\newblock \showarticletitle{Artificial life meets entertainment: lifelike autonomous agents}.
\newblock \bibinfo{journal}{\emph{Commun. ACM}} \bibinfo{volume}{38}, \bibinfo{number}{11} (\bibinfo{year}{1995}), \bibinfo{pages}{108--114}.
\newblock


\bibitem[{Maggie Harrison Dupré}(2024)]%
        {futurism2024}
\bibfield{author}{\bibinfo{person}{{Maggie Harrison Dupré}}.} \bibinfo{year}{2024}\natexlab{}.
\newblock \bibinfo{title}{{A Google-Backed AI Startup Is Hosting Chatbots Modeled After Real-Life School Shooters — and Their Victims}}.
\newblock \bibinfo{howpublished}{\url{https://futurism.com/character-ai-school-shooters-victims}}.
\newblock
\newblock
\shownote{[Online; last accessed January-2025]}.


\bibitem[Magnenat-Thalmann and Thalmann(2005)]%
        {magnenat2005virtual}
\bibfield{author}{\bibinfo{person}{Nadia Magnenat-Thalmann} {and} \bibinfo{person}{Daniel Thalmann}.} \bibinfo{year}{2005}\natexlab{}.
\newblock \showarticletitle{Virtual humans: thirty years of research, what next?}
\newblock \bibinfo{journal}{\emph{The Visual Computer}}  \bibinfo{volume}{21} (\bibinfo{year}{2005}), \bibinfo{pages}{997--1015}.
\newblock


\bibitem[Mann(2023)]%
        {businessinsider2023}
\bibfield{author}{\bibinfo{person}{Jyoti Mann}.} \bibinfo{year}{2023}\natexlab{}.
\newblock \bibinfo{title}{{An AI company brought back a feature to restore their chatbot's 'personalities' after an update separated users from their 'partners'}}.
\newblock \bibinfo{howpublished}{\url{https://www.businessinsider.com/ai-company-restoring-erotic-roleplay-chatbot-after-partners-cut-off-2023-3}}.
\newblock
\newblock
\shownote{[Online; last accessed January-2025]}.


\bibitem[Manzini et~al\mbox{.}(2024)]%
        {manzini2024code}
\bibfield{author}{\bibinfo{person}{Arianna Manzini}, \bibinfo{person}{Geoff Keeling}, \bibinfo{person}{Lize Alberts}, \bibinfo{person}{Shannon Vallor}, \bibinfo{person}{Meredith~Ringel Morris}, {and} \bibinfo{person}{Iason Gabriel}.} \bibinfo{year}{2024}\natexlab{}.
\newblock \showarticletitle{The Code That Binds Us: Navigating the Appropriateness of Human-AI Assistant Relationships}. In \bibinfo{booktitle}{\emph{Proceedings of the AAAI/ACM Conference on AI, Ethics, and Society}}, Vol.~\bibinfo{volume}{7}. \bibinfo{pages}{943--957}.
\newblock


\bibitem[Marcotte(2025)]%
        {salon2025AInurses}
\bibfield{author}{\bibinfo{person}{Amanda Marcotte}.} \bibinfo{year}{2025}\natexlab{}.
\newblock \bibinfo{title}{{"AI nurses" as "good as any doctor": RFK Jr. confirms he wants to take away people's health care}}.
\newblock \bibinfo{howpublished}{\url{https://www.salon.com/2025/01/30/ai-nurses-as-good-as-any-doctor-rfk-jr-confirms-he-wants-to-take-away-peoples-health-care/}}.
\newblock
\newblock
\shownote{[Online; last accessed February-2025]}.


\bibitem[{Maxwell Zeff}(2024)]%
        {techcrunch2024}
\bibfield{author}{\bibinfo{person}{{Maxwell Zeff}}.} \bibinfo{year}{2024}\natexlab{}.
\newblock \bibinfo{title}{{The abject weirdness of AI ads}}.
\newblock \bibinfo{howpublished}{\url{https://techcrunch.com/2024/12/03/the-abject-weirdness-of-ai-ads/}}.
\newblock
\newblock
\shownote{[Online; last accessed January-2025]}.


\bibitem[McIlroy-Young et~al\mbox{.}(2022a)]%
        {mcilroy2022mimetic}
\bibfield{author}{\bibinfo{person}{Reid McIlroy-Young}, \bibinfo{person}{Jon Kleinberg}, \bibinfo{person}{Siddhartha Sen}, \bibinfo{person}{Solon Barocas}, {and} \bibinfo{person}{Ashton Anderson}.} \bibinfo{year}{2022}\natexlab{a}.
\newblock \showarticletitle{Mimetic models: Ethical implications of ai that acts like you}. In \bibinfo{booktitle}{\emph{Proceedings of the 2022 AAAI/ACM Conference on AI, Ethics, and Society}}. \bibinfo{pages}{479--490}.
\newblock


\bibitem[McIlroy-Young et~al\mbox{.}(2022b)]%
        {mcilroy2022learning}
\bibfield{author}{\bibinfo{person}{Reid McIlroy-Young}, \bibinfo{person}{Russell Wang}, \bibinfo{person}{Siddhartha Sen}, \bibinfo{person}{Jon Kleinberg}, {and} \bibinfo{person}{Ashton Anderson}.} \bibinfo{year}{2022}\natexlab{b}.
\newblock \showarticletitle{Learning models of individual behavior in chess}. In \bibinfo{booktitle}{\emph{Proceedings of the 28th ACM SIGKDD Conference on Knowledge Discovery and Data Mining}}. \bibinfo{pages}{1253--1263}.
\newblock


\bibitem[{Megan Cerullo}(2024)]%
        {cbsnews2024-VoiceScam}
\bibfield{author}{\bibinfo{person}{{Megan Cerullo}}.} \bibinfo{year}{2024}\natexlab{}.
\newblock \bibinfo{title}{{AI voice scams are on the rise. Here's how to protect yourself}}.
\newblock \bibinfo{howpublished}{\url{https://www.cbsnews.com/news/elder-scams-family-safe-word/}}.
\newblock
\newblock
\shownote{[Online; last accessed January-2025]}.


\bibitem[Mende et~al\mbox{.}(2019)]%
        {mende2019service}
\bibfield{author}{\bibinfo{person}{Martin Mende}, \bibinfo{person}{Maura~L Scott}, \bibinfo{person}{Jenny Van~Doorn}, \bibinfo{person}{Dhruv Grewal}, {and} \bibinfo{person}{Ilana Shanks}.} \bibinfo{year}{2019}\natexlab{}.
\newblock \showarticletitle{Service robots rising: How humanoid robots influence service experiences and elicit compensatory consumer responses}.
\newblock \bibinfo{journal}{\emph{Journal of Marketing Research}} \bibinfo{volume}{56}, \bibinfo{number}{4} (\bibinfo{year}{2019}), \bibinfo{pages}{535--556}.
\newblock


\bibitem[Merriam-Webster({[n.\,d.]})]%
        {automaton}
\bibfield{author}{\bibinfo{person}{Merriam-Webster}.} \bibinfo{year}{[n.\,d.]}\natexlab{}.
\newblock \bibinfo{title}{Automaton}.
\newblock \bibinfo{howpublished}{\url{https://www.merriam-webster.com/dictionary/automaton}}.
\newblock
\newblock
\shownote{[Online; accessed December-2024]}.


\bibitem[Milani et~al\mbox{.}(2023)]%
        {milani2023navigates}
\bibfield{author}{\bibinfo{person}{Stephanie Milani}, \bibinfo{person}{Arthur Juliani}, \bibinfo{person}{Ida Momennejad}, \bibinfo{person}{Raluca Georgescu}, \bibinfo{person}{Jaroslaw Rzepecki}, \bibinfo{person}{Alison Shaw}, \bibinfo{person}{Gavin Costello}, \bibinfo{person}{Fei Fang}, \bibinfo{person}{Sam Devlin}, {and} \bibinfo{person}{Katja Hofmann}.} \bibinfo{year}{2023}\natexlab{}.
\newblock \showarticletitle{Navigates like me: Understanding how people evaluate human-like AI in video games}. In \bibinfo{booktitle}{\emph{Proceedings of the 2023 CHI Conference on Human Factors in Computing Systems}}. \bibinfo{pages}{1--18}.
\newblock


\bibitem[Mireshghallah et~al\mbox{.}(2024)]%
        {mireshghallah2024trust}
\bibfield{author}{\bibinfo{person}{Niloofar Mireshghallah}, \bibinfo{person}{Maria Antoniak}, \bibinfo{person}{Yash More}, \bibinfo{person}{Yejin Choi}, {and} \bibinfo{person}{Golnoosh Farnadi}.} \bibinfo{year}{2024}\natexlab{}.
\newblock \showarticletitle{Trust No Bot: Discovering Personal Disclosures in Human-{LLM} Conversations in the Wild}. In \bibinfo{booktitle}{\emph{First Conference on Language Modeling}}.
\newblock
\urldef\tempurl%
\url{https://openreview.net/forum?id=tIpWtMYkzU}
\showURL{%
\tempurl}


\bibitem[Mitchell et~al\mbox{.}(2025)]%
        {Mitchell2025AI}
\bibfield{author}{\bibinfo{person}{Margaret Mitchell}, \bibinfo{person}{Avijit Ghosh}, \bibinfo{person}{Sasha Luccioni}, {and} \bibinfo{person}{Giada Pistilli}.} \bibinfo{year}{2025}\natexlab{}.
\newblock \bibinfo{title}{{AI Agents Are Here. What Now?}}
\newblock \bibinfo{howpublished}{\url{https://huggingface.co/blog/ethics-soc-7}}.
\newblock
\newblock
\shownote{[Online; last accessed January-2025]}.


\bibitem[Morreale et~al\mbox{.}(2024)]%
        {morreale2024unwitting}
\bibfield{author}{\bibinfo{person}{Fabio Morreale}, \bibinfo{person}{Elham Bahmanteymouri}, \bibinfo{person}{Brent Burmester}, \bibinfo{person}{Andrew Chen}, {and} \bibinfo{person}{Michelle Thorp}.} \bibinfo{year}{2024}\natexlab{}.
\newblock \showarticletitle{The unwitting labourer: extracting humanness in AI training}.
\newblock \bibinfo{journal}{\emph{AI \& SOCIETY}} \bibinfo{volume}{39}, \bibinfo{number}{5} (\bibinfo{year}{2024}), \bibinfo{pages}{2389--2399}.
\newblock


\bibitem[Morris and Brubaker(2024)]%
        {morris2024generative}
\bibfield{author}{\bibinfo{person}{Meredith~Ringel Morris} {and} \bibinfo{person}{Jed~R Brubaker}.} \bibinfo{year}{2024}\natexlab{}.
\newblock \showarticletitle{Generative ghosts: Anticipating benefits and risks of AI afterlives}.
\newblock \bibinfo{journal}{\emph{arXiv preprint arXiv:2402.01662}} (\bibinfo{year}{2024}).
\newblock


\bibitem[Morrow(2025)]%
        {meta2025}
\bibfield{author}{\bibinfo{person}{Allison Morrow}.} \bibinfo{year}{2025}\natexlab{}.
\newblock \bibinfo{title}{{Meta scrambles to delete its own AI accounts after backlash intensifies}}.
\newblock \bibinfo{howpublished}{\url{https://www.cnn.com/2025/01/03/business/meta-ai-accounts-instagram-facebook/index.html}}.
\newblock
\newblock
\shownote{[Online; last accessed January-2025]}.


\bibitem[Mou et~al\mbox{.}(2024)]%
        {mou2024individual}
\bibfield{author}{\bibinfo{person}{Xinyi Mou}, \bibinfo{person}{Xuanwen Ding}, \bibinfo{person}{Qi He}, \bibinfo{person}{Liang Wang}, \bibinfo{person}{Jingcong Liang}, \bibinfo{person}{Xinnong Zhang}, \bibinfo{person}{Libo Sun}, \bibinfo{person}{Jiayu Lin}, \bibinfo{person}{Jie Zhou}, \bibinfo{person}{Xuanjing Huang}, {et~al\mbox{.}}} \bibinfo{year}{2024}\natexlab{}.
\newblock \showarticletitle{From Individual to Society: A Survey on Social Simulation Driven by Large Language Model-based Agents}.
\newblock \bibinfo{journal}{\emph{arXiv preprint arXiv:2412.03563}} (\bibinfo{year}{2024}).
\newblock


\bibitem[Muresan and Pohl(2019)]%
        {muresan2019chats}
\bibfield{author}{\bibinfo{person}{Andreea Muresan} {and} \bibinfo{person}{Henning Pohl}.} \bibinfo{year}{2019}\natexlab{}.
\newblock \showarticletitle{Chats with bots: Balancing imitation and engagement}. In \bibinfo{booktitle}{\emph{Extended abstracts of the 2019 CHI conference on human factors in computing systems}}. \bibinfo{pages}{1--6}.
\newblock


\bibitem[MyHeritage({[n.\,d.]})]%
        {DeepNostalgia}
\bibfield{author}{\bibinfo{person}{MyHeritage}.} \bibinfo{year}{[n.\,d.]}\natexlab{}.
\newblock \bibinfo{title}{Animate your family photos}.
\newblock \bibinfo{howpublished}{\url{https://www.myheritage.com/deep-nostalgia}}.
\newblock
\newblock
\shownote{[Online; last accessed January-2025]}.


\bibitem[Namvarpour and Razi(2024)]%
        {namvarpour2024uncovering}
\bibfield{author}{\bibinfo{person}{Mohammad Namvarpour} {and} \bibinfo{person}{Afsaneh Razi}.} \bibinfo{year}{2024}\natexlab{}.
\newblock \showarticletitle{Uncovering Contradictions in Human-AI Interactions: Lessons Learned from User Reviews of Replika}. In \bibinfo{booktitle}{\emph{Companion Publication of the 2024 Conference on Computer-Supported Cooperative Work and Social Computing}}. \bibinfo{pages}{579--586}.
\newblock


\bibitem[Nass and Lee(2000)]%
        {nass2000does}
\bibfield{author}{\bibinfo{person}{Clifford Nass} {and} \bibinfo{person}{Kwan~Min Lee}.} \bibinfo{year}{2000}\natexlab{}.
\newblock \showarticletitle{Does computer-generated speech manifest personality? An experimental test of similarity-attraction}. In \bibinfo{booktitle}{\emph{Proceedings of the SIGCHI conference on Human Factors in Computing Systems}}. \bibinfo{pages}{329--336}.
\newblock


\bibitem[Nass and Lee(2001)]%
        {nass2001does}
\bibfield{author}{\bibinfo{person}{Clifford Nass} {and} \bibinfo{person}{Kwan~Min Lee}.} \bibinfo{year}{2001}\natexlab{}.
\newblock \showarticletitle{Does computer-synthesized speech manifest personality? Experimental tests of recognition, similarity-attraction, and consistency-attraction.}
\newblock \bibinfo{journal}{\emph{Journal of experimental psychology: applied}} \bibinfo{volume}{7}, \bibinfo{number}{3} (\bibinfo{year}{2001}), \bibinfo{pages}{171}.
\newblock


\bibitem[Nass and Moon(2000)]%
        {nass2000machines}
\bibfield{author}{\bibinfo{person}{Clifford Nass} {and} \bibinfo{person}{Youngme Moon}.} \bibinfo{year}{2000}\natexlab{}.
\newblock \showarticletitle{Machines and mindlessness: Social responses to computers}.
\newblock \bibinfo{journal}{\emph{Journal of social issues}} \bibinfo{volume}{56}, \bibinfo{number}{1} (\bibinfo{year}{2000}), \bibinfo{pages}{81--103}.
\newblock


\bibitem[{Navkiran Dhaliwal}(2023)]%
        {goodereader2023}
\bibfield{author}{\bibinfo{person}{{Navkiran Dhaliwal}}.} \bibinfo{year}{2023}\natexlab{}.
\newblock \bibinfo{title}{{Audiobook Narrators and Authors Fear Apple Using Their Voices to Train AI}}.
\newblock \bibinfo{howpublished}{\url{https://goodereader.com/blog/audiobooks/audiobook-narrators-and-authors-fear-apple-using-their-voices-to-train-ai}}.
\newblock
\newblock
\shownote{[Online; last accessed January-2025]}.


\bibitem[Nguyen et~al\mbox{.}(2023)]%
        {nguyen2023chatbots}
\bibfield{author}{\bibinfo{person}{Mai Nguyen}, \bibinfo{person}{Lars-Erik Casper~Ferm}, \bibinfo{person}{Sara Quach}, \bibinfo{person}{Nicolas Pontes}, {and} \bibinfo{person}{Park Thaichon}.} \bibinfo{year}{2023}\natexlab{}.
\newblock \showarticletitle{Chatbots in frontline services and customer experience: An anthropomorphism perspective}.
\newblock \bibinfo{journal}{\emph{Psychology \& Marketing}} \bibinfo{volume}{40}, \bibinfo{number}{11} (\bibinfo{year}{2023}), \bibinfo{pages}{2201--2225}.
\newblock


\bibitem[Nicolescu and Tudorache(2022)]%
        {nicolescu2022human}
\bibfield{author}{\bibinfo{person}{Luminița Nicolescu} {and} \bibinfo{person}{Monica~Teodora Tudorache}.} \bibinfo{year}{2022}\natexlab{}.
\newblock \showarticletitle{Human-computer interaction in customer service: the experience with AI chatbots—a systematic literature review}.
\newblock \bibinfo{journal}{\emph{Electronics}} \bibinfo{volume}{11}, \bibinfo{number}{10} (\bibinfo{year}{2022}), \bibinfo{pages}{1579}.
\newblock


\bibitem[{Nitasha Tiku}(2024)]%
        {washingtonpost2024}
\bibfield{author}{\bibinfo{person}{{Nitasha Tiku}}.} \bibinfo{year}{2024}\natexlab{}.
\newblock \bibinfo{title}{{AI friendships claim to cure loneliness. Some are ending in suicide.}}
\newblock \bibinfo{howpublished}{\url{https://www.washingtonpost.com/technology/2024/12/06/ai-companion-chai-research-character-ai/}}.
\newblock
\newblock
\shownote{[Online; last accessed January-2025]}.


\bibitem[Olteanu et~al\mbox{.}(2020)]%
        {olteanu2020search}
\bibfield{author}{\bibinfo{person}{Alexandra Olteanu}, \bibinfo{person}{Fernando Diaz}, {and} \bibinfo{person}{Gabriella Kazai}.} \bibinfo{year}{2020}\natexlab{}.
\newblock \showarticletitle{When are search completion suggestions problematic?}
\newblock \bibinfo{journal}{\emph{Proceedings of the ACM on Human-Computer Interaction}} \bibinfo{volume}{4}, \bibinfo{number}{CSCW2} (\bibinfo{year}{2020}), \bibinfo{pages}{1--25}.
\newblock


\bibitem[Pan and Schwartz(2024)]%
        {pan2024multimodal}
\bibfield{author}{\bibinfo{person}{Xu Pan} {and} \bibinfo{person}{Odelia Schwartz}.} \bibinfo{year}{2024}\natexlab{}.
\newblock \showarticletitle{Multimodal AI needs active human interaction}.
\newblock \bibinfo{journal}{\emph{Nature Human Behaviour}} (\bibinfo{year}{2024}), \bibinfo{pages}{1--2}.
\newblock


\bibitem[Park et~al\mbox{.}(2024b)]%
        {park2024generative}
\bibfield{author}{\bibinfo{person}{Joon~Sung Park}, \bibinfo{person}{Carolyn~Q Zou}, \bibinfo{person}{Aaron Shaw}, \bibinfo{person}{Benjamin~Mako Hill}, \bibinfo{person}{Carrie Cai}, \bibinfo{person}{Meredith~Ringel Morris}, \bibinfo{person}{Robb Willer}, \bibinfo{person}{Percy Liang}, {and} \bibinfo{person}{Michael~S Bernstein}.} \bibinfo{year}{2024}\natexlab{b}.
\newblock \showarticletitle{Generative agent simulations of 1,000 people}.
\newblock \bibinfo{journal}{\emph{arXiv preprint arXiv:2411.10109}} (\bibinfo{year}{2024}).
\newblock


\bibitem[Park(2025)]%
        {techcrunch2025}
\bibfield{author}{\bibinfo{person}{Kate Park}.} \bibinfo{year}{2025}\natexlab{}.
\newblock \bibinfo{title}{{Nvidia backs MetAI, a Taiwanese startup that creates AI-powered digital twins}}.
\newblock \bibinfo{howpublished}{\url{https://techcrunch.com/2025/01/14/nvidia-backs-metai-a-taiwanese-startup-that-creates-ai-powered-digital-twins/}}.
\newblock
\newblock
\shownote{[Online; last accessed January-2025]}.


\bibitem[Park et~al\mbox{.}(2024a)]%
        {park2024empowering}
\bibfield{author}{\bibinfo{person}{Minju Park}, \bibinfo{person}{Sojung Kim}, \bibinfo{person}{Seunghyun Lee}, \bibinfo{person}{Soonwoo Kwon}, {and} \bibinfo{person}{Kyuseok Kim}.} \bibinfo{year}{2024}\natexlab{a}.
\newblock \showarticletitle{Empowering personalized learning through a conversation-based tutoring system with student modeling}. In \bibinfo{booktitle}{\emph{Extended Abstracts of the CHI Conference on Human Factors in Computing Systems}}. \bibinfo{pages}{1--10}.
\newblock


\bibitem[Perry(2023)]%
        {perry2023ai}
\bibfield{author}{\bibinfo{person}{Anat Perry}.} \bibinfo{year}{2023}\natexlab{}.
\newblock \showarticletitle{AI will never convey the essence of human empathy}.
\newblock \bibinfo{journal}{\emph{Nature Human Behaviour}} \bibinfo{volume}{7}, \bibinfo{number}{11} (\bibinfo{year}{2023}), \bibinfo{pages}{1808--1809}.
\newblock


\bibitem[Placani(2024)]%
        {placani2024anthropomorphism}
\bibfield{author}{\bibinfo{person}{Adriana Placani}.} \bibinfo{year}{2024}\natexlab{}.
\newblock \showarticletitle{Anthropomorphism in AI: Hype and Fallacy}.
\newblock \bibinfo{journal}{\emph{AI and Ethics}} \bibinfo{volume}{4}, \bibinfo{number}{1} (\bibinfo{date}{10} \bibinfo{year}{2024}), \bibinfo{pages}{691--698}.
\newblock
\urldef\tempurl%
\url{https://doi.org/10.1007/s43681-024-00419-4}
\showDOI{\tempurl}


\bibitem[Porra et~al\mbox{.}(2020)]%
        {porra2020can}
\bibfield{author}{\bibinfo{person}{Jaana Porra}, \bibinfo{person}{Mary Lacity}, {and} \bibinfo{person}{Michael~S Parks}.} \bibinfo{year}{2020}\natexlab{}.
\newblock \showarticletitle{Can computer based human-likeness endanger humanness?”--A philosophical and ethical perspective on digital assistants expressing feelings they can’t have}.
\newblock \bibinfo{journal}{\emph{Information Systems Frontiers}}  \bibinfo{volume}{22} (\bibinfo{year}{2020}), \bibinfo{pages}{533--547}.
\newblock


\bibitem[Prahl and Van~Swol(2021)]%
        {prahl2021out}
\bibfield{author}{\bibinfo{person}{Andrew Prahl} {and} \bibinfo{person}{Lyn~M Van~Swol}.} \bibinfo{year}{2021}\natexlab{}.
\newblock \showarticletitle{Out with the humans, in with the machines?: investigating the behavioral and psychological effects of replacing human advisors with a machine}.
\newblock \bibinfo{journal}{\emph{Human-Machine Communication}}  \bibinfo{volume}{2} (\bibinfo{year}{2021}), \bibinfo{pages}{209--234}.
\newblock


\bibitem[Price and Nicholson(2019)]%
        {price2019artificial}
\bibfield{author}{\bibinfo{person}{W Price} {and} \bibinfo{person}{II Nicholson}.} \bibinfo{year}{2019}\natexlab{}.
\newblock \showarticletitle{Artificial intelligence in the medical system: four roles for potential transformation}.
\newblock \bibinfo{journal}{\emph{Yale JL \& Tech.}}  \bibinfo{volume}{21} (\bibinfo{year}{2019}), \bibinfo{pages}{122}.
\newblock


\bibitem[Qian et~al\mbox{.}(2024)]%
        {qian-etal-2024-chatdev}
\bibfield{author}{\bibinfo{person}{Chen Qian}, \bibinfo{person}{Wei Liu}, \bibinfo{person}{Hongzhang Liu}, \bibinfo{person}{Nuo Chen}, \bibinfo{person}{Yufan Dang}, \bibinfo{person}{Jiahao Li}, \bibinfo{person}{Cheng Yang}, \bibinfo{person}{Weize Chen}, \bibinfo{person}{Yusheng Su}, \bibinfo{person}{Xin Cong}, \bibinfo{person}{Juyuan Xu}, \bibinfo{person}{Dahai Li}, \bibinfo{person}{Zhiyuan Liu}, {and} \bibinfo{person}{Maosong Sun}.} \bibinfo{year}{2024}\natexlab{}.
\newblock \showarticletitle{{C}hat{D}ev: Communicative Agents for Software Development}. In \bibinfo{booktitle}{\emph{Proceedings of the 62nd Annual Meeting of the Association for Computational Linguistics (Volume 1: Long Papers)}}, \bibfield{editor}{\bibinfo{person}{Lun-Wei Ku}, \bibinfo{person}{Andre Martins}, {and} \bibinfo{person}{Vivek Srikumar}} (Eds.). \bibinfo{publisher}{Association for Computational Linguistics}, \bibinfo{address}{Bangkok, Thailand}, \bibinfo{pages}{15174--15186}.
\newblock
\urldef\tempurl%
\url{https://doi.org/10.18653/v1/2024.acl-long.810}
\showDOI{\tempurl}


\bibitem[Rapp et~al\mbox{.}(2021)]%
        {rapp2021human}
\bibfield{author}{\bibinfo{person}{Amon Rapp}, \bibinfo{person}{Lorenzo Curti}, {and} \bibinfo{person}{Arianna Boldi}.} \bibinfo{year}{2021}\natexlab{}.
\newblock \showarticletitle{The human side of human-chatbot interaction: A systematic literature review of ten years of research on text-based chatbots}.
\newblock \bibinfo{journal}{\emph{International Journal of Human-Computer Studies}}  \bibinfo{volume}{151} (\bibinfo{year}{2021}), \bibinfo{pages}{102630}.
\newblock


\bibitem[Ren et~al\mbox{.}(2024)]%
        {ren2024emergence}
\bibfield{author}{\bibinfo{person}{Siyue Ren}, \bibinfo{person}{Zhiyao Cui}, \bibinfo{person}{Ruiqi Song}, \bibinfo{person}{Zhen Wang}, {and} \bibinfo{person}{Shuyue Hu}.} \bibinfo{year}{2024}\natexlab{}.
\newblock \showarticletitle{Emergence of Social Norms in Large Language Model-based Agent Societies}.
\newblock \bibinfo{journal}{\emph{Proceedings of the Thirty-Third International Joint Conference on Artificial Intelligence (IJCAI-24) Special Track on Human-Centred AI}} (\bibinfo{year}{2024}).
\newblock


\bibitem[Renzullo(2019)]%
        {renzullo2019anthropomorphized}
\bibfield{author}{\bibinfo{person}{Dalia Renzullo}.} \bibinfo{year}{2019}\natexlab{}.
\newblock \showarticletitle{Anthropomorphized AI as capitalist agents: the price we pay for familiarity}.
\newblock \bibinfo{journal}{\emph{Montreal AI Ethics Institute}} (\bibinfo{year}{2019}).
\newblock


\bibitem[Roberts(2022)]%
        {roberts2022you}
\bibfield{author}{\bibinfo{person}{Rebecca~J Roberts}.} \bibinfo{year}{2022}\natexlab{}.
\newblock \showarticletitle{You're Only Mostly Dead: Protecting Your Digital Ghost from Unauthorized Resurrection}.
\newblock \bibinfo{journal}{\emph{Fed. Comm. LJ}}  \bibinfo{volume}{75} (\bibinfo{year}{2022}), \bibinfo{pages}{273}.
\newblock


\bibitem[Robertson et~al\mbox{.}(2021)]%
        {robertson2021can}
\bibfield{author}{\bibinfo{person}{Ronald~E Robertson}, \bibinfo{person}{Alexandra Olteanu}, \bibinfo{person}{Fernando Diaz}, \bibinfo{person}{Milad Shokouhi}, {and} \bibinfo{person}{Peter Bailey}.} \bibinfo{year}{2021}\natexlab{}.
\newblock \showarticletitle{“I can’t reply with that”: Characterizing problematic email reply suggestions}. In \bibinfo{booktitle}{\emph{Proceedings of the 2021 CHI Conference on Human Factors in Computing Systems}}. \bibinfo{pages}{1--18}.
\newblock


\bibitem[Ropek(2024)]%
        {gizmodo2024}
\bibfield{author}{\bibinfo{person}{Lucas Ropek}.} \bibinfo{year}{2024}\natexlab{}.
\newblock \bibinfo{title}{{AI Firm’s ‘Stop Hiring Humans’ Billboard Campaign Sparks Outrage}}.
\newblock \bibinfo{howpublished}{\url{https://gizmodo.com/ai-firms-stop-hiring-humans-billboard-campaign-sparks-outrage-2000536368}}.
\newblock
\newblock
\shownote{[Online; last accessed January-2025]}.


\bibitem[Rossi et~al\mbox{.}(2018)]%
        {rossi2018socially}
\bibfield{author}{\bibinfo{person}{Silvia Rossi}, \bibinfo{person}{Mariacarla Staffa}, {and} \bibinfo{person}{Anna Tamburro}.} \bibinfo{year}{2018}\natexlab{}.
\newblock \showarticletitle{Socially assistive robot for providing recommendations: Comparing a humanoid robot with a mobile application}.
\newblock \bibinfo{journal}{\emph{International Journal of Social Robotics}}  \bibinfo{volume}{10} (\bibinfo{year}{2018}), \bibinfo{pages}{265--278}.
\newblock


\bibitem[Rothman(2018)]%
        {rothman2018right}
\bibfield{author}{\bibinfo{person}{Jennifer Rothman}.} \bibinfo{year}{2018}\natexlab{}.
\newblock \showarticletitle{The right of publicity: Privacy reimagined for New York}.
\newblock \bibinfo{journal}{\emph{Cardozo Arts \& Ent. LJ}}  \bibinfo{volume}{36} (\bibinfo{year}{2018}), \bibinfo{pages}{573}.
\newblock


\bibitem[Rothschild et~al\mbox{.}(2024)]%
        {rothschild2024opportunities}
\bibfield{author}{\bibinfo{person}{David~M Rothschild}, \bibinfo{person}{James Brand}, \bibinfo{person}{Hope Schroeder}, {and} \bibinfo{person}{Jenny Wang}.} \bibinfo{year}{2024}\natexlab{}.
\newblock \showarticletitle{Opportunities and risks of LLMs in survey research}.
\newblock \bibinfo{journal}{\emph{Available at SSRN}} (\bibinfo{year}{2024}).
\newblock


\bibitem[{Rowan Philp}(2024)]%
        {gijn2024}
\bibfield{author}{\bibinfo{person}{{Rowan Philp}}.} \bibinfo{year}{2024}\natexlab{}.
\newblock \bibinfo{title}{{How to Identify and Investigate AI Audio Deepfakes, a Major 2024 Election Threat}}.
\newblock \bibinfo{howpublished}{\url{https://gijn.org/resource/tipsheet-investigating-ai-audio-deepfakes/}}.
\newblock
\newblock
\shownote{[Online; last accessed January-2025]}.


\bibitem[{Ryan Morrison}(2023)]%
        {tomsguide2023}
\bibfield{author}{\bibinfo{person}{{Ryan Morrison}}.} \bibinfo{year}{2023}\natexlab{}.
\newblock \bibinfo{title}{{Breaking the news — AI avatars entirely replace human newscasters for the first time}}.
\newblock \bibinfo{howpublished}{\url{https://www.tomsguide.com/news/breaking-news-ai-avatars-entirely-replace-human-newscasters-for-the-first-time}}.
\newblock
\newblock
\shownote{[Online; last accessed January-2025]}.


\bibitem[Seeber et~al\mbox{.}(2020)]%
        {seeber2020collaborating}
\bibfield{author}{\bibinfo{person}{Isabella Seeber}, \bibinfo{person}{Lena Waizenegger}, \bibinfo{person}{Stefan Seidel}, \bibinfo{person}{Stefan Morana}, \bibinfo{person}{Izak Benbasat}, {and} \bibinfo{person}{Paul~Benjamin Lowry}.} \bibinfo{year}{2020}\natexlab{}.
\newblock \showarticletitle{Collaborating with technology-based autonomous agents: Issues and research opportunities}.
\newblock \bibinfo{journal}{\emph{Internet Research}} \bibinfo{volume}{30}, \bibinfo{number}{1} (\bibinfo{year}{2020}), \bibinfo{pages}{1--18}.
\newblock


\bibitem[Sellen and Horvitz(2024)]%
        {sellen2024rise}
\bibfield{author}{\bibinfo{person}{Abigail Sellen} {and} \bibinfo{person}{Eric Horvitz}.} \bibinfo{year}{2024}\natexlab{}.
\newblock \showarticletitle{The rise of the ai co-pilot: Lessons for design from aviation and beyond}.
\newblock \bibinfo{journal}{\emph{Commun. ACM}} \bibinfo{volume}{67}, \bibinfo{number}{7} (\bibinfo{year}{2024}), \bibinfo{pages}{18--23}.
\newblock


\bibitem[Shanahan(2024)]%
        {shanahan2024talking}
\bibfield{author}{\bibinfo{person}{Murray Shanahan}.} \bibinfo{year}{2024}\natexlab{}.
\newblock \showarticletitle{Talking about Large Language Models}.
\newblock \bibinfo{journal}{\emph{Commun. ACM}} \bibinfo{volume}{67}, \bibinfo{number}{2} (\bibinfo{date}{Jan.} \bibinfo{year}{2024}), \bibinfo{pages}{68–79}.
\newblock
\showISSN{0001-0782}
\urldef\tempurl%
\url{https://doi.org/10.1145/3624724}
\showDOI{\tempurl}


\bibitem[Shin et~al\mbox{.}(2019)]%
        {shin2019uncanny}
\bibfield{author}{\bibinfo{person}{Mincheol Shin}, \bibinfo{person}{Se~Jung Kim}, {and} \bibinfo{person}{Frank Biocca}.} \bibinfo{year}{2019}\natexlab{}.
\newblock \showarticletitle{The uncanny valley: No need for any further judgments when an avatar looks eerie}.
\newblock \bibinfo{journal}{\emph{Computers in Human Behavior}}  \bibinfo{volume}{94} (\bibinfo{year}{2019}), \bibinfo{pages}{100--109}.
\newblock


\bibitem[Sidner et~al\mbox{.}(2018)]%
        {sidner2018creating}
\bibfield{author}{\bibinfo{person}{Candace~L Sidner}, \bibinfo{person}{Timothy Bickmore}, \bibinfo{person}{Bahador Nooraie}, \bibinfo{person}{Charles Rich}, \bibinfo{person}{Lazlo Ring}, \bibinfo{person}{Mahni Shayganfar}, {and} \bibinfo{person}{Laura Vardoulakis}.} \bibinfo{year}{2018}\natexlab{}.
\newblock \showarticletitle{Creating new technologies for companionable agents to support isolated older adults}.
\newblock \bibinfo{journal}{\emph{ACM Transactions on Interactive Intelligent Systems (TiiS)}} \bibinfo{volume}{8}, \bibinfo{number}{3} (\bibinfo{year}{2018}), \bibinfo{pages}{1--27}.
\newblock


\bibitem[Sinatra et~al\mbox{.}(2021)]%
        {sinatra2021social}
\bibfield{author}{\bibinfo{person}{Anne~M Sinatra}, \bibinfo{person}{Kimberly~A Pollard}, \bibinfo{person}{Benjamin~T Files}, \bibinfo{person}{Ashley~H Oiknine}, \bibinfo{person}{Mark Ericson}, {and} \bibinfo{person}{Peter Khooshabeh}.} \bibinfo{year}{2021}\natexlab{}.
\newblock \showarticletitle{Social fidelity in virtual agents: Impacts on presence and learning}.
\newblock \bibinfo{journal}{\emph{Computers in Human Behavior}}  \bibinfo{volume}{114} (\bibinfo{year}{2021}), \bibinfo{pages}{106562}.
\newblock


\bibitem[Slater et~al\mbox{.}(2019)]%
        {slater2019experimental}
\bibfield{author}{\bibinfo{person}{Mel Slater}, \bibinfo{person}{Sol{\`e}ne Neyret}, \bibinfo{person}{Tania Johnston}, \bibinfo{person}{Guillermo Iruretagoyena}, \bibinfo{person}{Merc{\`e} {\'A}lvarez de la~Campa Crespo}, \bibinfo{person}{Miquel Alab{\`e}rnia-Segura}, \bibinfo{person}{Bernhard Spanlang}, {and} \bibinfo{person}{Guillem Feixas}.} \bibinfo{year}{2019}\natexlab{}.
\newblock \showarticletitle{An experimental study of a virtual reality counselling paradigm using embodied self-dialogue}.
\newblock \bibinfo{journal}{\emph{Scientific reports}} \bibinfo{volume}{9}, \bibinfo{number}{1} (\bibinfo{year}{2019}), \bibinfo{pages}{10903}.
\newblock


\bibitem[Small et~al\mbox{.}(1999)]%
        {small1999demonstration}
\bibfield{author}{\bibinfo{person}{Stephen~D Small}, \bibinfo{person}{Richard~C Wuerz}, \bibinfo{person}{Robert Simon}, \bibinfo{person}{Nathan Shapiro}, \bibinfo{person}{Alasdair Conn}, {and} \bibinfo{person}{Gary Setnik}.} \bibinfo{year}{1999}\natexlab{}.
\newblock \showarticletitle{Demonstration of high-fidelity simulation team training for emergency medicine}.
\newblock \bibinfo{journal}{\emph{Academic Emergency Medicine}} \bibinfo{volume}{6}, \bibinfo{number}{4} (\bibinfo{year}{1999}), \bibinfo{pages}{312--323}.
\newblock


\bibitem[Smith(2024)]%
        {wsj2024AI}
\bibfield{author}{\bibinfo{person}{Ray~A. Smith}.} \bibinfo{year}{2024}\natexlab{}.
\newblock \bibinfo{title}{{AI Is Starting to Threaten White-Collar Jobs. Few Industries Are Immune}}.
\newblock \bibinfo{howpublished}{\url{https://www.wsj.com/lifestyle/careers/ai-is-starting-to-threaten-white-collar-jobs-few-industries-are-immune-9cdbcb90}}.
\newblock
\newblock
\shownote{[Online; last accessed February-2025]}.


\bibitem[Song and Luximon(2020)]%
        {song2020trust}
\bibfield{author}{\bibinfo{person}{Yao Song} {and} \bibinfo{person}{Yan Luximon}.} \bibinfo{year}{2020}\natexlab{}.
\newblock \showarticletitle{Trust in AI agent: A systematic review of facial anthropomorphic trustworthiness for social robot design}.
\newblock \bibinfo{journal}{\emph{Sensors}} \bibinfo{volume}{20}, \bibinfo{number}{18} (\bibinfo{year}{2020}), \bibinfo{pages}{5087}.
\newblock


\bibitem[Spargo-Ryan(2024)]%
        {theguardian2024Job}
\bibfield{author}{\bibinfo{person}{Anna Spargo-Ryan}.} \bibinfo{year}{2024}\natexlab{}.
\newblock \bibinfo{title}{{Job hunting is demoralising enough without having my personality eviscerated by an AI interviewer}}.
\newblock \bibinfo{howpublished}{\url{https://www.theguardian.com/commentisfree/2024/oct/17/job-hunting-ai-robot-interview-personality-test-ntwnfb}}.
\newblock
\newblock
\shownote{[Online; last accessed February-2025]}.


\bibitem[Stark(2024)]%
        {stark2024animation}
\bibfield{author}{\bibinfo{person}{Luke Stark}.} \bibinfo{year}{2024}\natexlab{}.
\newblock \showarticletitle{Animation and Artificial Intelligence}. In \bibinfo{booktitle}{\emph{The 2024 ACM Conference on Fairness, Accountability, and Transparency}}. \bibinfo{pages}{1663--1671}.
\newblock


\bibitem[{Stefano Pozzebon}(2024)]%
        {cnn2024}
\bibfield{author}{\bibinfo{person}{{Stefano Pozzebon}}.} \bibinfo{year}{2024}\natexlab{}.
\newblock \bibinfo{title}{{In Venezuela, AI news anchors aren’t replacing journalists. They’re protecting them}}.
\newblock \bibinfo{howpublished}{\url{https://www.cnn.com/2024/09/18/americas/venezuela-retweets-ai-news-maduro-intl-latam/index.html}}.
\newblock
\newblock
\shownote{[Online; last accessed January-2025]}.


\bibitem[Strohmann et~al\mbox{.}(2023)]%
        {strohmann2023toward}
\bibfield{author}{\bibinfo{person}{Timo Strohmann}, \bibinfo{person}{Dominik Siemon}, \bibinfo{person}{Bijan Khosrawi-Rad}, {and} \bibinfo{person}{Susanne Robra-Bissantz}.} \bibinfo{year}{2023}\natexlab{}.
\newblock \showarticletitle{Toward a design theory for virtual companionship}.
\newblock \bibinfo{journal}{\emph{Human--Computer Interaction}} \bibinfo{volume}{38}, \bibinfo{number}{3-4} (\bibinfo{year}{2023}), \bibinfo{pages}{194--234}.
\newblock


\bibitem[Studios({[n.\,d.]})]%
        {AIStudios}
\bibfield{author}{\bibinfo{person}{AI Studios}.} \bibinfo{year}{[n.\,d.]}\natexlab{}.
\newblock \bibinfo{title}{{AI Avatars}}.
\newblock \bibinfo{howpublished}{\url{https://www.aistudios.com/ai-avatars}}.
\newblock
\newblock
\shownote{[Online; last accessed January-2025]}.


\bibitem[Suchman(2023)]%
        {suchman2023uncontroversial}
\bibfield{author}{\bibinfo{person}{Lucy Suchman}.} \bibinfo{year}{2023}\natexlab{}.
\newblock \showarticletitle{The uncontroversial ‘thingness’ of AI}.
\newblock \bibinfo{journal}{\emph{Big Data \& Society}} \bibinfo{volume}{10}, \bibinfo{number}{2} (\bibinfo{year}{2023}), \bibinfo{pages}{20539517231206794}.
\newblock


\bibitem[Suh et~al\mbox{.}(2011)]%
        {suh2011if}
\bibfield{author}{\bibinfo{person}{Kil-Soo Suh}, \bibinfo{person}{Hongki Kim}, {and} \bibinfo{person}{Eung~Kyo Suh}.} \bibinfo{year}{2011}\natexlab{}.
\newblock \showarticletitle{What if your avatar looks like you? Dual-congruity perspectives for avatar use}.
\newblock \bibinfo{journal}{\emph{MIs Quarterly}} (\bibinfo{year}{2011}), \bibinfo{pages}{711--729}.
\newblock


\bibitem[Sun et~al\mbox{.}(2024)]%
        {sun2024lawluo}
\bibfield{author}{\bibinfo{person}{Jingyun Sun}, \bibinfo{person}{Chengxiao Dai}, \bibinfo{person}{Zhongze Luo}, \bibinfo{person}{Yangbo Chang}, {and} \bibinfo{person}{Yang Li}.} \bibinfo{year}{2024}\natexlab{}.
\newblock \showarticletitle{Lawluo: A chinese law firm co-run by llm agents}.
\newblock \bibinfo{journal}{\emph{arXiv preprint arXiv:2407.16252}} (\bibinfo{year}{2024}).
\newblock


\bibitem[{Superhuman AI}(2024)]%
        {SuperhumanAI}
\bibfield{author}{\bibinfo{person}{{Superhuman AI}}.} \bibinfo{year}{2024}\natexlab{}.
\newblock \bibinfo{howpublished}{\url{https://https://superhuman.com/}}.
\newblock
\newblock
\shownote{[Online; last accessed January-2025]}.


\bibitem[Suresh et~al\mbox{.}(2024)]%
        {suresh2024participation}
\bibfield{author}{\bibinfo{person}{Harini Suresh}, \bibinfo{person}{Emily Tseng}, \bibinfo{person}{Meg Young}, \bibinfo{person}{Mary Gray}, \bibinfo{person}{Emma Pierson}, {and} \bibinfo{person}{Karen Levy}.} \bibinfo{year}{2024}\natexlab{}.
\newblock \showarticletitle{Participation in the age of foundation models}. In \bibinfo{booktitle}{\emph{The 2024 ACM Conference on Fairness, Accountability, and Transparency}}. \bibinfo{pages}{1609--1621}.
\newblock


\bibitem[Szolin et~al\mbox{.}(2023)]%
        {szolin2023exploring}
\bibfield{author}{\bibinfo{person}{Kim Szolin}, \bibinfo{person}{Daria~J Kuss}, \bibinfo{person}{Filip~M Nuyens}, {and} \bibinfo{person}{Mark~D Griffiths}.} \bibinfo{year}{2023}\natexlab{}.
\newblock \showarticletitle{Exploring the user-avatar relationship in videogames: A systematic review of the Proteus effect}.
\newblock \bibinfo{journal}{\emph{Human--Computer Interaction}} \bibinfo{volume}{38}, \bibinfo{number}{5-6} (\bibinfo{year}{2023}), \bibinfo{pages}{374--399}.
\newblock


\bibitem[{Thom Waite}(2023)]%
        {dazeddigital2023}
\bibfield{author}{\bibinfo{person}{{Thom Waite}}.} \bibinfo{year}{2023}\natexlab{}.
\newblock \bibinfo{title}{{Are we entering a new age of AI-powered narcissism}}.
\newblock \bibinfo{howpublished}{\url{https://www.dazeddigital.com/life-culture/article/60754/1/entering-a-new-age-of-ai-powered-narcissism-grimes-clone-chatbot-michelle-huang}}.
\newblock
\newblock
\shownote{[Online; last accessed January-2025]}.


\bibitem[T{\"o}rnberg et~al\mbox{.}(2023)]%
        {tornberg2023simulating}
\bibfield{author}{\bibinfo{person}{Petter T{\"o}rnberg}, \bibinfo{person}{Diliara Valeeva}, \bibinfo{person}{Justus Uitermark}, {and} \bibinfo{person}{Christopher Bail}.} \bibinfo{year}{2023}\natexlab{}.
\newblock \showarticletitle{Simulating social media using large language models to evaluate alternative news feed algorithms}.
\newblock \bibinfo{journal}{\emph{arXiv preprint arXiv:2310.05984}} (\bibinfo{year}{2023}).
\newblock


\bibitem[Troshani et~al\mbox{.}(2021)]%
        {troshani2021we}
\bibfield{author}{\bibinfo{person}{Indrit Troshani}, \bibinfo{person}{Sally Rao~Hill}, \bibinfo{person}{Claire Sherman}, {and} \bibinfo{person}{Damien Arthur}.} \bibinfo{year}{2021}\natexlab{}.
\newblock \showarticletitle{Do we trust in AI? Role of anthropomorphism and intelligence}.
\newblock \bibinfo{journal}{\emph{Journal of Computer Information Systems}} \bibinfo{volume}{61}, \bibinfo{number}{5} (\bibinfo{year}{2021}), \bibinfo{pages}{481--491}.
\newblock


\bibitem[Turkle(2007)]%
        {turkle2007authenticity}
\bibfield{author}{\bibinfo{person}{Sherry Turkle}.} \bibinfo{year}{2007}\natexlab{}.
\newblock \showarticletitle{Authenticity in the age of digital companions}.
\newblock \bibinfo{journal}{\emph{Interaction Studies}} \bibinfo{volume}{8}, \bibinfo{number}{3} (\bibinfo{year}{2007}), \bibinfo{pages}{501--517}.
\newblock
\showISSN{1572-0373}
\urldef\tempurl%
\url{https://doi.org/10.1075/is.8.3.11tur}
\showDOI{\tempurl}


\bibitem[Valenzuela et~al\mbox{.}(2024)]%
        {valenzuela2024artificial}
\bibfield{author}{\bibinfo{person}{Ana Valenzuela}, \bibinfo{person}{Stefano Puntoni}, \bibinfo{person}{Donna Hoffman}, \bibinfo{person}{Noah Castelo}, \bibinfo{person}{Julian De~Freitas}, \bibinfo{person}{Berkeley Dietvorst}, \bibinfo{person}{Christian Hildebrand}, \bibinfo{person}{Young~Eun Huh}, \bibinfo{person}{Robert Meyer}, \bibinfo{person}{Miriam~E Sweeney}, {et~al\mbox{.}}} \bibinfo{year}{2024}\natexlab{}.
\newblock \showarticletitle{How artificial intelligence constrains the human experience}.
\newblock \bibinfo{journal}{\emph{Journal of the Association for Consumer Research}} \bibinfo{volume}{9}, \bibinfo{number}{3} (\bibinfo{year}{2024}), \bibinfo{pages}{000--000}.
\newblock


\bibitem[Volante et~al\mbox{.}(2016)]%
        {volante2016effects}
\bibfield{author}{\bibinfo{person}{Matias Volante}, \bibinfo{person}{Sabarish~V Babu}, \bibinfo{person}{Himanshu Chaturvedi}, \bibinfo{person}{Nathan Newsome}, \bibinfo{person}{Elham Ebrahimi}, \bibinfo{person}{Tania Roy}, \bibinfo{person}{Shaundra~B Daily}, {and} \bibinfo{person}{Tracy Fasolino}.} \bibinfo{year}{2016}\natexlab{}.
\newblock \showarticletitle{Effects of virtual human appearance fidelity on emotion contagion in affective inter-personal simulations}.
\newblock \bibinfo{journal}{\emph{IEEE transactions on visualization and computer graphics}} \bibinfo{volume}{22}, \bibinfo{number}{4} (\bibinfo{year}{2016}), \bibinfo{pages}{1326--1335}.
\newblock


\bibitem[Vozenilek et~al\mbox{.}(2004)]%
        {vozenilek2004see}
\bibfield{author}{\bibinfo{person}{John Vozenilek}, \bibinfo{person}{J~Stephen Huff}, \bibinfo{person}{Martin Reznek}, {and} \bibinfo{person}{James~A Gordon}.} \bibinfo{year}{2004}\natexlab{}.
\newblock \showarticletitle{See one, do one, teach one: advanced technology in medical education}.
\newblock \bibinfo{journal}{\emph{Academic Emergency Medicine}} \bibinfo{volume}{11}, \bibinfo{number}{11} (\bibinfo{year}{2004}), \bibinfo{pages}{1149--1154}.
\newblock


\bibitem[Wang et~al\mbox{.}(2024a)]%
        {wang2024large}
\bibfield{author}{\bibinfo{person}{Angelina Wang}, \bibinfo{person}{Jamie Morgenstern}, {and} \bibinfo{person}{John~P. Dickerson}.} \bibinfo{year}{2024}\natexlab{a}.
\newblock \showarticletitle{Large language models should not replace human participants because they can misportray and flatten identity groups}.
\newblock
\urldef\tempurl%
\url{https://api.semanticscholar.org/CorpusID:267412455}
\showURL{%
\tempurl}


\bibitem[Wang et~al\mbox{.}(2025)]%
        {wang2024decoding}
\bibfield{author}{\bibinfo{person}{Chenxi Wang}, \bibinfo{person}{Zongfang Liu}, \bibinfo{person}{Dequan Yang}, {and} \bibinfo{person}{Xiuying Chen}.} \bibinfo{year}{2025}\natexlab{}.
\newblock \showarticletitle{Decoding Echo Chambers: {LLM}-Powered Simulations Revealing Polarization in Social Networks}. In \bibinfo{booktitle}{\emph{Proceedings of the 31st International Conference on Computational Linguistics}}, \bibfield{editor}{\bibinfo{person}{Owen Rambow}, \bibinfo{person}{Leo Wanner}, \bibinfo{person}{Marianna Apidianaki}, \bibinfo{person}{Hend Al-Khalifa}, \bibinfo{person}{Barbara~Di Eugenio}, {and} \bibinfo{person}{Steven Schockaert}} (Eds.). \bibinfo{publisher}{Association for Computational Linguistics}, \bibinfo{address}{Abu Dhabi, UAE}, \bibinfo{pages}{3913--3923}.
\newblock
\urldef\tempurl%
\url{https://aclanthology.org/2025.coling-main.264/}
\showURL{%
\tempurl}


\bibitem[Wang et~al\mbox{.}(2024b)]%
        {wang-etal-2024-rolellm}
\bibfield{author}{\bibinfo{person}{Noah Wang}, \bibinfo{person}{Z.y. Peng}, \bibinfo{person}{Haoran Que}, \bibinfo{person}{Jiaheng Liu}, \bibinfo{person}{Wangchunshu Zhou}, \bibinfo{person}{Yuhan Wu}, \bibinfo{person}{Hongcheng Guo}, \bibinfo{person}{Ruitong Gan}, \bibinfo{person}{Zehao Ni}, \bibinfo{person}{Jian Yang}, \bibinfo{person}{Man Zhang}, \bibinfo{person}{Zhaoxiang Zhang}, \bibinfo{person}{Wanli Ouyang}, \bibinfo{person}{Ke Xu}, \bibinfo{person}{Wenhao Huang}, \bibinfo{person}{Jie Fu}, {and} \bibinfo{person}{Junran Peng}.} \bibinfo{year}{2024}\natexlab{b}.
\newblock \showarticletitle{{R}ole{LLM}: Benchmarking, Eliciting, and Enhancing Role-Playing Abilities of Large Language Models}. In \bibinfo{booktitle}{\emph{Findings of the Association for Computational Linguistics: ACL 2024}}, \bibfield{editor}{\bibinfo{person}{Lun-Wei Ku}, \bibinfo{person}{Andre Martins}, {and} \bibinfo{person}{Vivek Srikumar}} (Eds.). \bibinfo{publisher}{Association for Computational Linguistics}, \bibinfo{address}{Bangkok, Thailand}, \bibinfo{pages}{14743--14777}.
\newblock
\urldef\tempurl%
\url{https://doi.org/10.18653/v1/2024.findings-acl.878}
\showDOI{\tempurl}


\bibitem[Wang et~al\mbox{.}(2021)]%
        {wang2021want}
\bibfield{author}{\bibinfo{person}{Shuohang Wang}, \bibinfo{person}{Yang Liu}, \bibinfo{person}{Yichong Xu}, \bibinfo{person}{Chenguang Zhu}, {and} \bibinfo{person}{Michael Zeng}.} \bibinfo{year}{2021}\natexlab{}.
\newblock \showarticletitle{Want To Reduce Labeling Cost? {GPT}-3 Can Help}. In \bibinfo{booktitle}{\emph{Findings of the Association for Computational Linguistics: EMNLP 2021}}, \bibfield{editor}{\bibinfo{person}{Marie-Francine Moens}, \bibinfo{person}{Xuanjing Huang}, \bibinfo{person}{Lucia Specia}, {and} \bibinfo{person}{Scott Wen-tau Yih}} (Eds.). \bibinfo{publisher}{Association for Computational Linguistics}, \bibinfo{address}{Punta Cana, Dominican Republic}, \bibinfo{pages}{4195--4205}.
\newblock
\urldef\tempurl%
\url{https://doi.org/10.18653/v1/2021.findings-emnlp.354}
\showDOI{\tempurl}


\bibitem[Weber-Guskar(2022)]%
        {weber2022reflecting}
\bibfield{author}{\bibinfo{person}{Eva Weber-Guskar}.} \bibinfo{year}{2022}\natexlab{}.
\newblock \showarticletitle{Reflecting (on) Replika}.
\newblock \bibinfo{journal}{\emph{Social Robotics and the Good Life}} (\bibinfo{year}{2022}), \bibinfo{pages}{103}.
\newblock


\bibitem[Weidinger et~al\mbox{.}(2022)]%
        {weidinger2022taxonomy}
\bibfield{author}{\bibinfo{person}{Laura Weidinger}, \bibinfo{person}{Jonathan Uesato}, \bibinfo{person}{Maribeth Rauh}, \bibinfo{person}{Conor Griffin}, \bibinfo{person}{Po-Sen Huang}, \bibinfo{person}{John Mellor}, \bibinfo{person}{Amelia Glaese}, \bibinfo{person}{Myra Cheng}, \bibinfo{person}{Borja Balle}, \bibinfo{person}{Atoosa Kasirzadeh}, \bibinfo{person}{Courtney Biles}, \bibinfo{person}{Sasha Brown}, \bibinfo{person}{Zac Kenton}, \bibinfo{person}{Will Hawkins}, \bibinfo{person}{Tom Stepleton}, \bibinfo{person}{Abeba Birhane}, \bibinfo{person}{Lisa~Anne Hendricks}, \bibinfo{person}{Laura Rimell}, \bibinfo{person}{William Isaac}, \bibinfo{person}{Julia Haas}, \bibinfo{person}{Sean Legassick}, \bibinfo{person}{Geoffrey Irving}, {and} \bibinfo{person}{Iason Gabriel}.} \bibinfo{year}{2022}\natexlab{}.
\newblock \showarticletitle{Taxonomy of Risks posed by Language Models}. In \bibinfo{booktitle}{\emph{Proceedings of the 2022 ACM Conference on Fairness, Accountability, and Transparency}} (Seoul, Republic of Korea) \emph{(\bibinfo{series}{FAccT '22})}. \bibinfo{publisher}{Association for Computing Machinery}, \bibinfo{address}{New York, NY, USA}, \bibinfo{pages}{214–229}.
\newblock
\showISBNx{9781450393522}
\urldef\tempurl%
\url{https://doi.org/10.1145/3531146.3533088}
\showDOI{\tempurl}


\bibitem[Wells-Edwards(2022)]%
        {wells2022s}
\bibfield{author}{\bibinfo{person}{Bryn Wells-Edwards}.} \bibinfo{year}{2022}\natexlab{}.
\newblock \showarticletitle{What's in a Voice? The Legal Implications of Voice Cloning}.
\newblock \bibinfo{journal}{\emph{Ariz. L. Rev.}}  \bibinfo{volume}{64} (\bibinfo{year}{2022}), \bibinfo{pages}{1213}.
\newblock


\bibitem[Whittaker et~al\mbox{.}(2021)]%
        {whittaker2021rise}
\bibfield{author}{\bibinfo{person}{Lucas Whittaker}, \bibinfo{person}{Kate Letheren}, {and} \bibinfo{person}{Rory Mulcahy}.} \bibinfo{year}{2021}\natexlab{}.
\newblock \showarticletitle{The rise of deepfakes: A conceptual framework and research agenda for marketing}.
\newblock \bibinfo{journal}{\emph{Australasian Marketing Journal}} \bibinfo{volume}{29}, \bibinfo{number}{3} (\bibinfo{year}{2021}), \bibinfo{pages}{204--214}.
\newblock


\bibitem[Widder et~al\mbox{.}(2022)]%
        {widder2022limits}
\bibfield{author}{\bibinfo{person}{David~Gray Widder}, \bibinfo{person}{Dawn Nafus}, \bibinfo{person}{Laura Dabbish}, {and} \bibinfo{person}{James Herbsleb}.} \bibinfo{year}{2022}\natexlab{}.
\newblock \showarticletitle{Limits and possibilities for “Ethical AI” in open source: A study of deepfakes}. In \bibinfo{booktitle}{\emph{Proceedings of the 2022 ACM Conference on Fairness, Accountability, and Transparency}}. \bibinfo{pages}{2035--2046}.
\newblock


\bibitem[Wilder et~al\mbox{.}(2021)]%
        {wilder2021learning}
\bibfield{author}{\bibinfo{person}{Bryan Wilder}, \bibinfo{person}{Eric Horvitz}, {and} \bibinfo{person}{Ece Kamar}.} \bibinfo{year}{2021}\natexlab{}.
\newblock \showarticletitle{Learning to complement humans}. In \bibinfo{booktitle}{\emph{Proceedings of the Twenty-Ninth International Conference on International Joint Conferences on Artificial Intelligence}}. \bibinfo{pages}{1526--1533}.
\newblock


\bibitem[{Will Knight}(2022)]%
        {wired2022}
\bibfield{author}{\bibinfo{person}{{Will Knight}}.} \bibinfo{year}{2022}\natexlab{}.
\newblock \bibinfo{title}{{Algorithms Can Now Mimic Any Artist. Some Artists Hate It}}.
\newblock \bibinfo{howpublished}{\url{https://www.wired.com/story/artists-rage-against-machines-that-mimic-their-work/}}.
\newblock
\newblock
\shownote{[Online; last accessed January-2025]}.


\bibitem[Winkle et~al\mbox{.}(2021)]%
        {winkle2021assessing}
\bibfield{author}{\bibinfo{person}{Katie Winkle}, \bibinfo{person}{Praminda Caleb-Solly}, \bibinfo{person}{Ute Leonards}, \bibinfo{person}{Ailie Turton}, {and} \bibinfo{person}{Paul Bremner}.} \bibinfo{year}{2021}\natexlab{}.
\newblock \showarticletitle{Assessing and addressing ethical risk from anthropomorphism and deception in socially assistive robots}. In \bibinfo{booktitle}{\emph{Proceedings of the 2021 ACM/IEEE International Conference on Human-Robot Interaction}}. \bibinfo{pages}{101--109}.
\newblock


\bibitem[Wong(2025)]%
        {latimes2025AIdarkthoughts}
\bibfield{author}{\bibinfo{person}{Queenie Wong}.} \bibinfo{year}{2025}\natexlab{}.
\newblock \bibinfo{title}{{Teens are spilling dark thoughts to AI chatbots. Who’s to blame when something goes wrong?}}
\newblock \bibinfo{howpublished}{\url{https://www.latimes.com/business/story/2025-02-25/teens-are-spilling-dark-thoughts-to-ai-chatbots-whos-to-blame-when-something-goes-wrong}}.
\newblock
\newblock
\shownote{[Online; last accessed February-2025]}.


\bibitem[Wu et~al\mbox{.}(2023a)]%
        {wu2023deep}
\bibfield{author}{\bibinfo{person}{Min Wu}, \bibinfo{person}{Nanxi Wang}, {and} \bibinfo{person}{Kum~Fai Yuen}.} \bibinfo{year}{2023}\natexlab{a}.
\newblock \showarticletitle{Deep versus superficial anthropomorphism: Exploring their effects on human trust in shared autonomous vehicles}.
\newblock \bibinfo{journal}{\emph{Computers in Human Behavior}}  \bibinfo{volume}{141} (\bibinfo{year}{2023}), \bibinfo{pages}{107614}.
\newblock


\bibitem[Wu et~al\mbox{.}(2023b)]%
        {wu2023llms}
\bibfield{author}{\bibinfo{person}{Tongshuang Wu}, \bibinfo{person}{Haiyi Zhu}, \bibinfo{person}{Maya Albayrak}, \bibinfo{person}{Alexis Axon}, \bibinfo{person}{Amanda Bertsch}, \bibinfo{person}{Wenxing Deng}, \bibinfo{person}{Ziqi Ding}, \bibinfo{person}{Bill Guo}, \bibinfo{person}{Sireesh Gururaja}, \bibinfo{person}{Tzu-Sheng Kuo}, {et~al\mbox{.}}} \bibinfo{year}{2023}\natexlab{b}.
\newblock \showarticletitle{Llms as workers in human-computational algorithms? replicating crowdsourcing pipelines with llms}.
\newblock \bibinfo{journal}{\emph{arXiv preprint arXiv:2307.10168}} (\bibinfo{year}{2023}).
\newblock


\bibitem[Xiao et~al\mbox{.}(2023)]%
        {xiao2023simulating}
\bibfield{author}{\bibinfo{person}{Bushi Xiao}, \bibinfo{person}{Ziyuan Yin}, {and} \bibinfo{person}{Zixuan Shan}.} \bibinfo{year}{2023}\natexlab{}.
\newblock \showarticletitle{Simulating public administration crisis: A novel generative agent-based simulation system to lower technology barriers in social science research}.
\newblock \bibinfo{journal}{\emph{arXiv preprint arXiv:2311.06957}} (\bibinfo{year}{2023}).
\newblock


\bibitem[Yee and Bailenson(2007)]%
        {yee2007proteus}
\bibfield{author}{\bibinfo{person}{Nick Yee} {and} \bibinfo{person}{Jeremy Bailenson}.} \bibinfo{year}{2007}\natexlab{}.
\newblock \showarticletitle{The Proteus effect: The effect of transformed self-representation on behavior}.
\newblock \bibinfo{journal}{\emph{Human communication research}} \bibinfo{volume}{33}, \bibinfo{number}{3} (\bibinfo{year}{2007}), \bibinfo{pages}{271--290}.
\newblock


\bibitem[Young et~al\mbox{.}(2024)]%
        {young2024participation}
\bibfield{author}{\bibinfo{person}{Meg Young}, \bibinfo{person}{Upol Ehsan}, \bibinfo{person}{Ranjit Singh}, \bibinfo{person}{Emnet Tafesse}, \bibinfo{person}{Michele Gilman}, \bibinfo{person}{Christina Harrington}, {and} \bibinfo{person}{Jacob Metcalf}.} \bibinfo{year}{2024}\natexlab{}.
\newblock \showarticletitle{Participation versus scale: Tensions in the practical demands on participatory AI}.
\newblock \bibinfo{journal}{\emph{First Monday}} (\bibinfo{year}{2024}).
\newblock


\bibitem[Yu et~al\mbox{.}(2024)]%
        {yu2024researchtown}
\bibfield{author}{\bibinfo{person}{Haofei Yu}, \bibinfo{person}{Zhaochen Hong}, \bibinfo{person}{Zirui Cheng}, \bibinfo{person}{Kunlun Zhu}, \bibinfo{person}{Keyang Xuan}, \bibinfo{person}{Jinwei Yao}, \bibinfo{person}{Tao Feng}, {and} \bibinfo{person}{Jiaxuan You}.} \bibinfo{year}{2024}\natexlab{}.
\newblock \showarticletitle{ResearchTown: Simulator of Human Research Community}.
\newblock \bibinfo{journal}{\emph{arXiv preprint arXiv:2412.17767}} (\bibinfo{year}{2024}).
\newblock


\bibitem[Zhan et~al\mbox{.}(2024)]%
        {zhan2024healthcare}
\bibfield{author}{\bibinfo{person}{Xiao Zhan}, \bibinfo{person}{Noura Abdi}, \bibinfo{person}{William Seymour}, {and} \bibinfo{person}{Jose Such}.} \bibinfo{year}{2024}\natexlab{}.
\newblock \showarticletitle{Healthcare Voice AI Assistants: Factors Influencing Trust and Intention to Use}.
\newblock \bibinfo{journal}{\emph{Proceedings of the ACM on Human-Computer Interaction}} \bibinfo{volume}{8}, \bibinfo{number}{CSCW1} (\bibinfo{year}{2024}), \bibinfo{pages}{1--37}.
\newblock


\bibitem[Zhang et~al\mbox{.}(2024a)]%
        {zhang2024my}
\bibfield{author}{\bibinfo{person}{Renwen Zhang}, \bibinfo{person}{Han Li}, \bibinfo{person}{Han Meng}, \bibinfo{person}{Jinyuan Zhan}, \bibinfo{person}{Hongyuan Gan}, {and} \bibinfo{person}{Yi-Chieh Lee}.} \bibinfo{year}{2024}\natexlab{a}.
\newblock \showarticletitle{"My Replika Cheated on Me and She Liked It": A Taxonomy of Algorithmic Harms in Human-AI Relationships}.
\newblock \bibinfo{journal}{\emph{arXiv preprint arXiv:2410.20130}} (\bibinfo{year}{2024}).
\newblock


\bibitem[Zhang et~al\mbox{.}(2021)]%
        {zhang2021ideal}
\bibfield{author}{\bibinfo{person}{Rui Zhang}, \bibinfo{person}{Nathan~J McNeese}, \bibinfo{person}{Guo Freeman}, {and} \bibinfo{person}{Geoff Musick}.} \bibinfo{year}{2021}\natexlab{}.
\newblock \showarticletitle{" An ideal human" expectations of AI teammates in human-AI teaming}.
\newblock \bibinfo{journal}{\emph{Proceedings of the ACM on Human-Computer Interaction}} \bibinfo{volume}{4}, \bibinfo{number}{CSCW3} (\bibinfo{year}{2021}), \bibinfo{pages}{1--25}.
\newblock


\bibitem[Zhang et~al\mbox{.}(2016)]%
        {zhang2016data}
\bibfield{author}{\bibinfo{person}{Xiang Zhang}, \bibinfo{person}{Hans-Frederick Brown}, {and} \bibinfo{person}{Anil Shankar}.} \bibinfo{year}{2016}\natexlab{}.
\newblock \showarticletitle{Data-driven personas: Constructing archetypal users with clickstreams and user telemetry}. In \bibinfo{booktitle}{\emph{Proceedings of the 2016 CHI conference on human factors in computing systems}}. \bibinfo{pages}{5350--5359}.
\newblock


\bibitem[Zhang et~al\mbox{.}(2024b)]%
        {zhang2024electionsim}
\bibfield{author}{\bibinfo{person}{Xinnong Zhang}, \bibinfo{person}{Jiayu Lin}, \bibinfo{person}{Libo Sun}, \bibinfo{person}{Weihong Qi}, \bibinfo{person}{Yihang Yang}, \bibinfo{person}{Yue Chen}, \bibinfo{person}{Hanjia Lyu}, \bibinfo{person}{Xinyi Mou}, \bibinfo{person}{Siming Chen}, \bibinfo{person}{Jiebo Luo}, {et~al\mbox{.}}} \bibinfo{year}{2024}\natexlab{b}.
\newblock \showarticletitle{Electionsim: Massive population election simulation powered by large language model driven agents}.
\newblock \bibinfo{journal}{\emph{arXiv preprint arXiv:2410.20746}} (\bibinfo{year}{2024}).
\newblock


\bibitem[Zheng and Huang(2023)]%
        {zheng2023self}
\bibfield{author}{\bibinfo{person}{Qingxiao Zheng} {and} \bibinfo{person}{Yun Huang}.} \bibinfo{year}{2023}\natexlab{}.
\newblock \showarticletitle{The Self 2.0: How AI-Enhanced Self-Clones Transform Self-Perception and Improve Presentation Skills}.
\newblock \bibinfo{journal}{\emph{arXiv preprint arXiv:2310.15112}} (\bibinfo{year}{2023}).
\newblock


\bibitem[Zou and Topol(2025)]%
        {zou2025rise}
\bibfield{author}{\bibinfo{person}{James Zou} {and} \bibinfo{person}{Eric~J Topol}.} \bibinfo{year}{2025}\natexlab{}.
\newblock \showarticletitle{The rise of agentic AI teammates in medicine}.
\newblock \bibinfo{journal}{\emph{The Lancet}} \bibinfo{volume}{405}, \bibinfo{number}{10477} (\bibinfo{year}{2025}), \bibinfo{pages}{457}.
\newblock


\end{thebibliography}
\end{document}